\documentclass[preprintnumbers,amsmath,amssymb,floatfix,10pt,prd,onecolumn,superscriptaddress,nofootinbib]{revtex4}
\usepackage{latexsym}
\usepackage{epsfig}
\usepackage{epstopdf}
\usepackage{graphicx}
\usepackage{amssymb}
\usepackage{amsmath}
\usepackage{amsfonts}
\usepackage{subfigure}
\usepackage{dcolumn}
\usepackage{bm}
\usepackage{color}
\usepackage{comment}
\usepackage[utf8]{inputenc}
\usepackage[dvipsnames]{xcolor}
\usepackage{hyperref}
\hypersetup{colorlinks=true,linkcolor=blue,citecolor=red,urlcolor=magenta}
\usepackage{mathtools}

\begin{document}
\title{\bf Influence of plasma on the optical appearance of spinning black hole in Kalb-Ramond gravity and its Existence around M87* and Sgr A*}
\author{Muhammad Ali Raza}
\email{maliraza01234@gmail.com}\affiliation{Department of Mathematics, COMSATS University Islamabad, Lahore Campus, Lahore, Pakistan}
\author{M. Zubair}
\email{mzubairkk@gmail.com;drmzubair@cuilahore.edu.pk}\affiliation{Department of Mathematics, COMSATS University Islamabad, Lahore Campus, Lahore, Pakistan}
\author{Eiman Maqsood}
\email{eiman1482000@gmail.com}\affiliation{Department of Mathematics, COMSATS University Islamabad, Lahore Campus, Lahore, Pakistan}

\begin{abstract}
The visible universe is filled with different types of plasma media in the form of stars, nebulas and other forms of excited gases. These matter fields have a high influence on the gravity and are likely to be present around the black holes due to the immense gravitational attraction. Since a plasma medium affects the speed of light, therefore we investigated the null geodesics and various optical features around the rotating black hole in Kalb-Ramond gravity immersed in plasma medium. Various plasma distributions are considered to develop a comparative study for their influence on unstable null circular orbits, shadows and evaporation rate of the black hole in the presence of a plasma medium. Moreover, the shadow results are also compared with Event Horizon Telescope data for M78* and Sgr A* in order to estimate the parametric bounds for which the rotating black hole in Kalb-Ramond gravity is considered either M87* or Sgr A* under the different values of plasma parameters. From this analysis, we also found the distribution of plasma that has a significant impact on the above mentioned features and is most likely to be present around M87* and Sgr A*.\\
\textbf{Keywords:} Black hole, Kalb-Ramond field, plasma, shadow, M87*, Sgr A*.
\end{abstract}
\maketitle
\date{\today}

\section{Introduction}
According to Newton's concept, the gravity being a force of attraction is instantaneous and eventually becomes weaker as the separation between the two objects increases. This concept did not support the cosmic catastrophe problem, that is, the Earth would leave its orbit immediately as the Sun vanishes. This implies that a signal traveled to Earth from the Sun faster than the cosmic speed. To answer such complicated mysteries of the universe, Albert Einstein laid the foundation of a new concept of gravity, in 1915, based on the concept of spacetime curvature known as General Relativity (GR) \cite{1,2,3,4}. It describes how the force of gravity actually works. Soon after this discovery, Schwarzschild \cite{5} proposed the pioneering solution of the Einstein field equations in the most ideal scenario resulting into a black hole (BH) known as Schwarzschild BH. Later, various other BH solutions were presented based on different characteristics and properties such as Reissner-Nordstr\"{o}m \cite{6,7}, Kerr \cite{8} and Kerr-Newman \cite{9,10} BHs being the most fundamental solutions.

Besides the fundamental background of the theory, another group of scientists was looking to unify gravity with the other subatomic fundamental forces of nature. In this context, Kaluza \cite{11} and Klein \cite{12,13} proposed Kaluza-Klein theory that describes the unification of the electromagnetic and gravitational theories based on 5D structure with quantum interpretation. Over the decades, several attempt were made by the scientists in order to construct a theory based on the principles of GR and quantum mechanics. Several theories have been proposed, however the technological deficiencies restrict us to verify some of these theories. However, we are still capable enough to use the advancement of the technology to a certain extent. In this regard, GR has been tested observationally at high resolutions and obtained great precision \cite{14,15}. For the theoretical description of the results, it was required to construct the modified and alternative gravitational theories. A modified theory can be constructed by modifying the Einstein-Hilbert action that can be accomplished by introducing the Kalb-Ramond (KR) field as a possible example \cite{16}. Since, the hybridization of a superstring and a bosonic string generates a closed string that gives rise to heterotic string theory. Hence, the KR field, being inherently a quantum field, exhibits a strong connection to the excitation of closed strings within the framework of heterotic string theory \cite{17}. The KR field is described by an antisymmetric tensor of rank two whose vacuum expectation value (VEV) remains non-zero. Spontaneous Lorentz symmetry breaking (LSB) manifests as a consequence of a nonminimal coupling between gravity and a non-zero VEV \cite{18}. Khodadi \cite{19} considered a BH with spin in Einstein-bumblebee gravity for which he studied the effect of the spontaneous LSB on superradiance scattering and instability. He found the weakening of superradiance scattering for scalar wave with low frequencies and the positive values of LSB parameter. However, for the negative values, the scattering becomes stronger. Moreover, Khodadi \cite{20} also investigated that LSB parameter gives rise to a BH with fast rotation that is distinct from Kerr BH. For this particular BH, the investigation focused on analyzing the energy extraction arising from magnetic reconnection occurring within the ergoregion. He found when the BH is surrounded by a plasma region with a weak magnetization, the energy extraction due to magnetic reconnection increases for negative values of LSB parameter. Khodadi et al. \cite{21} explored the rate of energy deposition for gamma-rays induced by the annihilation of neutrino pairs for a slowly rotating BH, with modifications due to spontaneous LSB, specifically considering the plane passing through the equator. It was found that the energy deposition rate is increased for positive values of LSB parameter. The cosmological aspects of LSB has also been investigated, see Refs. \cite{22,23}. Some properties are deduced for the KR field such as a tensor of third rank which is antisymmetric and is proposed as the source of spacetime torsion \cite{24}, intrinsic angular momentum and different massive structures in the universe arising from the topological defects \cite{25}. The KR field has also been considered to study its gravitational properties and its impact on particles \cite{26,27,28,29,30}. Recently, Lessa et al. \cite{29} developed a metric described as power law hairy BH in KR gravity that encompasses some interesting properties and reduces to fundamental BHs, that is, Schwarzschild and Reissner-Nordstr\"{o}m BHs under various restrictions of the BH parameters. Later, Kumar et al. \cite{30} converted this static hairy BH into its rotating counterpart and investigated its shadow and strong lensing. The coupling of gravity with KR field may be useful to obtain some desired and feasible quantum field theoretic results in curved spacetime, quantum geometrodynamics and quantum gravity.

It remains a fundamental question that what do these BHs appear to a human eye. Such a visual image of a BH is actually not the image of the BH itself, instead it is a 2D silhouette known as shadow of the BH. The photon sphere is made up of photon orbits that are circular in which the photons are captured in the stable and unstable orbits. In the unstable orbits, the light particles cannot reside for a long time and thus scatter away or fall into the horizon generating a bright circular ring or a dark 2D image, respectively \cite{31}. Some pioneering and early shadow studies are given in Refs. \cite{32,33,34,35,36}. Later, various BH solutions were considered to calculate their shadows, see \cite{37,38,39,40,41,42,43,44,45,46,47,48,49}. Sharif and Iftikhar \cite{50} considered a noncommutative rotating BH comprising a charge. They found that the shadow deviates from a circular shape with decrease in the noncommutative charge. Lee et al. \cite{51} considered the anisotropic matter surrounding a rotating BH for which the shadows have been calculated. It was found that the shadow observables are highly influenced by the anisotropy of the matter. Amarilla and Eiroa \cite{52} investigated the shadows of a braneworld BH in the Randall-Sundrum model. Their study revealed that the shadow undergoes significant distortion due to the influence of angular momentum and terms associated with tidal charge.

To probe the physical reality of BHs, the collaborative efforts of the Event Horizon Telescope (EHT) have yielded valuable observational data. This dataset encompasses shadow images capturing M87* \cite{53,54,55,56,57,58} at the galactic center of Messier 87 and Sgr A* \cite{59} at the galactic center of the Milky Way, providing essential insights into the existence and characteristics of BHs. Utilization of the EHT observational data, scientists have undertaken studies to compare theoretical BH shadow sizes with those observed for M87* and Sgr A*. This analysis enables us to determine those values of the BH parameters for which the shadow size lies in agreement with that of M87* and Sgr A*. Consequently, the theoretically considered BH may be regarded as M87* or Sgr A*. For details, we refer to \cite{60,61,62,63,64,65}.

A plasma is the fourth fundamental state of matter that comprises charged particles in the form of electrons or ions. The largest amount of ordinary matter in the visible universe is plasma that mostly exists in stars, intracluster and intergalactic media. Therefore, a BH may also be surrounded by a material media such as a plasma which is observationally proposed in the Ref. \cite{66}. Therefore, in the context of shadows and null geodesics around a BH, if the BH is surrounded by a plasma medium, it may affect the motion of light particles as compared to that in the non-plasma medium. It is quite certain that a material media such as plasma lowers the speed of light due to which the photons do not travel along the null geodesics and as a result the shadows get affected. The speed of light in a plasma is dependent on its electron frequency and hence the index of refraction. The study of light propagation in plasma having no pressure but is magnetized, is governed by the Hamiltonian formalism that was derived precisely by Breuer and Ehlers \cite{67,68}. Later, Perlick \cite{69} derived the Hamiltonian formulation for a rather simpler case of a plasma without pressure and magnetization. Perlick et al. \cite{70} also explored the shadow of static and spherically symmetric BHs immersed in a plasma medium. Perlick and Tsupko \cite{71} applied the Hamiltonian and Hamilton-Jacobi formulations to determine the geodesic equations for studying the shadow of Kerr BH in plasma. They considered several cases of plasma distributions and proposed a separating function in $r$ and $\theta$. Recently, various BH solutions have been considered to investigate their shadows in the presence of plasma, see Refs. \cite{72,73,74,75,76,77}.

Since, the composition of a plasma is mostly excited gaseous matter and charged particles such as electrons. Therefore, the quantum mechanical principles may apply in understanding various phenomena and the interaction of these particles may also give rise to quantum aspects at the subatomic regime. For instance, an electron being an elementary particle is an element in the standard model of physics, obeys various principles of quantum field theory. Therefore, it is quite certain that a plasma would also follow such quantum theoretic principles. Since, due to quantum nature of KR field, a BH generated in KR gravity may give rise to certain quantum effects in the nearby curved spacetime. Therefore, such a BH when surrounded by a plasma is more likely to give rise to quantum aspects in its vicinity and it is quite possible to give rise to new physics at the quantum and subatomic level. Therefore, with this motivation, we consider the rotating BH in KR gravity that is surrounded by plasma described by different distribution functions and investigate the optical features and the impact of plasma on the photon motion.

The manuscript is divided into the sections as: The following section comprise the action of the theory and BH metrics. We also discussed the horizon structure briefly. The third section is related to the null geodesics and effective potential. This section is further divided into three subsections corresponding to the plasma distribution functions. Moreover, the first subsection has further two subcases. The fourth section is related to the shadows and is also divided into further subsections corresponding to the cases discussed in the third section. Similarly, corresponding to the plasma cases, the distortion and the rate of emitted energy is discussed in fifth section. The comparison of shadow sizes with observational EHT data is given in sixth section and finally in seventh section, we conclude the manuscript. Note that throughout the manuscript, we consider $G=c=M=1$, where $G$, $c$ and $M$ are Newton’s constant, speed of light and mass of BH, respectively.

\section{The Black Holes in Kalb-Ramond Gravity}
The mathematical formulation representing the nonminimal coupling between a self-interacting KR field and gravity is succinctly expressed by the action described as \cite{18,29}
\begin{eqnarray}
S=\int\sqrt{-g}d^4x\bigg[\frac{R}{16\pi G}-\frac{1}{12}H_{\alpha\mu\nu}H^{\alpha\mu\nu}-V\big(B_{\mu\nu}B^{\mu\nu}\pm b_{\mu\nu}b^{\mu\nu}\big)+\frac{1}{16\pi G}\big(\xi_2B^{\mu\lambda}B^\nu_\lambda R_{\mu\nu}+\xi_3B_{\mu\nu}B^{\mu\nu}R\big)\bigg], \label{1}
\end{eqnarray}
where the symbol $R$ denotes the Ricci scalar and $g=det(g_{\mu\nu})$. The second rank tensor $B_{\mu\nu}$ is anti-symmetric and defines the KR field that is responsible for the LSB vacuum expectation value (VEV) $b_{\mu\nu}\neq0$, where this VEV has some relation with the potential $V$ \cite{16}. With this VEV, we can obtain two spacelike vectors and a timelike vector by decomposing the tensor field $B_{\mu\nu}$ in the same way as we decompose the Maxwell's linear field tensor $F_{\mu\nu}$ \cite{18}. The tensor field $H_{\alpha\beta\delta}=\partial_{\alpha[}B_{\beta\delta]}$ being antisymmetric, is analogical to the Maxwell's linear field tensor $F_{\mu\nu}$, while the KR field tensor $B_{\mu\nu}$ has an analogy with the vector potential. Therefore, we can write the KR action in analogy with the electrodynamics. The symbols $\xi_i$ are the nonminimal coupling constants, where $i=2,3$. Lessa et al. \cite{29} considered the vacuum KR field as $B_{\mu\nu}B^{\mu\nu}=b_{\mu\nu}b^{\mu\nu}$ to study how the KR VEV affects the gravitational field. In a flat spacetime, the LSB VEV follows $\partial_\sigma b_{\mu\nu}=0$ and $j^2=\eta^{\alpha\beta}\eta^{\mu\nu}b_{\alpha\mu}b_{\beta\nu}$. Furthermore, if the VEV is constant, then the strength of KR field vanishes \cite{18}. In this regard, we can consider the KR VEV $b_{\mu\nu}$ as a constant whose Hamiltonian vanishes. Under this analogy in the curved spacetime background, we can also consider the VEV as a constant, that is, $\nabla_\sigma b_{\mu\nu}=0$ such that the KR field strength and Hamiltonian also vanish. To derive a spacetime metric describing a BH in KR gravity, the norm of VEV $b_{\mu\nu}$ must be a constant whose Hamiltonian vanishes. The gravitational field equations are modified to be
\begin{eqnarray}
R_{\mu\nu}-\frac{1}{2}Rg_{\mu\nu}=\kappa T_{\mu\nu}^{\xi_2}, \label{2}
\end{eqnarray}
where $R_{\mu\nu}$ and $T_{\mu\nu}^{\xi_2}$ are the Ricci and the energy-momentum tensors, respectively. The line element for a 4D static and spherically symmetric metric is given as
\begin{eqnarray}
ds^2=-\chi_t(r)dt^2 +\chi_r(r)^{-1}dr^2+r^2d\Omega_2^2 \label{3}
\end{eqnarray}
with $d\Omega_2^2$ defining the 2-sphere. The KR VEV ansatz are
\begin{eqnarray}
b_2=-\epsilon(r)dt\wedge dr, \label{4}
\end{eqnarray}
where $\epsilon=-b_{tr}$. As said earlier that the norm of the ansatz of KR VEV is $b^2$ which is a constant, then for the metric (\ref{3}) we obtain a pseudo-electric field
\begin{eqnarray}
\epsilon(r)=|b|\sqrt{\frac{\chi_t(r)}{2\chi_r(r)}} \label{5}
\end{eqnarray}
that is static and is in radial direction. Therefore, the values of the metric functions come out to be
\begin{equation}
\chi_t(r)=\chi_r(r)=1-\frac{R_{sc}}{r}+\gamma r^{-\frac{2}{\lambda}}, \label{6}
\end{equation}
where $R_{sc}=2M$ is the Schwarzschild radius. Hence, the metric (\ref{3}) becomes
\begin{eqnarray}
ds^2=-\bigg(1-\frac{R_{sc}}{r}+\gamma r^{-\frac{2}{\lambda}}\bigg)dt^2 +\bigg(1-\frac{R_{sc}}{r}+\gamma r^{-\frac{2}{\lambda}}\bigg)^{-\frac{1}{2}}dr^2+r^2d\Omega_2^2 \label{7}
\end{eqnarray}
that describes BH in KR gravity. Here, $\gamma$ and $\lambda$ are considered to be the corresponding spontaneous LSB parameters. The parameter $\lambda$ is defined in terms of nonminimal coupling parameter and VEV as $\lambda=\xi_2|b|^2$, where $b^2=b_{\mu\nu}b^{\mu\nu}$. Whereas, the parameter $\gamma$ is a constant of integration. The parameters $M$ and $\gamma$ has dimensions of $[length]$ and $[length]^{\frac{2}{\lambda}}$, respectively. The metric (\ref{7}) can also be termed as power-law hairy BH solution. For a fixed $\gamma$ and taking either $|b|^2=0$ or $\xi_2=0$, we get $\lambda\rightarrow0$ and thus the BH solution (\ref{7}) becomes Schwarzschild solution. Therefore, for $\lambda\in(0,\infty)$ the BH in KR gravity deviates from the Schwarzschild BH. Since the gravitational field is not significantly affected by the LSB, so the necessity arises for the coupling constant to possess a small value. However, it is anticipated that the LSB occurs at Planck scale. Therefore, the VEV $b^2$ that gives rise to the LSB coupling, might also occur at Planck scale. In this context, some systems are generated by the spontaneous LSB that are described by small value of coupling constant but a large value of VEV. It can also be noted that the metric (\ref{7}) reduces to Schwarzschild de-Sitter BH corresponding to $\lambda=-1$ defining an asymptotically non-flat BH. However, it reduces to asymptotically flat Reissner-Nordstr\"{o}m BH for $\lambda=1$. Therefore, BH metric (\ref{7}) is asymptotically non-flat metric for $\lambda\leq0$ and hence, we consider only positive values of $\lambda$.

The rotating BHs are defined by a spin parameter and encompasses various interesting features, especially, the particle motion and optical features. Since, the BHs located at the center of galaxies are usually spinning BHs. Therefore, it is more convenient and feasible to consider a spinning BH for a rigorous comparative analysis for the shadows of the theoretical and supermassive BHs. The Newman-Janis algorithm is a concise and useful method to convert a static BH metric into its rotating counterpart. Therefore, by incorporating this algorithm, the BH metric (\ref{7}) was transformed into the rotating metric by Kumar et al. \cite{30} that is given by
\begin{eqnarray}
ds^{2}&=&-\bigg(\frac{\Delta(r)-a^2\sin^2{\theta}}{\zeta^2}\bigg)dt^{2}+\frac{\zeta^2}{\Delta(r)}dr^2+\zeta^2d\theta^{2}+\frac{\sin^{2}\theta}{\zeta^2}\bigg(\mathcal{U}(r;a)^2-\Delta(r) a^2\sin^{2}\theta\bigg)d\phi^{2} \nonumber\\
&&+\frac{2a\sin^{2}\theta}{\zeta^2}\bigg(\Delta(r)-\mathcal{U}(r;a)\bigg)dtd\phi, \label{8}
\end{eqnarray}
with the metric functions defined as
\begin{eqnarray}
\Delta(r)&=&a^2+r^2\chi_t(r)=\mathcal{U}(r;a)-R_{sc}r+\gamma r^{\frac{2(\lambda-1)}{\lambda}}, \label{9}\\
\zeta^2&=&r^2+a^2\cos^2\theta, \label{10}
\end{eqnarray}
with $a$ as spin parameter and $\mathcal{U}(r;a)=r^2+a^2$. By supposing $\lambda=0$ and $\lambda=1$ in (\ref{8}), we can recover the Kerr and Kerr-Newman BHs, respectively. In addition to this, the static metric (\ref{7}) is recovered by supposing $a=0$ in the metric (\ref{8}). Since the metric coefficients in (\ref{8}) are independent of the coordinates $t$ and $\phi$. Therefore, the Killing vectors $(\partial_t)^\mu$ and $(\partial_\phi)^\mu$ exist that define the rotational and time symmetries. We focus on the study of the behavior of various plasma distributions surrounding the rotating BH in KR gravity. We  also find out that what distribution of plasma is more influential on the optical features of the BH (\ref{8}). It will also enable us to find that the BH (\ref{8}) surrounded by a particular plasma medium is whether M87* or Sgr A*. Hence, we draw a comparison between the angular radius of the shadows of the BH (\ref{8}) with M87* and Sgr A*.

\begin{figure}[t!]
	\begin{center}
		\includegraphics[width=0.41\textwidth]{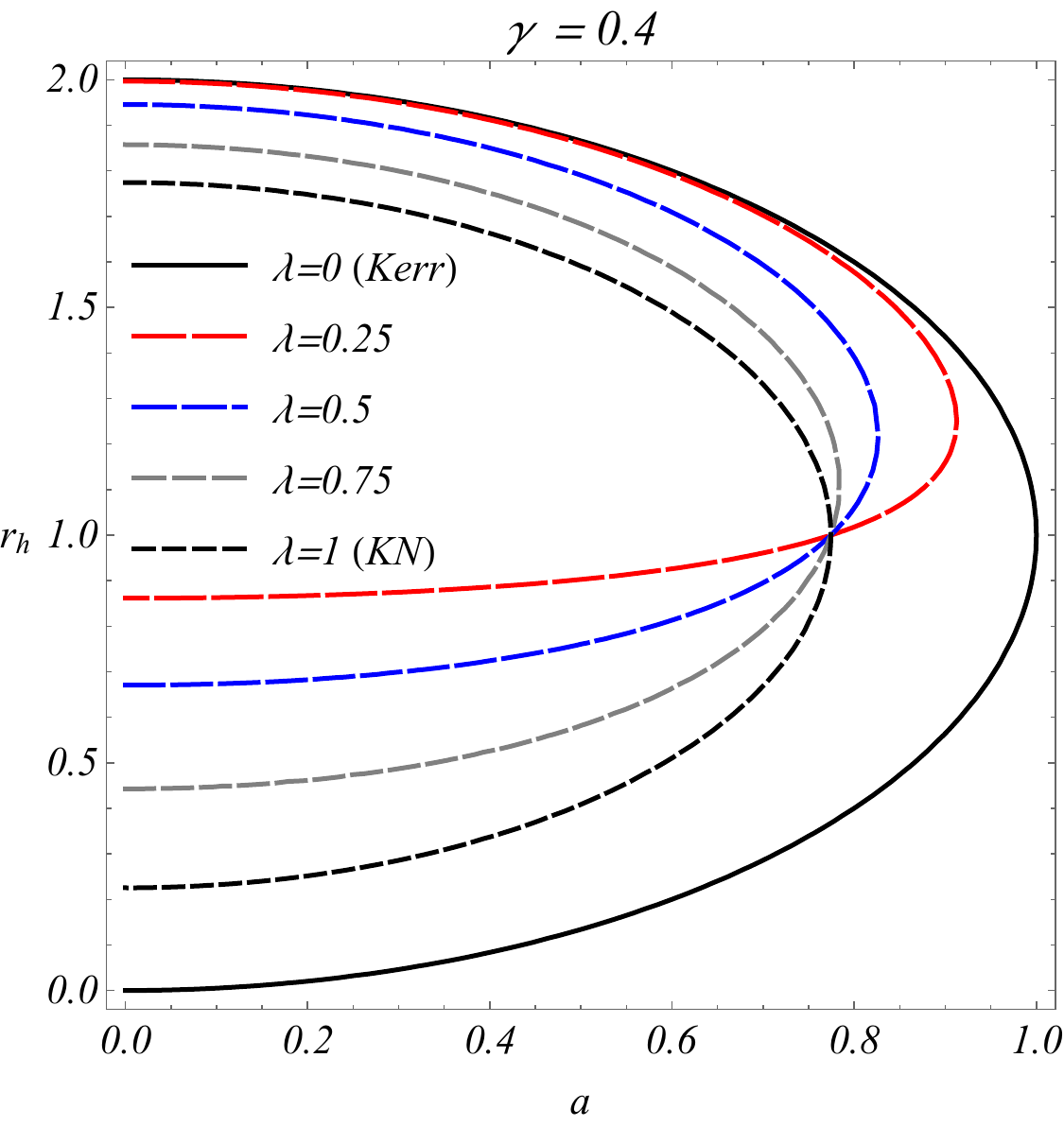}~~~~~~
		\includegraphics[width=0.41\textwidth]{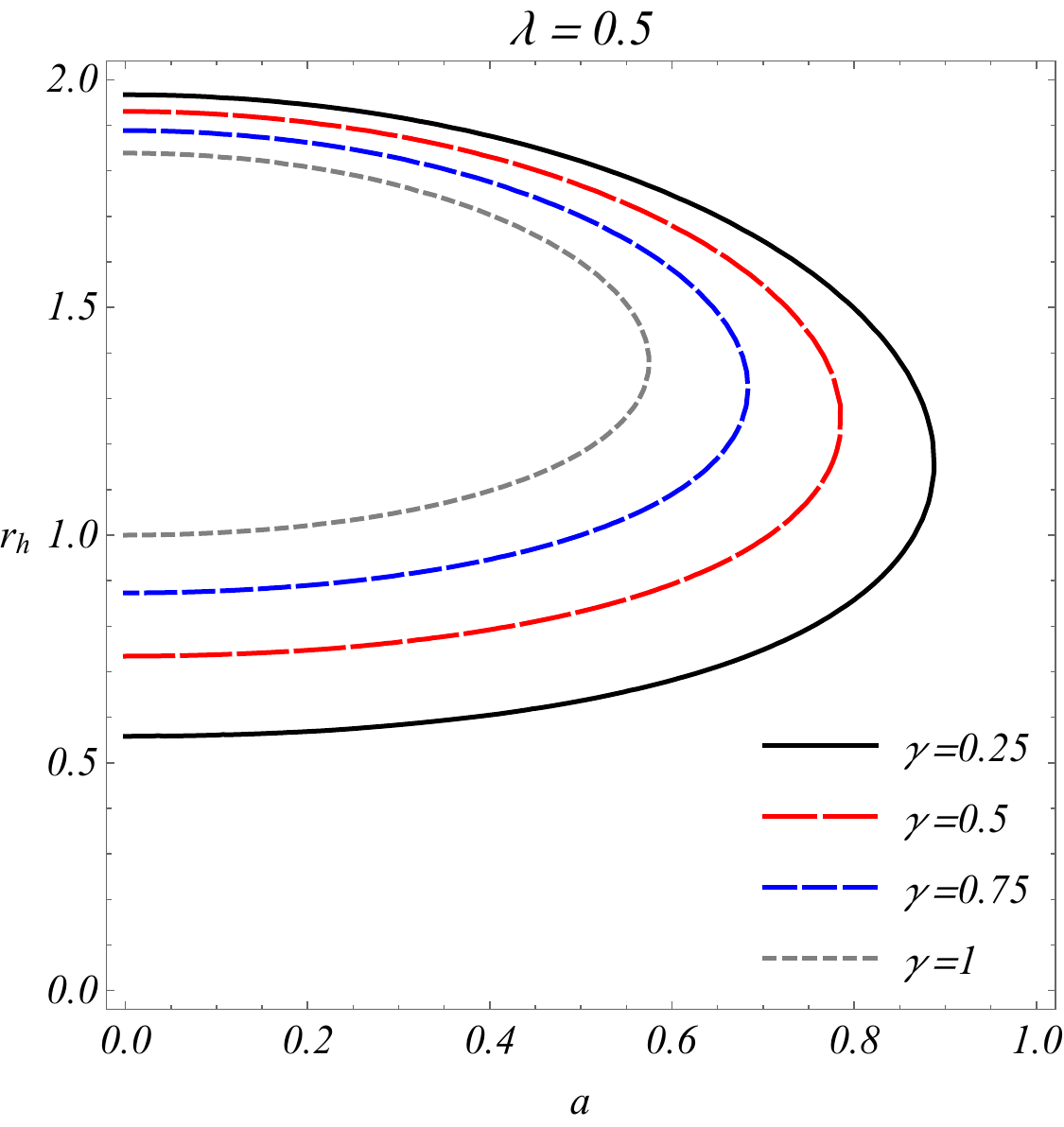}
	\end{center}
	\caption{Plots illustrating the variation of horizons $r_h$ vs $a$ for various values of $\lambda$ and $\gamma$. \label{hr}}
\end{figure}

The outer horizon, also known as event horizon defines the size of the BH that has been plotted along with Cauchy horizon for with respect to the spin for different values of $\lambda$ and $\gamma$ in Fig. \ref{hr}. To study the horizon structure, we need to determine the roots $r_h$ by solving the metric function $\Delta(r)\big|_{r_h}=0$. For each curve in the figure, it is evident that the Cauchy horizon keeps on increasing and the event horizon keeps on decreasing with respect to $a$, until the extremal value of $a$ is achieved where the metric function $\delta$ has a double root and both horizons become identical. Moreover, the extreme value of $a$ decreases as $\lambda$ and $\gamma$ increase in both plots. The event and Cauchy horizons in the left plot decrease for the fixed value of $a$ as $\lambda$ increases except for $\lambda=0$. For $\lambda=0$, the Kerr BH that shows the largest extremal value of $a$ and has the largest difference of horizons. For $\lambda\neq0$, it deviates from the Kerr and approaches the Kerr-Newman BH for $\lambda=1$. For a particular value of $a$ as $\gamma$ is elevated in the right plot, it can be seen that the value of event horizon drops but the Cauchy horizon is elevated.

\section{Null Geodesics} \label{nullgeod}
There is a critical region in the neighborhood of the BH where the light rays are trapped that emerge out of a bright source. This region is known as photon sphere that comprise the unstable circular null orbits. Some photons in these orbits fall inside the event horizon, whereas the remaining photons scatter away to the infinity. As a result of this phenomenon, the shadow of the BH is formed as a 2D image. We know that a photon in the unstable circular null orbit may revolve around the BH for more than one time before leaving the orbit. Whereas, a photon resides forever in the stable circular null orbit. The effective potential describes these circular null orbits. Since, it is quite certain that the astrophysical BHs are mostly immersed in material media, especially a plasma medium. Hence, our study delves into the characteristics of null geodesics and the effective potential governing the motion of photons around a rotating black hole within the framework of KR gravity. This investigation is conducted in the presence of a plasma medium, and various distribution functions are considered to describe its properties. Hence, the properties of particle orbits around the BH are governed by the effective potential that can be determined by incorporating the Hamilton-Jacobi \cite{78} and Hamiltonian formulations. Since, we encounter four unknown variables for which we need four constants of motion in order to solve the system of equations completely. Therefore, we obtain two constants by using the Hamiltonian formulation, namely, the angular momentum $L$ along $\phi$ direction  and energy $E$ of the particle. The particle's mass being the third constant, we employ the Hamilton-Jacobi method that generates a constant termed as Carter constant \cite{78} which is treated as the fourth constant. This formalism also ensures the separability of terms as a function of $r$ and $\theta$ coordinates. We divide this analysis into different subsections based on various plasma distributions.

\subsection{Case I}\label{n1}
We begin with a general case in which the frequency of electrons in plasma is assumed a function of the $r$ and $\theta$ \cite{31,71,72}. The Hamiltonian that describes the photon motion in a plasma medium without pressure and magnetization is given as
\begin{equation}
\mathcal{H}=\frac{1}{2}\Big[g^{\mu\nu}p_\mu p_\nu+\omega_p(r,\theta)^2\Big], \label{11}
\end{equation}
where, $p_\mu$ are the generalized momenta. The plasma electron frequency $\omega_p(r,\theta)$ is related to the refractive index $n(r,\theta)$ as
\begin{equation}
n(r,\theta)=\sqrt{1-\frac{\omega_p(r,\theta)^2}{\omega(r,\theta)^2}}, \label{12}
\end{equation}
where, $\omega(r,\theta)$ is the frequency of photon measured by a static observer outside the event horizon. The frequency of electrons in plasma is written in terms of the number density of electrons as
\begin{equation}
\omega_p(r,\theta)=\frac{4\pi e^2}{m_e}N_e(r,\theta), \label{13}
\end{equation}
where, $e$, $m_e$ and $N_e$ are the charge, mass and number density of the electron, respectively. The Hamiltonian corresponds to the Hamilton's equations
\begin{equation}
\dot{x}^\mu=\partial_\tau x^\mu=\partial_{p_\mu}\mathcal{H}, \qquad \dot{p}_\mu=\partial_\tau p_\mu=-\partial_{x^\mu}\mathcal{H} \label{14}
\end{equation}
with $\tau$ being an affine parameter. Since, the plasma medium is dense and dispersive, therefore, the motion of photons is affected by the frequencies of electrons in the plasma and photons in a light ray. Therefore, the frequency of the propagating photon must be greater than the frequency of the electrons in plasma, that is, $\omega(r,\theta)^2\geq\omega_p(r,\theta)^2$. Since the observer is static with the four-velocity in comoving coordinates being $U^\mu(r,\theta)=(-g_{tt}(r,\theta))^{-1/2}$ and the photon frequency is $\omega(r,\theta)=-p_\mu U^\mu(r,\theta)=-p_t U^t(r,\theta)$. The Planck's relation relates the energy and the angular frequency as $E=\hbar\omega_0$ which gives $p_t=-\omega_0$ by setting the units for $\hbar=1$. Therefore, we obtain
\begin{equation}
\omega(r,\theta)=\omega_0\big(-g_{tt}(r,\theta)\big)^{-\frac{1}{2}}. \label{15}
\end{equation}
Using the Hamiltonian approach, we obtain the following two equations
\begin{eqnarray}
\zeta^2\dot{t}&=&\frac{1}{\Delta(r)}\Big[\Big(\mathcal{U}(r;a)^2-a^2\Delta(r)\sin^2\theta\Big)E+a\big(\Delta(r)-\mathcal{U}(r;a)\big)L\Big], \label{16} \\
\zeta^2\dot{\phi}&=&\frac{1}{\Delta(r)}\Big[\big(\Delta(r)\csc^2\theta-a^2\big)L-a\big(\Delta(r)-\mathcal{U}(r;a)\big)E\Big]. \label{17}
\end{eqnarray}
Here, we obtained the two constants; the energy $E$ and angular momentum $L$, where the third constant is taken as $\mathcal{H}=0$. Therefore, by virtue of the expression $p_\mu=\partial_{x^\mu} S_J$, the Hamilton-Jacobi equation can be expressed in the form
\begin{equation}
g^{\mu\nu}\partial_{x^\mu}\mathcal{S_J}\partial_{x^\nu}\mathcal{S_J}=0 \label{18}
\end{equation}
with Jacobi action
\begin{equation}
\mathcal{S_J}=-Et+L\phi+\mathcal{A}_r(r)+\mathcal{A}_\theta(\theta) \label{19}
\end{equation}
with $\mathcal{A}_r(r)$ and $\mathcal{A}_\theta(\theta)$ as unknown arbitrary functions. We obtain as a result of solving the Hamilton-Jacobi equation,
\begin{equation}
\Delta(r)\big(\partial_r\mathcal{A}_r(r)\big)^2-\frac{\Big(\mathcal{U}(r;a)E-aL\Big)^2}{\Delta(r)}+(L-aE)^2+\big(\partial_\theta\mathcal{A}_\theta(\theta)\big)^2+\big(L^2\csc^2\theta-a^2E^2\big)\cos^2\theta+\zeta^2\omega_p^2(r,\theta)=0. \label{20}
\end{equation}
Since, the last term on the left hand side of Eq. (\ref{20}) is mixed in $r$ and $\theta$ coordinates, so we cannot separate this equation. For this, we assume \cite{31,71,72}
\begin{equation}
\omega_p(r,\theta)=\frac{\sqrt{f_r(r)+f_\theta(\theta)}}{\zeta}, \label{21}
\end{equation}
where, $f_r(r)$ and $f_\theta(\theta)$ are arbitrary functions that ensures the separability of the Eq. (\ref{20}). Therefore, the separation gives the following equations:
\begin{eqnarray}
\partial_r\mathcal{A}_r(r)&=&\bigg[\frac{\Big(\mathcal{U}(r;a)E-aL\Big)^2}{\Delta(r)^2}-\frac{f_r(r)+(aE-L)^2+\mathcal{Z}}{\Delta(r)}\bigg]^{\frac{1}{2}}, \label{22}\\
\partial_\theta\mathcal{A}_\theta(\theta)&=&\sqrt{\mathcal{Z}-\big(L^2\csc^2\theta-a^2E^2\big)\cos^2\theta-f_\theta(\theta)}, \label{23}
\end{eqnarray}
where, $\mathcal{Z}$ is the Carter constant \cite{78} that is considered as the fourth constant of motion. Simplifying further, the null geodesic equations for the photon motion can be written as
\begin{eqnarray}
\zeta^2\dot{t}&=&a\big(L-aE\sin^2{\theta}\big)+\frac{\mathcal{U}(r;a)}{\Delta(r)}\big(\mathcal{U}(r;a)E-aL\big), \label{24} \\
\zeta^2\dot{r}&=&\pm\sqrt{\mathcal{R}(r)}, \label{25} \\
\zeta^2\dot{\theta}&=&\pm\sqrt{\Theta(\theta)}, \label{26} \\
\zeta^2\dot{\phi}&=&\big(L\csc^2\theta-aE\big)-\frac{a}{\Delta(r)}\big(aL-\mathcal{U}(r;a)E\big), \label{27}
\end{eqnarray}
where
\begin{eqnarray}
\mathcal{R}(r)&=&\big(aL-\mathcal{U}(r;a)E\big)^2-\Delta(r)\big(f_r(r)+\mathcal{Z}+(aE-L)^2\big), \label{28} \\
\Theta(\theta)&=&\mathcal{Z}+a^2E^2\cos^2\theta-L^2\cot^2\theta-f_\theta(\theta). \label{29}
\end{eqnarray}
The constants of motion are renamed as $L=E\Gamma$ and $\mathcal{Z}=E^2\Sigma$ and thus
\begin{eqnarray}
\mathcal{R}(r)&=&\big(\mathcal{U}(r;a)-a\Gamma\big)^2-\Delta(r)\big(\Sigma+(\Gamma-a)^2+f_r(r)\big). \label{30}
\end{eqnarray}
The radial equation is related to the effective potential $V_{eff}(r)$ as $\dot{r}^2+2V_{eff}(r)=0$. Then, the effective potential at the equator $\Big(\theta=\frac{\pi}{2}\Big)$ becomes
\begin{eqnarray}
V_{eff}(r)=-\frac{\mathcal{R}(r)}{2r^4}=-\frac{\big(\mathcal{U}(r;a)-a\Gamma\big)^2-\Delta(r)\big(\Sigma+(\Gamma-a)^2+f_r(r)\big)}{2r^4}. \label{31}
\end{eqnarray}
Our major aim is to determine the size and shape of the shadows that are dependent upon the unstable circular light orbits in the neighborhood of the BH. Since these orbits form a sphere, therefore, the photons revolving in these orbits around the BH must stay at the surface of Euclidean sphere described by the equation $r=A_0$, where $A_0$ is a constant. The circular null orbits are described by $\dot{r}=0=\ddot{r}$ that implies $V_{eff}(r_p)=0=\partial_rV_{eff}(r_p)$, where $r_p$ is the radius of photon sphere. These conditions are connected to the function $\mathcal{R}(r)$ as $\mathcal{R}(r_p)=0=\partial_r\mathcal{R}(r_p)$. The unstable orbits are described by the local maxima in effective potential curve and is mathematically given by the condition $\partial^2_rV_{eff}(r_p)<0$. Next, we consider two subcases in which we assume different values of the functions $f_r(r)$ and $f_\theta(\theta)$ to study the size of photon sphere.

\subsubsection{Case Ia}\label{Ia}
In this case, we consider \cite{71}
\begin{eqnarray}
f_r(r)=\omega_c^2\sqrt{M^3r}, \qquad f_\theta(\theta)=0, \label{32}
\end{eqnarray}
where, the constant $\omega_c$ has the same dimensions as the frequency. By using the values of functions given in Eq. (\ref{32}) into (\ref{31}), we plotted the effective potential $V_{eff}(r)$ vs $r$ in the upper panel of Fig. \ref{veff}. In each plot, we varied the values of $\omega_c$ for each curve with fixed $\gamma$, $\lambda$ and $a$. The peaks observed in the effective potential curves signify the positions of unstable circular orbits for light trajectories. By increasing the value of $\omega_c$, there is no effect on the size of the BH as $\omega_c$ is not a BH parameter. However, the size of the unstable circular null orbits increased by a small amount. This is because the function $f_r(r)$ appears in the relation (\ref{31}) linearly such that $f_r(r)\propto\omega_c^2$ and $f_r(r)\propto r$. Eventually, the Eqs. (\ref{13}) and (\ref{21}) show that $N_e\propto\omega_c^2$ and $N_e\propto r$. Therefore, the increase in $\omega_c$ causes increase in $N_e$ and thus strengthening the plasma medium. Moreover, as $N_e\propto r$, so the number of electrons are further increased raising the mass of the plasma field because electrons are massive particles. This causes an extra increase in gravity due to plasma by a small amount. Therefore, with increase in $\omega_c$ causing the increase in gravity, the photons are captured slightly away from the BH, raising the size of unstable null orbits.

\subsubsection{Case Ib}\label{Ib}
In this case, we consider \cite{71}
\begin{eqnarray}
f_r(r)=0, \qquad f_\theta(\theta)=\omega_c^2M^2\big(1+2\sin^2\theta\big). \label{33}
\end{eqnarray}
Because of the symmetry on the equatorial plane, the effective potential does not depend on $\theta$ and hence $f_\theta(\theta)$. Therefore, for this case, the unstable circular null orbits and photon sphere behave as if there is no plasma medium. This behavior of photon sphere is quite usual and has been studied in detail in Ref. \cite{79}.

\subsection{Case II}\label{n2}
Now, we consider the case in which the refractive index $n$ is a function of the coordinate $r$ only and possesses no dependence on $\theta$ \cite{73}. Therefore, we can write $n=n(r)$ and thus the plasma frequency becomes
\begin{equation}
\omega_p^2(r)=\big(n(r)^2-1\big)g^{tt}p_t^2. \label{34}
\end{equation}
The Hamiltonian (\ref{11}) for this case becomes
\begin{equation}
\mathcal{H}=\frac{1}{2}\Big[g^{\mu\nu}p_\mu p_\nu+\big(n(r)^2-1\big)g^{tt}p_t^2\Big]. \label{35}
\end{equation}
Solving for the $t$ and $\phi$ coordinates, we obtain
\begin{eqnarray}
\zeta^2\dot{t}&=&\frac{1}{\Delta(r)}\Big[n(r)^2\Big(\mathcal{U}(r;a)^2-a^2\Delta(r)\sin^2\theta\Big)E+a\big(\Delta(r)-\mathcal{U}(r;a)\big)L\Big], \label{36} \\
\zeta^2\dot{\phi}&=&\frac{1}{\Delta(r)}\Big[\big(\Delta(r)\csc^2\theta-a^2\big)L-a\big(\Delta(r)-\mathcal{U}(r;a)\big)E\Big]. \label{37}
\end{eqnarray}
In the same way as in the case \ref{n1}, $E$ and $L$ are the two constants of motion with $\mathcal{H}=0$ as the third constant of motion. Therefore, for the fourth constant, we solve the Hamilton-Jacobi equation (\ref{18}) with Jacobi function (\ref{19}) that comes out to be
\begin{eqnarray}
\Delta(r)\big(\partial_r\mathcal{A}_r(r)\big)^2-\frac{\Big(\mathcal{U}(r;a)E-aL\Big)^2}{\Delta(r)}+(L-aE)^2+\big(\partial_\theta\mathcal{A}_\theta(\theta)\big)^2+\big(L^2\csc^2\theta-a^2E^2\big)\cos^2\theta \nonumber\\-\frac{\big(n(r)^2-1\big)E^2\mathcal{U}(r;a)^2}{\Delta(r)}+\big(n(r)^2-1\big)a^2E^2\sin^2\theta=0. \label{38}
\end{eqnarray}
Since, the last term on the left hand side of Eq. (\ref{38}) comprise $n(r)^2a^2E^2\sin^2\theta$ which is mixed in $r$ and $\theta$ coordinates and is generally non-separable equation. For this, we assume that the orientation of the system at the equatorial plane is slightly perturbed by shifting off the plane by a fractionally small angle $\psi$ such that $\theta=\frac{\pi}{2}+\psi$. It is known that the unstable circular orbits of light are the part of photon sphere and can be found at any plane. Therefore, a distant observer located at equatorial plane can observe the photon orbits off the equatorial plane. The approximation $\theta=\frac{\pi}{2}+\psi$ is valid because $\psi$ is very small and we may obtain the desirable results for BH shadow. Hence, we place the observer at the equatorial plane. However, the null geodesics will be modified under this assumption. By taking $\theta=\frac{\pi}{2}+\psi$ in Eq. (\ref{38}), we obtain
\begin{eqnarray}
(L-aE)^2-\frac{\Big(\mathcal{U}(r;a)E-aL\Big)^2}{\Delta(r)}+\Delta(r)\big(\partial_r\mathcal{A}_r(r)\big)^2+\big(\partial_\psi\mathcal{A}_\psi(\psi)\big)^2+\big(n(r)^2-1\big)a^2E^2 \nonumber\\-\frac{\big(n(r)^2-1\big)E^2\mathcal{U}(r;a)^2}{\Delta(r)}=0, \label{39}
\end{eqnarray}
which is separated into:
\begin{eqnarray}
\Delta(r)\big(\partial_r\mathcal{A}_r(r)\big)^2&=&\frac{\Big(\mathcal{U}(r;a)E-aL\Big)^2}{\Delta(r)}-(L-aE)^2-\big(n(r)^2-1\big)a^2E^2+\frac{\big(n(r)^2-1\big)E^2\mathcal{U}(r;a)^2}{\Delta(r)}-\mathcal{Z}, \label{40}\\
\big(\partial_\psi\mathcal{A}_\psi(\psi)\big)^2&=&\mathcal{Z}, \label{41}
\end{eqnarray}
where, the Carter constant $\mathcal{Z}$ is considered as the fourth constant of motion. Further simplifying, the null geodesic equations governing photon motion can be expressed as
\begin{eqnarray}
r^2\dot{t}&=&a\big(L-n(r)^2aE\cos^2\psi\big)+\frac{\mathcal{U}(r;a)}{\Delta(r)}\big(n(r)^2\mathcal{U}(r;a)E-aL\big), \label{42} \\
r^2\dot{r}&=&\pm\sqrt{\mathcal{R}(r)}, \label{43} \\
r^2\dot{\psi}&=&\pm\sqrt{\Theta(\psi)}, \label{44} \\
r^2\dot{\phi}&=&\big(L\sec^2\psi-aE\big)-\frac{a}{\Delta(r)}\big(aL-\mathcal{U}(r;a)E\big), \label{45}
\end{eqnarray}
where
\begin{eqnarray}
\mathcal{R}(r)&=&\big(n(r)^2-1\big)\mathcal{U}(r;a)^2E^2+\big(\mathcal{U}(r;a)E-aL\big)^2-\Delta(r)\big(\mathcal{Z}+(L-aE)^2+\big(n(r)^2-1\big)a^2E^2\big), \label{46} \\
\Theta(\psi)&=&\mathcal{Z}. \label{47}
\end{eqnarray}
In terms of $\Gamma$ and $\Sigma$, the function $\mathcal{R}(r)$ in Eq. (\ref{46}) becomes
\begin{eqnarray}
\mathcal{R}(r)=\big(n(r)^2-1\big)\mathcal{U}(r;a)^2+\big(\mathcal{U}(r;a)-a\Gamma\big)^2-\Delta(r)\big(\Sigma+(\Gamma-a)^2+\big(n(r)^2-1\big)a^2\big). \label{48}
\end{eqnarray}
From the radial equation, the effective potential $V_{eff}(r)$ near the equatorial plane becomes
\begin{eqnarray}
V_{eff}(r)=-\frac{\mathcal{R}(r)}{2r^4}=-\frac{\big(n(r)^2-1\big)\mathcal{U}(r;a)^2+\big(\mathcal{U}(r;a)-a\Gamma\big)^2-\Delta(r)\big(\Sigma+(\Gamma-a)^2+\big(n(r)^2-1\big)a^2\big)}{2r^4}. \label{49}
\end{eqnarray}
We suppose the function
\begin{eqnarray}
n(r)^2=1-\frac{k}{r}, \label{49a}
\end{eqnarray}
where $k$ is a constant that describes the strength of plasma around the BH. Assuming this value of $n(r)$ in Eq. (\ref{49}), we plot the effective potential $V_{eff}(r)$ vs $r$ in the middle panel of Fig. \ref{veff}. In both plots, we varied the values of $k$ for each curve and kept $\gamma$, $\lambda$ and $a$ fixed. By increasing the value of $k$, there is no effect on the size of the BH as it defines the plasma medium only. However, the size of the unstable circular null orbits decreased by a small amount, in converse to the upper panel. This is because the function $n(r)^2-1=-\frac{k}{r}$ appears in the relation (\ref{49}) linearly such that $N_e\propto k$ and $N_e\propto \frac{1}{r}$. Therefore, the increase in $k$ causes an increase in $N_e$ but the inverse relation with $r$ affects the strength of the plasma medium. As a result, the increase in mass of the plasma medium is affected by $N_e\propto \frac{1}{r}$, unlike the case \ref{Ia}. Thus, the gravity does not increase significantly due to plasma itself. Hence, with the increase in $k$, the photons are captured slightly closer to the BH, reducing the size of unstable null orbits.

\subsection{Case III}\label{n3}
In this case, we consider that the refractive index $n$ is a constant \cite{73}. Therefore, we may write
\begin{eqnarray}
n^2=1-k, \label{49b}
\end{eqnarray}
where $k$ is the same parameter as in the case \ref{n2}. Since, the last term on the left hand side of Eq. (\ref{38}) comprise $n(r)^2a^2E^2\sin^2\theta$ which is mixed in $r$ and $\theta$ coordinates corresponding to the case \ref{n2}. However, in this case, this term is only function of $\theta$ because $n\neq n(r)$. Therefore, the Eq. (\ref{38}) is separable. We obtain
\begin{eqnarray}
\Delta(r)\big(\partial_r\mathcal{A}_r(r)\big)^2&=&\frac{\Big(\mathcal{U}(r;a)E-aL\Big)^2}{\Delta(r)}-(L-aE)^2+\frac{\big(n(r)^2-1\big)E^2\mathcal{U}(r;a)^2}{\Delta(r)}-\mathcal{Z}, \label{50}\\
\big(\partial_\theta\mathcal{A}_\theta(\theta)\big)^2&=&\mathcal{Z}+a^2E^2\cos^2\theta-L^2\cot^2\theta-\big(n^2-1\big)a^2E^2\sin^2\theta. \label{51}
\end{eqnarray}
Therefore, following Eqs. (\ref{36}) and (\ref{37}), and simplifying further, the Eqs. (\ref{50}) and (\ref{51}), the null geodesic equations for the photon motion can be written as
\begin{eqnarray}
\zeta^2\dot{t}&=&a\big(L-n^2aE\sin^2\theta\big)+\frac{\mathcal{U}(r;a)}{\Delta(r)}\big(n^2\mathcal{U}(r;a)E-aL\big), \label{52} \\
\zeta^2\dot{r}&=&\pm\sqrt{\mathcal{R}(r)}, \label{53} \\
\zeta^2\dot{\theta}&=&\pm\sqrt{\Theta(\theta)}, \label{54} \\
\zeta^2\dot{\phi}&=&\big(L\csc^2\theta-aE\big)-\frac{a}{\Delta(r)}\big(aL-\mathcal{U}(r;a)E\big), \label{55}
\end{eqnarray}
where
\begin{eqnarray}
\mathcal{R}(r)&=&\big(n^2-1\big)\mathcal{U}(r;a)^2E^2+\big(\mathcal{U}(r;a)E-aL\big)^2-\Delta(r)\big(\mathcal{Z}+(L-aE)^2\big), \label{56} \\
\Theta(\theta)&=&\mathcal{Z}-\big(n^2-1\big)a^2E^2\sin^2\theta+a^2E^2\cos^2\theta-L^2\cot^2\theta. \label{57}
\end{eqnarray}
In terms of $\Gamma$ and $\Sigma$, the function $\mathcal{R}(r)$ in Eq. (\ref{56}) becomes
\begin{figure}[t!]
	\begin{center}
		\subfigure{\includegraphics[width=0.41\textwidth]{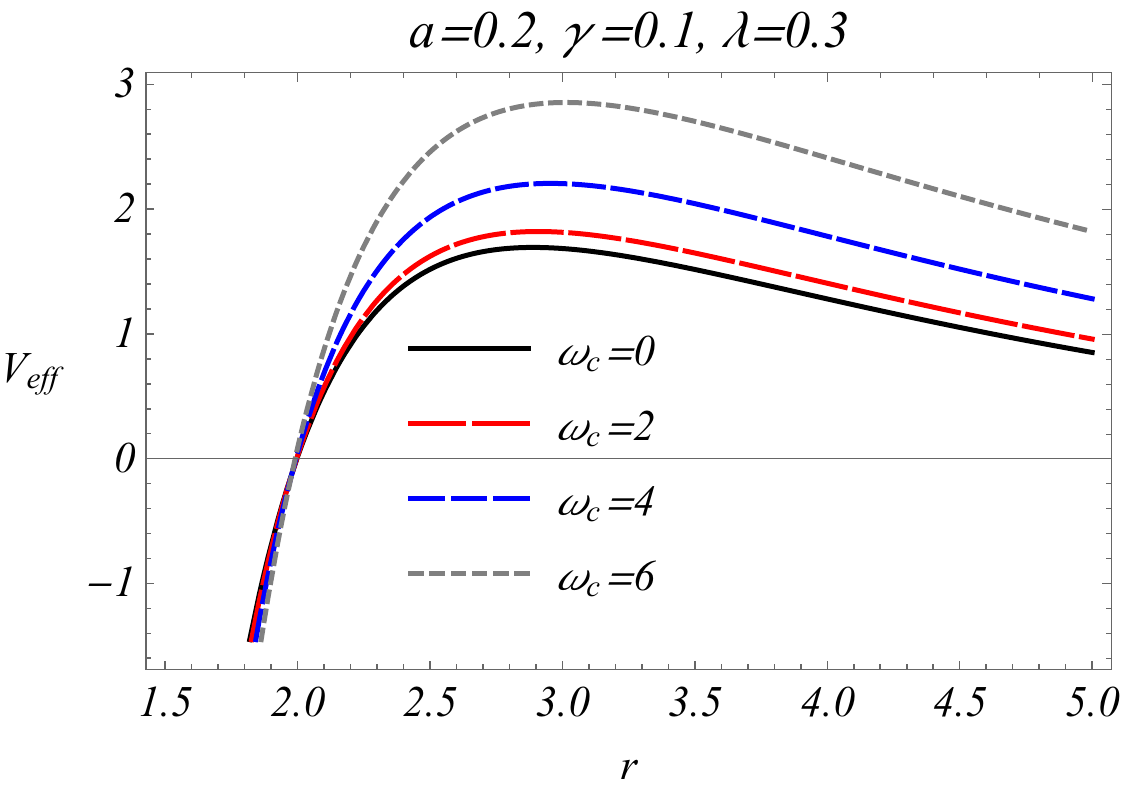}}~~~
		\subfigure{\includegraphics[width=0.41\textwidth]{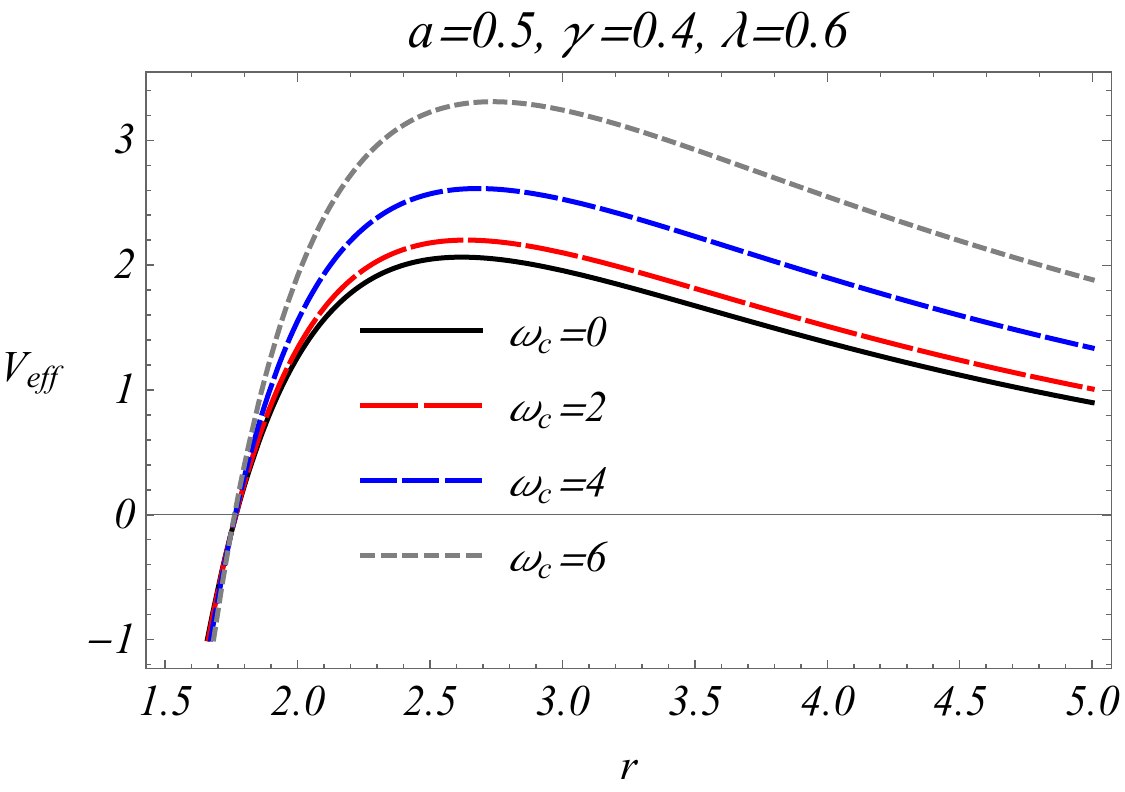}}
		\subfigure{\includegraphics[width=0.41\textwidth]{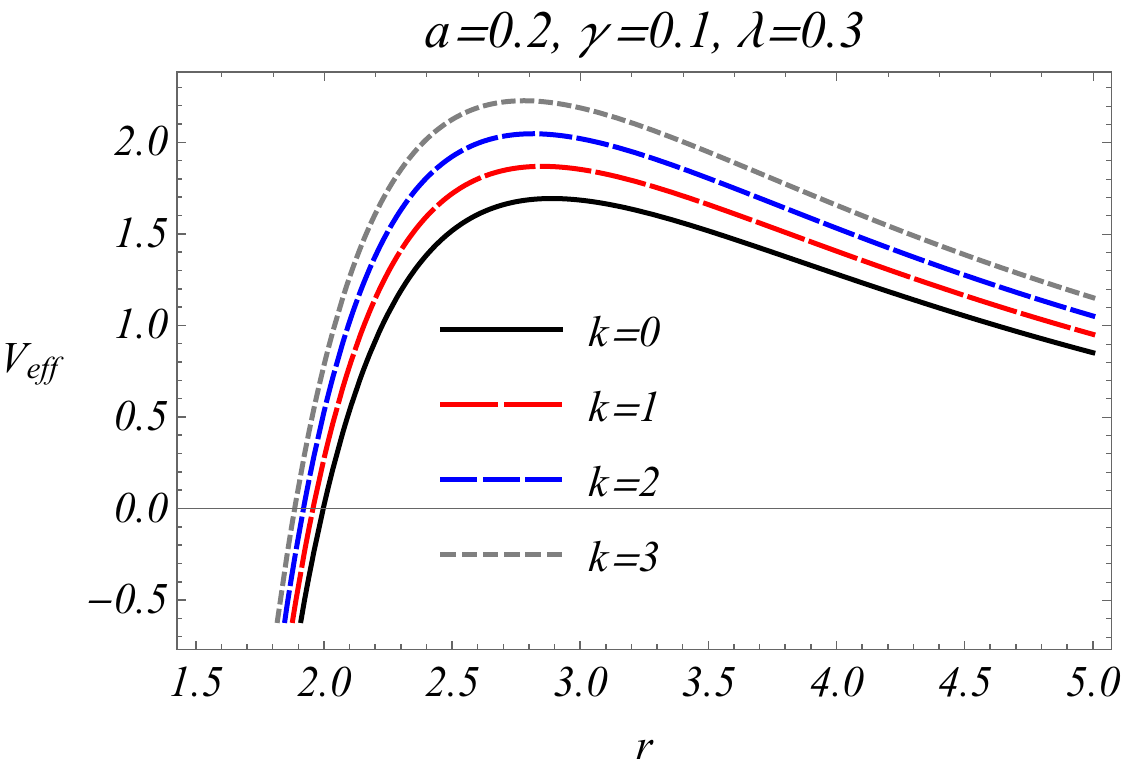}}~~~
		\subfigure{\includegraphics[width=0.41\textwidth]{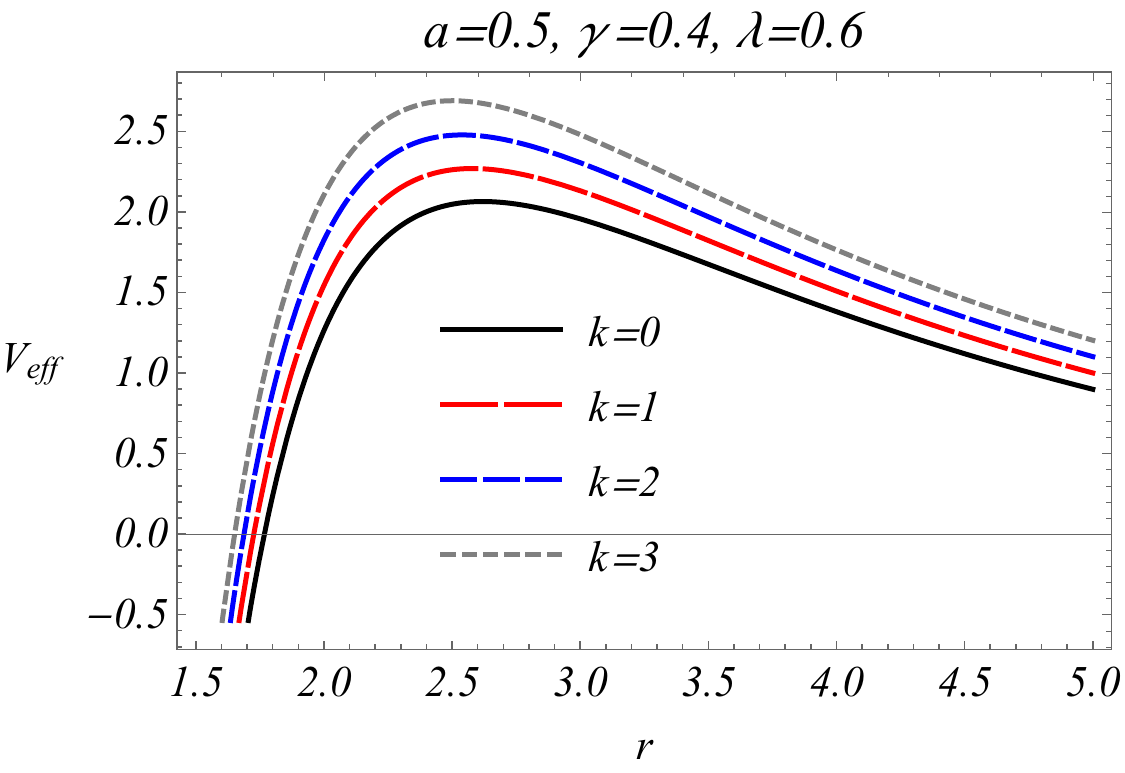}}
		\subfigure{\includegraphics[width=0.41\textwidth]{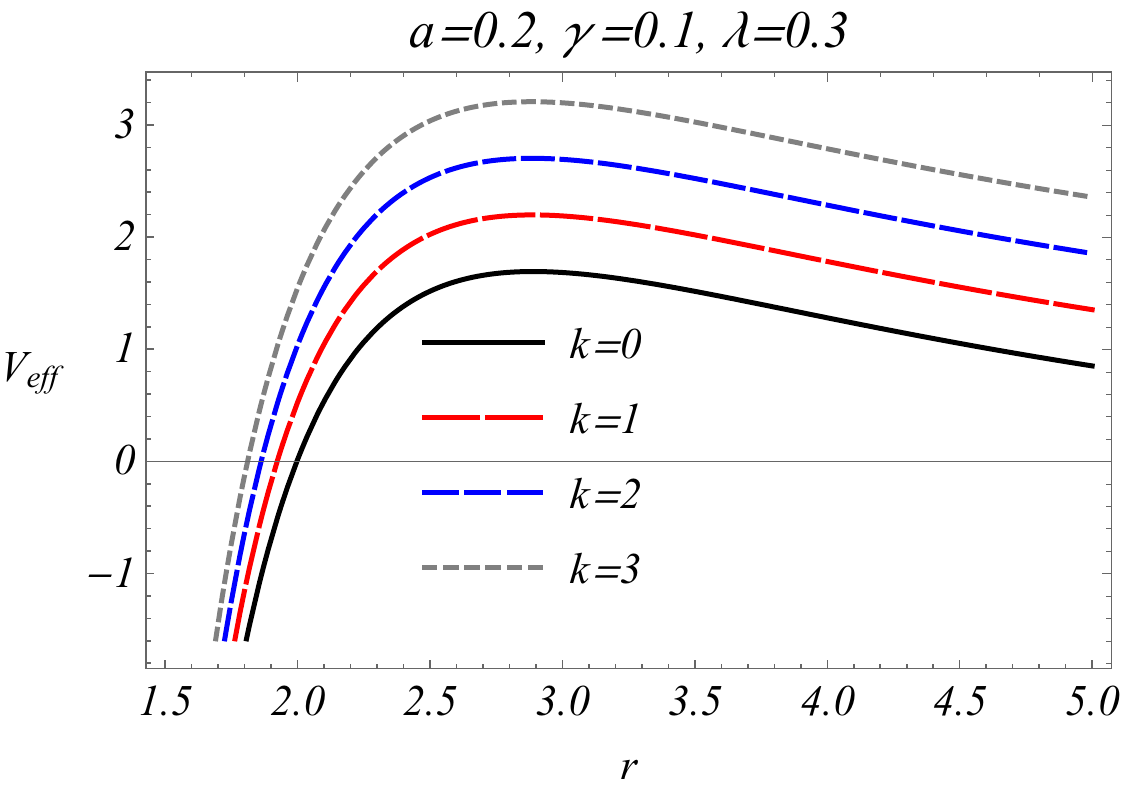}}~~~
		\subfigure{\includegraphics[width=0.41\textwidth]{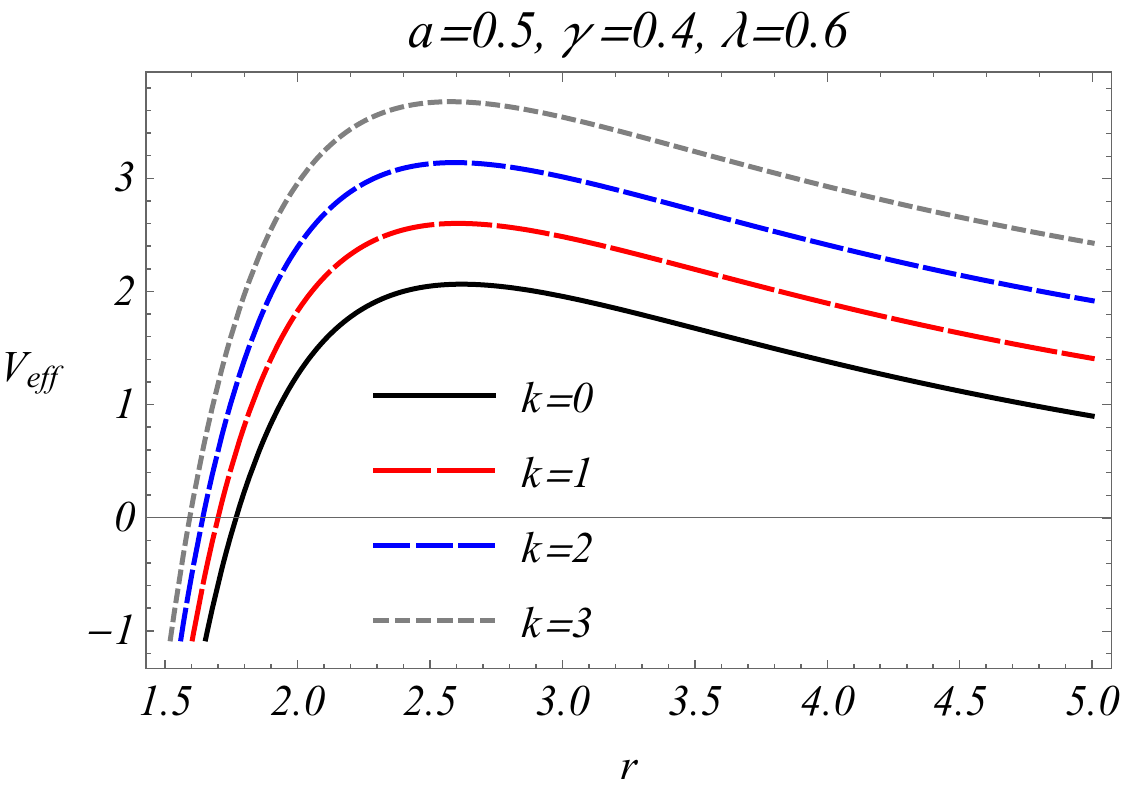}}
	\end{center}
	\caption{The effective potential $V_{eff}$ is sketched versus $r$ to analyze the variation for different values of $a$, $\lambda$ and $\gamma$. The top, middle and bottom panels correspond to the case \ref{Ia}, \ref{n2} and \ref{n3}, respectively. \label{veff}}
\end{figure}
\begin{eqnarray}
\mathcal{R}(r)=\big(n^2-1\big)\mathcal{U}(r;a)^2+\big(\mathcal{U}(r;a)-a\Gamma\big)^2-\Delta(r)\big(\Sigma+(\Gamma-a)^2\big). \label{58}
\end{eqnarray}
From the radial equation, the effective potential $V_{eff}(r)$ near the equatorial plane becomes
\begin{eqnarray}
V_{eff}(r)=-\frac{\mathcal{R}(r)}{2r^4}=-\frac{\big(n^2-1\big)\mathcal{U}(r;a)^2+\big(\mathcal{U}(r;a)-a\Gamma\big)^2-\Delta(r)\big(\Sigma+(\Gamma-a)^2\big)}{2r^4}. \label{59}
\end{eqnarray}
By using the given value of $n$ in Eq. (\ref{49b}) into (\ref{59}), we plot the effective potential $V_{eff}(r)$ vs $r$ in the lower panel of Fig. \ref{veff}. In both plots, we fixed the values of $\gamma$, $\lambda$ and $a$ as in the upper and middle panels and varied $k$ for each curve. By increasing the value of $k$, there is no effect on the size of the BH as it defines the plasma medium only. However, the size of the unstable circular null orbits decreased by a small amount, as in the case of middle panel. This is because the function $n^2-1=-k$ appears in the relation (\ref{59}) linearly such that $N_e\propto k$. Therefore, the increase in $k$ causes an increase in $N_e$ but there is no relation with $r$. As a result, the increase in mass of the plasma medium is not enhanced by the relation $N_e\propto r$ or diminished by the relation $N_e\propto \frac{1}{r}$. Hence, with the increase in $k$, the photons are captured slightly closer to the BH by a very small amount, reducing the size of unstable null orbits.

\section{Shadows}
This section comprise the optical images of the BH (\ref{8}) immersed in plasma medium corresponding to the cases discussed in the section \ref{nullgeod}. Since, out of all circular null orbits, the unstable orbits contain the photons that form the shadow of the BH. The BH (\ref{8}) constructed by Kumar et al. \cite{30} was further analyzed in the same paper for which they calculated the shadows as viewed from the equator in the absence of plasma. However, in Ref. \cite{79}, we studied the shadows as viewed by an observer away from the equator varying different BH parameters in non-plasma medium for the same BH (\ref{8}) in this work. Hence, we only calculate the shadows by varying the plasma parameters, keeping the other BH parameters fixed. The 2D shadow image of a spatially 3D BH is projected in terms of coordinates $\alpha$ and $\beta$ as
\begin{eqnarray}
\alpha&=&-\lim\limits_{r_0\rightarrow\infty}\bigg(r_0^2\sin\theta_0\bigg[\frac{d\phi}{dr}\bigg]_{\theta\rightarrow\theta_0}\bigg), \label{60}\\
\beta&=&\lim\limits_{r_0\rightarrow\infty}\bigg(r_0^2\bigg[\frac{d\theta}{dr}\bigg]_{\theta\rightarrow\theta_0}\bigg), \label{61}
\end{eqnarray}
where, the location of the observer is $(r_0\rightarrow\infty,\theta_0)$. The mathematical formulation for calculating the shadows is derived by considering the circular photon orbits conditions for different cases of plasma distributions.

\subsection{Case I}\label{aa1}
Using Eq. (\ref{30}) in the conditions $\mathcal{R}(r_p)=0=\partial_r\mathcal{R}(r_p)$, we get the values of $\Gamma$ and $\Sigma$ given by
\begin{eqnarray}
\Gamma(r_p)&=&\frac{\mathcal{U}(r;a)\Delta'(r)-\Delta(r)\big[2r+\sqrt{4r^2-f_r'(r)\Delta'(r)}\big]}{a\Delta'(r)}\bigg|_{r=r_p}, \label{62}\\
\Sigma(r_p)&=&\bigg[\frac{1}{a^2\Delta'(r)^2}\bigg(8r^2\Delta(r)\big[a^2-\Delta(r)\big]-\big[r^4+a^2f_r(r)\big]\Delta'(r)^2 \nonumber\\
&&+\big[4r^3-\big(a^2-\Delta(r)\big)f_r'(r)\big]\Delta(r)\Delta'(r)+2r\Delta(r)\big[2\big(a^2-\Delta(r)\big)+r\Delta'(r)\big]\sqrt{4r^2-f_r'(r)\Delta'(r)}\bigg)\bigg]\bigg|_{r=r_p}. \label{63}
\end{eqnarray}
Now, we consider two different combinations of $f_r(r)$ and $f_\theta(\theta)$ as in Eqs. (\ref{32}) and (\ref{33}) to obtain the exact shadow equations.

\subsubsection{Case Ia}\label{s1a}
We simplify the differentials $\frac{d\phi}{dr}$ and $\frac{d\theta}{dr}$ in Eqs. (\ref{60}) and (\ref{61}) by using Eqs. (\ref{25})-(\ref{29}) and (\ref{32}). Then, by applying the limits, we get
\begin{eqnarray}
\alpha(r_p)&=&-\Gamma(r_p)\csc\theta_0, \label{64} \\
\beta(r_p)&=&\pm\sqrt{\Sigma(r_p)+a^2\cos^2\theta_0-\Gamma(r_p)^2\cot^2\theta_0}. \label{65}
\end{eqnarray}
The null sphere around the rotating BH has a finite width with a variable radial distance $r_p\in\big[r_{p,min},r_{p,max}\big]$, where $r_{p,min}$ and $r_{p,max}$ denote the minimum and maximum radii of the null sphere. Therefore, $r_p$ behaves as a parameter in calculating shadows. The values $r_{p,min}$ and $r_{p,max}$ are the real and positive roots of the equation $\beta(r_p)=0$. We consider the observer's location $\big(\infty,\frac{\pi}{2}\big)$, the celestial coordinates reduce to
\begin{eqnarray}
\alpha(r_p)&=&-\Gamma(r_p), \label{66} \\
\beta(r_p)&=&\pm\sqrt{\Sigma(r_p)}. \label{67}
\end{eqnarray}
The shadows have been computed and graphed, holding values of $a$, $\lambda$ and $\gamma$ constant, while varying $\omega_c$ for each individual curve in Fig. \ref{Sh}. Clearly, the shadow size decreases with increases in $\omega_c$. Apparently, there is no difference in both plots. However, with the change in $\gamma$ and $\lambda$, there is a small change in sizes of the shadows. Moreover, we observe an obvious horizontal shift in the shadow due to the increase in $a$. The increase in $\omega_c$ causes increase in $f_r(r)$ and thus $\omega_p^2$. Therefore, with increase in $\omega_p^2$, the light slows down in plasma due to which the spiral paths formed in the domain of outer communication become smaller. This causes the shadows to shrink.

\subsubsection{Case Ib}\label{S1b}
We evaluate the differentials $\frac{d\phi}{dr}$ and $\frac{d\theta}{dr}$ in Eqs. (\ref{60}) and (\ref{61}) by using Eqs. (\ref{25})-(\ref{29}) and (\ref{33}). Then, by applying the limits, we get
\begin{eqnarray}
\alpha(r_p)&=&-\Gamma(r_p)\csc\theta_0, \label{68} \\
\beta(r_p)&=&\pm\sqrt{\Sigma(r_p)+a^2\cos^2\theta_0-\Gamma(r_p)^2\cot^2\theta_0-\omega_c^2M^2\big(1+2\sin^2\theta_0\big)}. \label{69}
\end{eqnarray}
For the observer to be located at $\big(\infty,\frac{\pi}{2}\big)$, the celestial coordinates become
\begin{eqnarray}
\alpha(r_p)&=&-\Gamma(r_p), \label{70} \\
\beta(r_p)&=&\pm\sqrt{\Sigma(r_p)-3\omega_c^2M^2}. \label{71}
\end{eqnarray}
By varying the value of $\omega_c$ for each curve and fixing $a$, $\lambda$ and $\gamma$, we plotted the shadows in the Fig. \ref{Sha}. Clearly, the shadow shrinks by increasing the value of $\omega_c$. The variation in shrinking of shadow is higher than that in the previous subcase. There is a very small difference in shadows in both plots with the change in $\gamma$ and $\lambda$. Moreover, the increase in $a$ causes the shadow to shift towards the right. As in the previous subcase, the shadows shrink due to the strengthening of the plasma medium that affects the speed of light in plasma. As a result, the spirals outside the photon sphere become smaller.

\subsection{Case II}\label{S2}
\begin{figure}[t!]
	\begin{center}
		\subfigure{\includegraphics[width=0.41\textwidth]{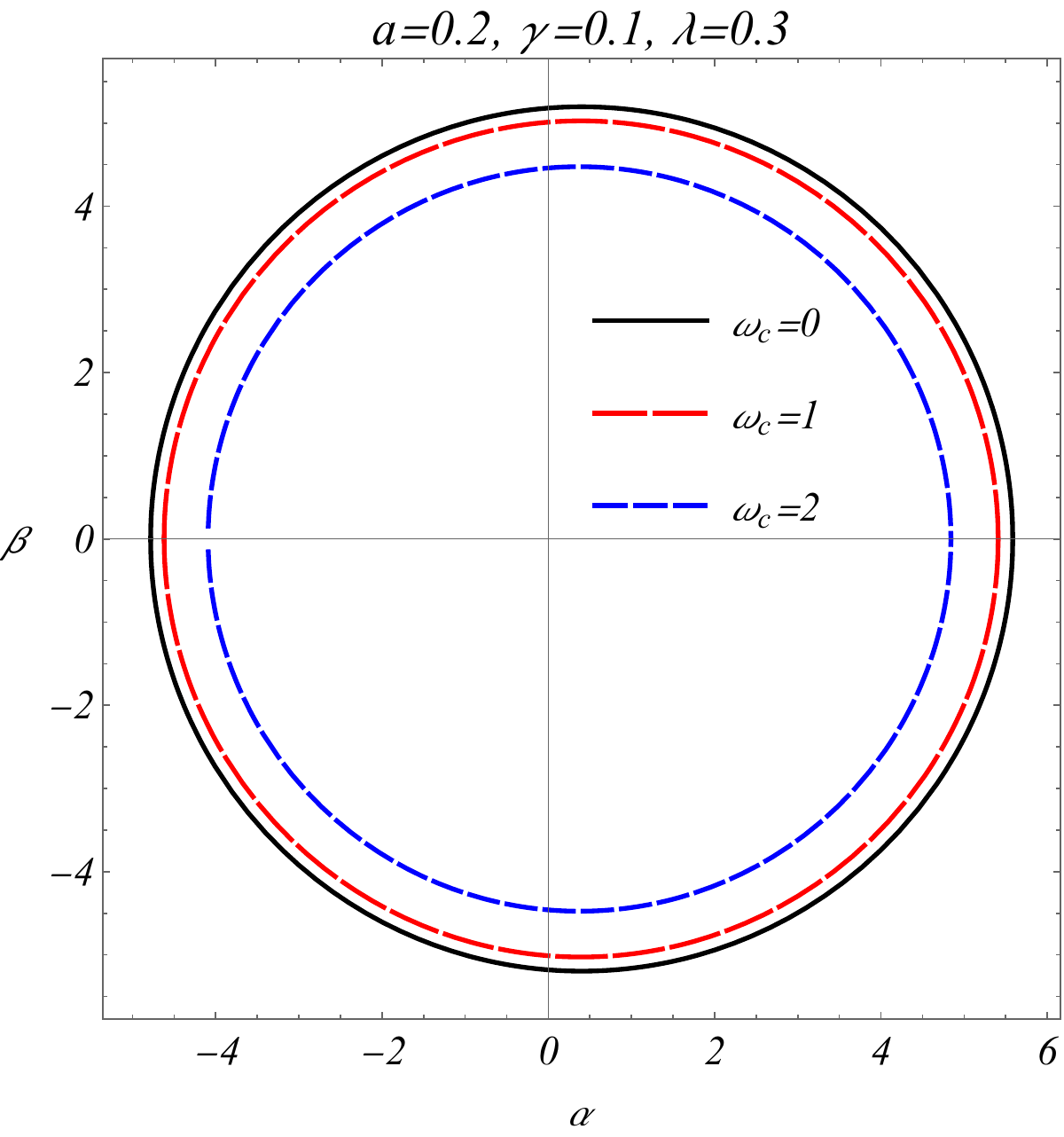}}~~~
		\subfigure{\includegraphics[width=0.41\textwidth]{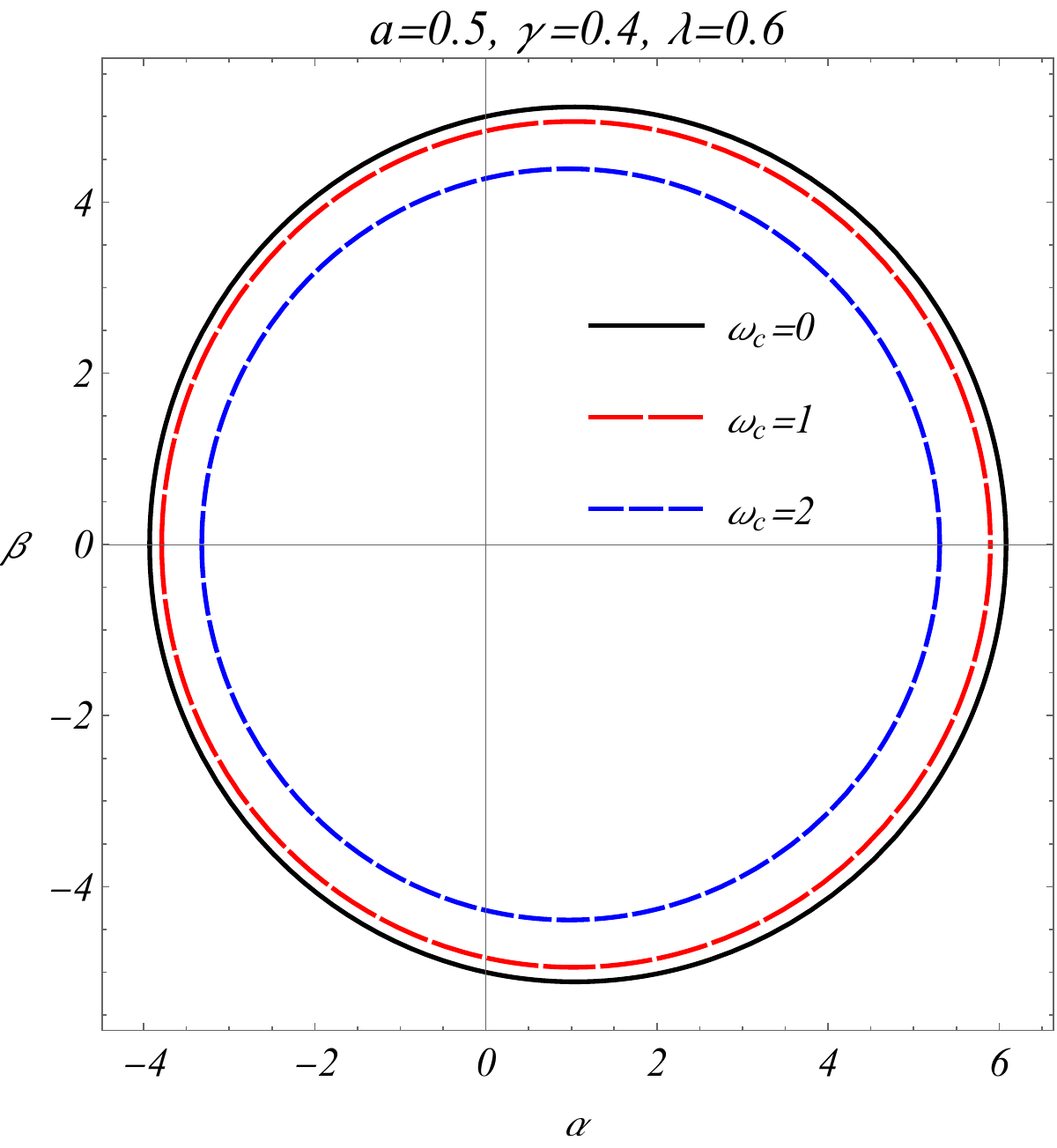}}
	\end{center}
	\caption{Plots showing the behavior of shadows for different values of $\omega_c$ and fixed $a$, $\lambda$ and $\gamma$ corresponding to the case \ref{s1a}. \label{Sh}}
\end{figure}

\begin{figure}[t!]
	\begin{center}
		\subfigure{\includegraphics[width=0.41\textwidth]{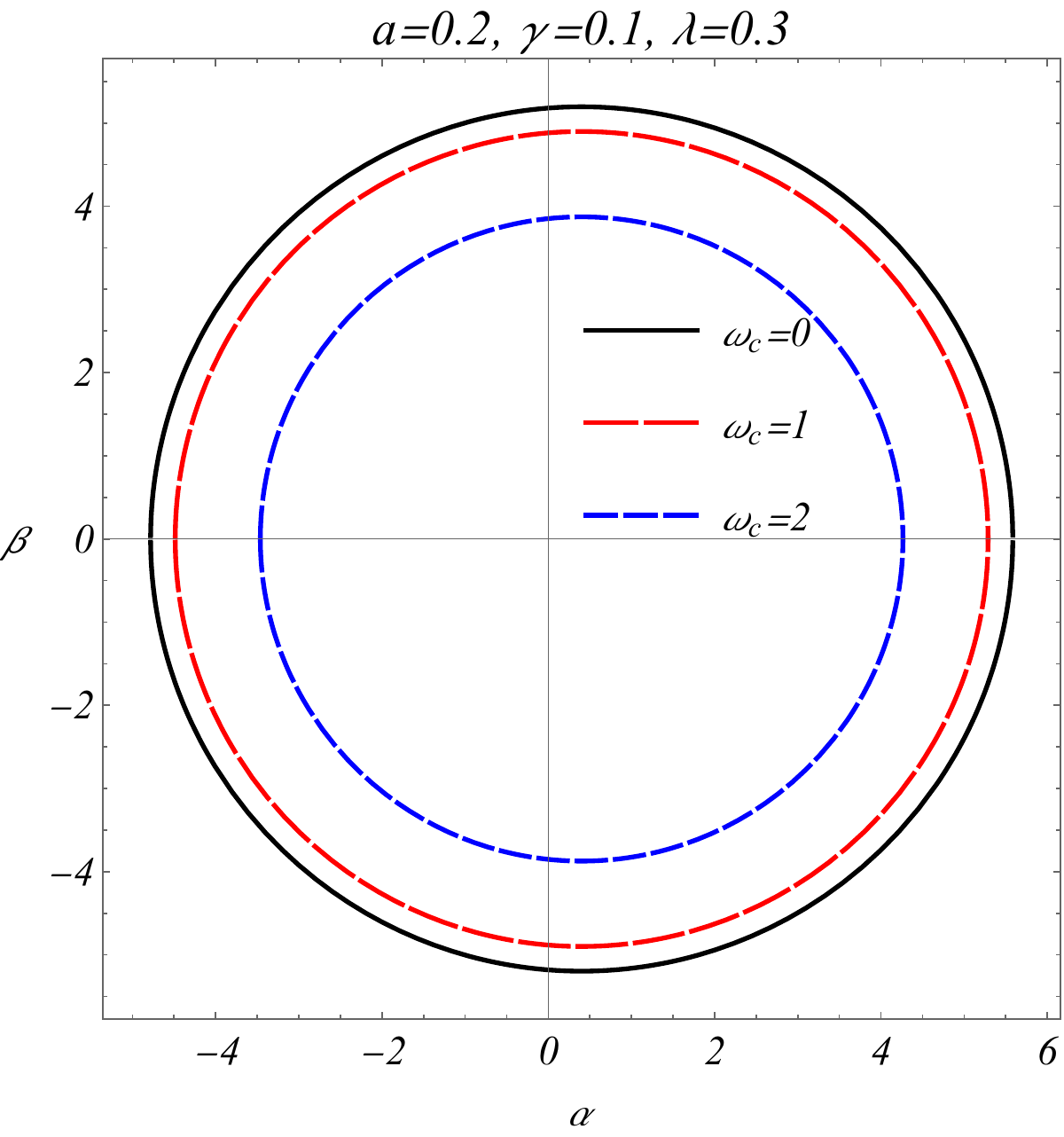}}~~~
		\subfigure{\includegraphics[width=0.41\textwidth]{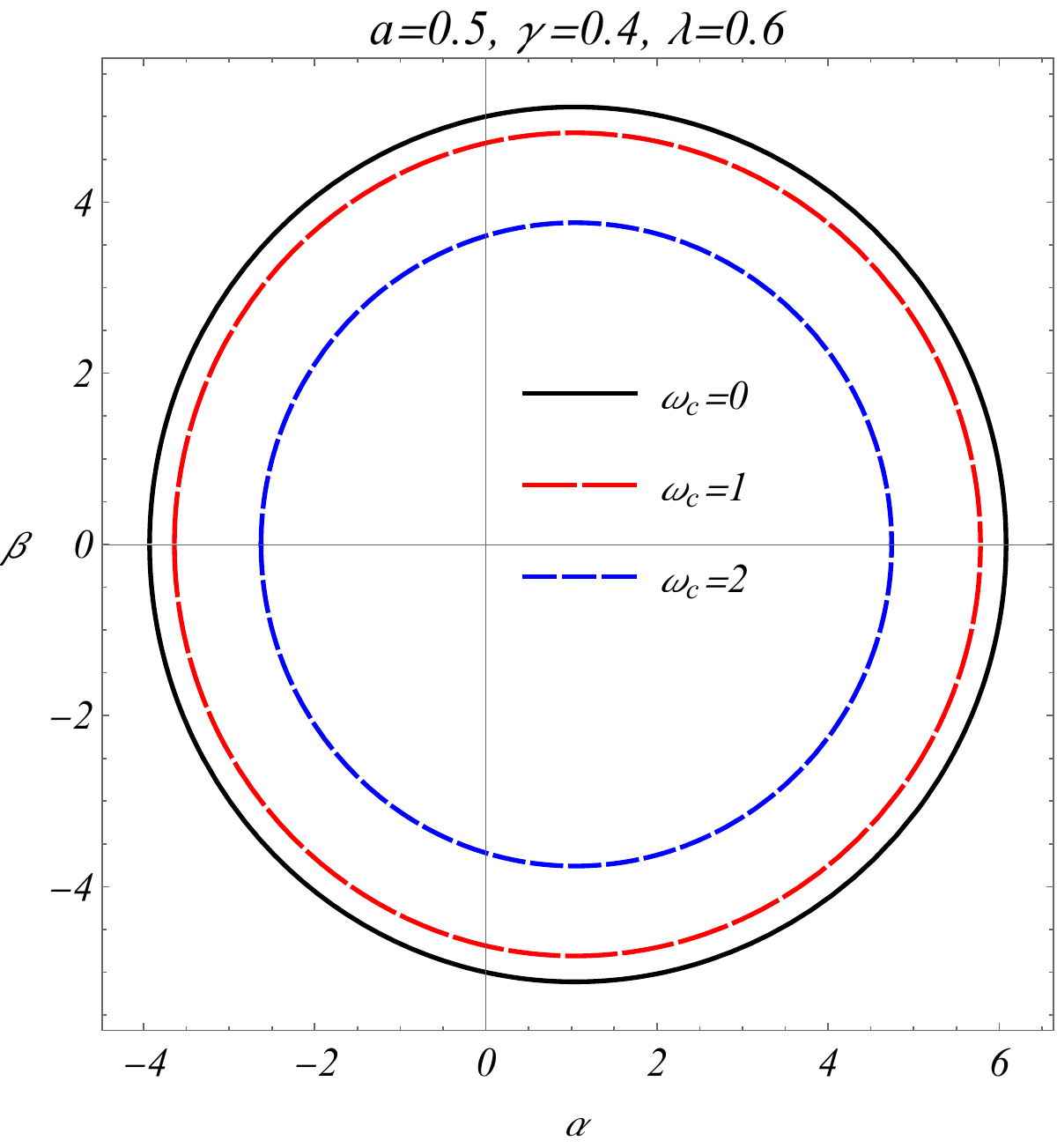}}
	\end{center}
	\caption{The behavior of shadows for different values of $\omega_c$ and fixed $a$, $\lambda$ and $\gamma$ corresponding to the case \ref{S1b}. \label{Sha}}
\end{figure}

\begin{figure}[t!]
	\begin{center}
		\subfigure{\includegraphics[width=0.41\textwidth]{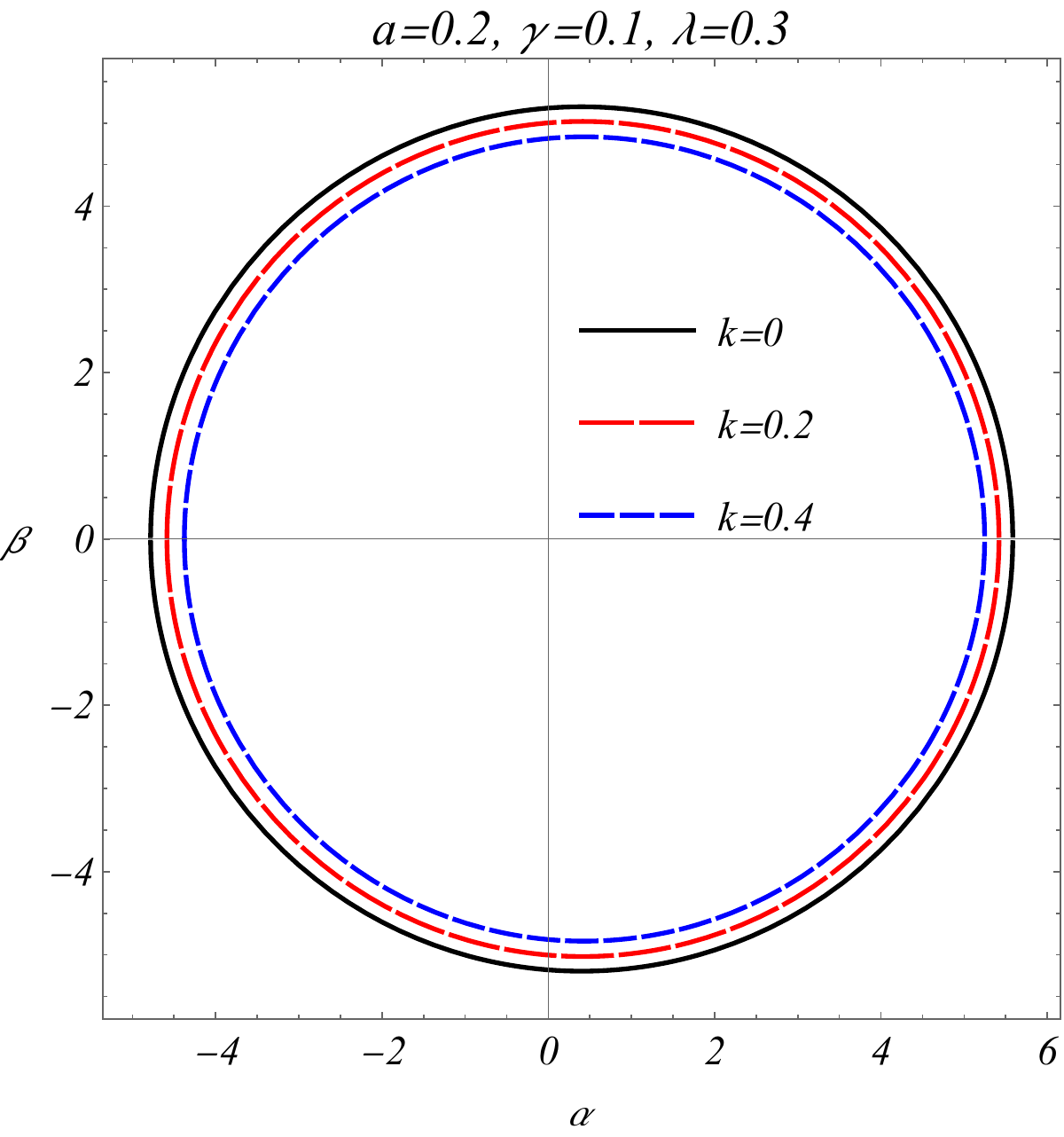}}~~~
		\subfigure{\includegraphics[width=0.41\textwidth]{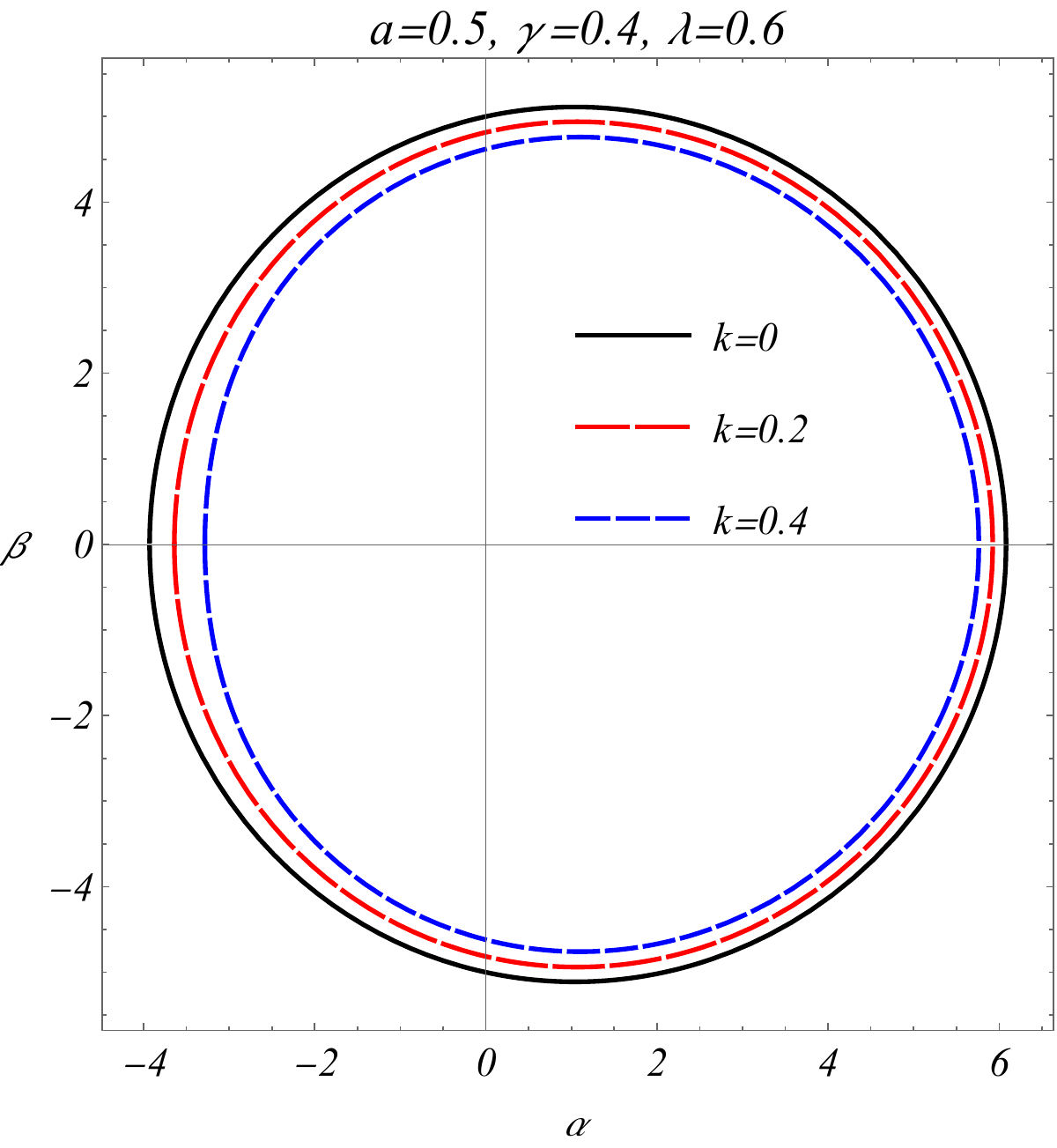}}
	\end{center}
	\caption{Plots illustrating the variation of shadows for different values of $k$ and fixed $a$, $\lambda$ and $\gamma$ corresponding to the case \ref{S2}. \label{Shb}}
\end{figure}

Using Eq. (\ref{48}) in the conditions $\mathcal{R}(r_p)=0=\partial_r\mathcal{R}(r_p)$, we get
\begin{eqnarray}
\Gamma(r_p)&=&\bigg[\frac{1}{a\Delta'(r)}\bigg(\mathcal{U}(r;a)\Delta'(r)-2r\Delta(r)-\mathcal{T}\bigg)\bigg]\bigg|_{r=r_p}, \label{72}\\
\Sigma(r_p)&=&\bigg[\frac{1}{a^2\Delta'(r)^2}\bigg(8r^2\big[a^2-\Delta(r)\big]\Delta(r)+2\Delta'(r)\Big(2r\Big[a^2\mathcal{U}(r;a)\big(n(r)^2-1\big) \nonumber\\&&+\big[a^2+2r^2-\mathcal{U}(r;a)n(r)^2\big]\Delta(r)\Big]+n(r)\big(a^2-\Delta(r)\big)n'(r)\Big[\mathcal{U}(r;a)^2-a^2\Delta(r)\Big]\Big)\nonumber\\&&+r^2\Delta'(r)^2\Big(\Big(2a^2+r^2\Big)n(r)^2-2\mathcal{U}(r;a)\Big)+2r\mathcal{T}\big[2\big(a^2-\Delta(r)\big)+r\Delta'(r)\big]\bigg)\bigg]\bigg|_{r=r_p}, \label{73}
\end{eqnarray}
where,
\begin{eqnarray}
\mathcal{T}&=&\Big[4r^2\Delta(r)^2+2\Delta(r)\Delta'(r)\Big(2r\mathcal{U}(r;a)\big(n(r)^2-1\big)\nonumber\\&&+n(r)n'(r)\big[\mathcal{U}(r;a)^2-a^2\Delta(r)\big]\Big)-\Delta'(r)^2\mathcal{U}(r;a)^2\big(n(r)^2-1\big)\Big]^{\frac{1}{2}}. \label{74}
\end{eqnarray}
Evaluating the differentials $\frac{d\phi}{dr}$ and $\frac{d\theta}{dr}$ in Eqs. (\ref{60}) and (\ref{61}) by using Eqs. (\ref{43})-(\ref{47}) and inserting the value of the function $n(r)^2=1-\frac{k}{r}$ and solving the limits, we get
\begin{eqnarray}
\alpha(r_p)&=&-\Gamma(r_p), \label{75} \\
\beta(r_p)&=&\pm\sqrt{\Sigma(r_p)}. \label{76}
\end{eqnarray}
This case directly corresponds to the null geodesics in a near equatorial plane observed by an observer located at the equatorial plane. In the Fig. \ref{Shb}, we have plotted the shadows by varying the parameter $k$ for each curve and kept $a$, $\lambda$ and $\gamma$ fixed. According to the figure, the shadows shrink with increases in $k$. In this case, a small variation in $k$ shows an enormous variation in the shadow size. Whereas, we had to consider a larger variation in $\omega_c$ for a visible change in the shadow size in the previous cases given in Figs. \ref{Sh} and \ref{Sha}. Therefore, the parameter $k$ is more sensitive than $\omega_c$. For higher values of $a$, $\gamma$ and $\lambda$, the flatness in the shadow increases with increase in $k$.
\begin{figure}[t!]
	\begin{center}
		\subfigure{\includegraphics[width=0.41\textwidth]{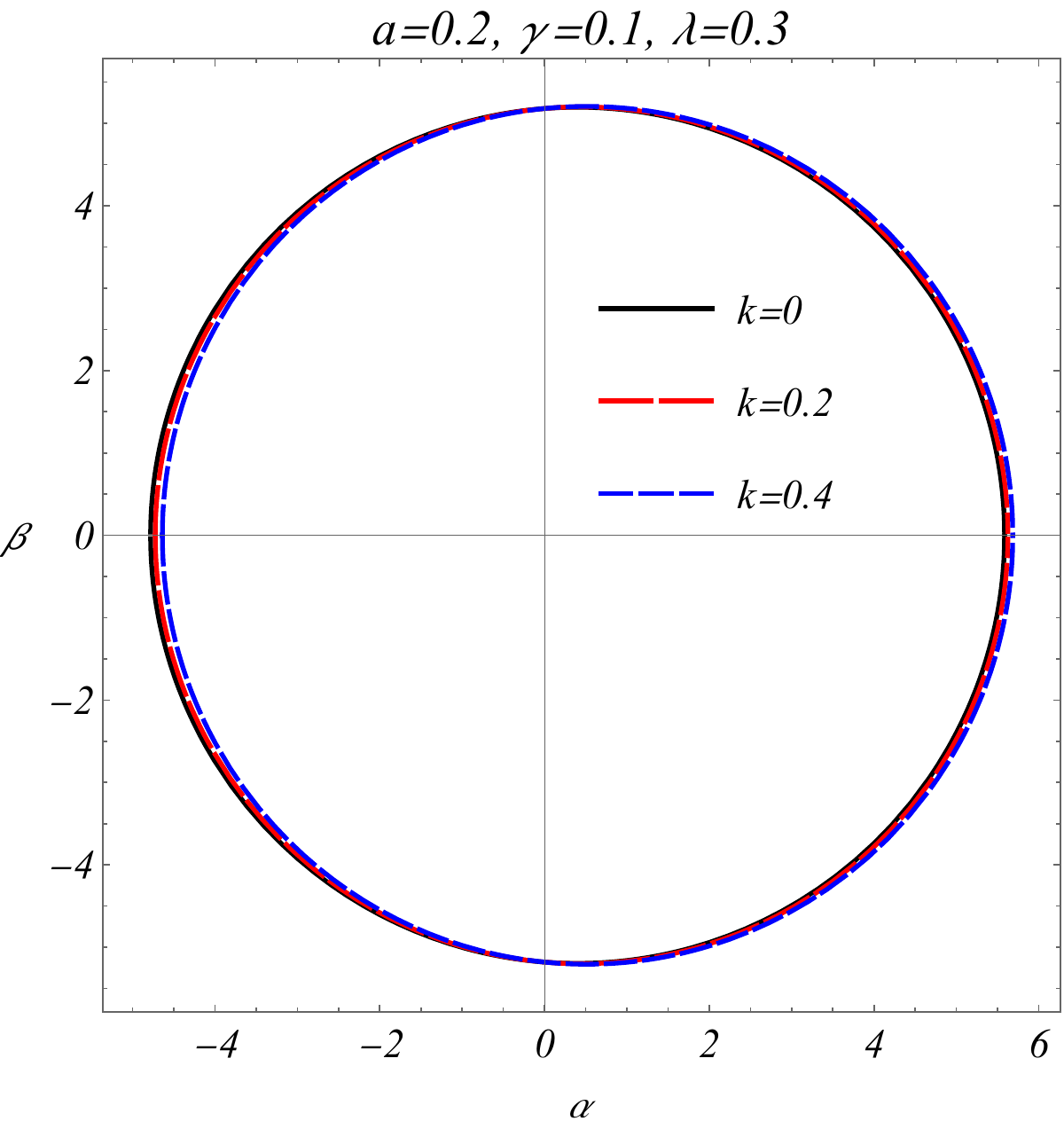}}~~~
		\subfigure{\includegraphics[width=0.41\textwidth]{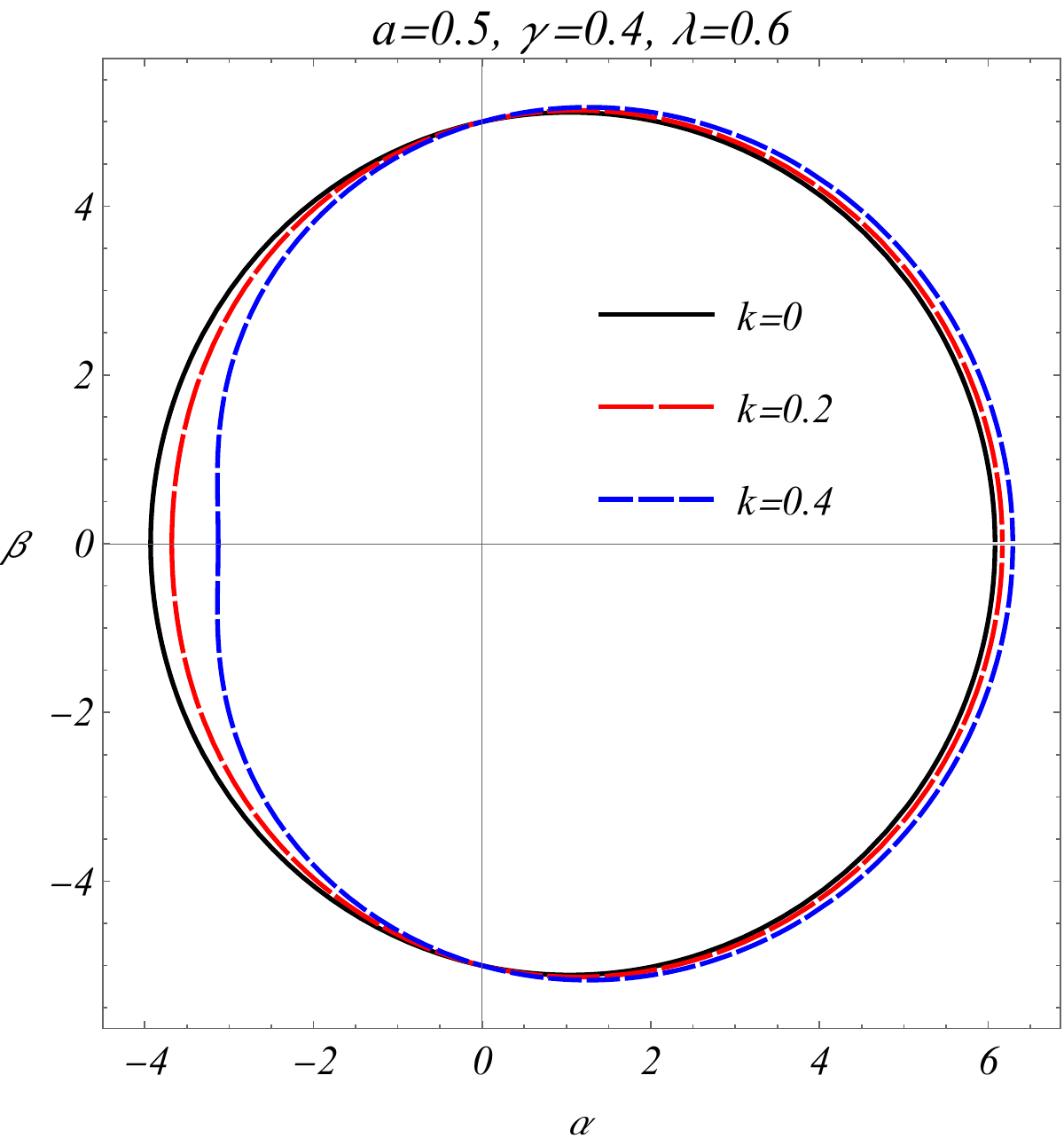}}
	\end{center}
	\caption{Shadow plots for various values of $k$ and fixed $a$, $\lambda$ and $\gamma$ corresponding to the case \ref{S3}. \label{Shc}}
\end{figure}

\subsection{Case III}\label{S3}
Using Eq. (\ref{58}) in the conditions $\mathcal{R}(r_p)=0=\partial_r\mathcal{R}(r_p)$, we get
\begin{eqnarray}
\Gamma(r_p)&=&\bigg[\frac{1}{a\Delta'(r)}\bigg(\mathcal{U}(r;a)\Delta'(r)-2r\Delta(r)-\mathcal{P}\bigg)\bigg]\bigg|_{r=r_p}, \label{77}\\
\Sigma(r_p)&=&\bigg[\frac{1}{a^2\Delta'(r)^2}\bigg(r^4\Delta'(r)^2\big[n^2-2\big]+4r\Delta(r)\Big(2r\big[a^2-\Delta(r)\big]-\Delta'(r)\big[a^2\big(n^2-1\big)+r^2\big(n^2-2\big)\big]\Big)\nonumber\\&&+2r\Big(2\big[a^2-\Delta(r)\big]+r\Delta'(r)\Big)\mathcal{P}+a^2\big(n^2-1\big)\Delta'(r)\Big[2r^2\big(2r+\Delta'(r)\big)+a^2\big(4r+\Delta'(r)\big)\Big]\bigg)\bigg]\bigg|_{r=r_p}, \label{78}
\end{eqnarray}
where,
\begin{eqnarray}
\mathcal{P}&=&\sqrt{4r^2\Delta(r)^2+\Delta'(r)\big(n^2-1\big)\mathcal{U}(r;a)\big[4r\Delta(r)-\mathcal{U}(r;a)\Delta'(r)\big]}. \label{79}
\end{eqnarray}
The differentials $\frac{d\phi}{dr}$ and $\frac{d\theta}{dr}$ in Eqs. (\ref{60}) and (\ref{61}) are simplified by using Eqs. (\ref{53})-(\ref{57}) and (\ref{49b}). Then solving the limits, we get
\begin{eqnarray}
\alpha(r_p)&=&-\frac{\Gamma(r_p)\csc\theta_0}{n}, \label{80} \\
\beta(r_p)&=&\pm\frac{\sqrt{\Sigma(r_p)+a^2\big(1-n^2\sin^2\theta_0\big)-\Gamma(r_p)^2\cot^2\theta_0}}{n}. \label{81}
\end{eqnarray}
Whereas, for the equatorial observer,
\begin{eqnarray}
\alpha(r_p)&=&-\frac{\Gamma(r_p)}{n}, \label{82} \\
\beta(r_p)&=&\pm\frac{\sqrt{\Sigma(r_p)-a^2\big(n^2-1\big)}}{n}. \label{83}
\end{eqnarray}
We have plotted the shadows in the Fig. \ref{Shc} by varying $k$ representing each curve and fixing the values of $a$, $\lambda$ and $\gamma$. With smaller values assigned to the BH parameters, an increase in $k$ results in minimal changes in the size of the shadow. Whereas, for higher values of BH parameters, the shape and size of the shadow show a significant variation. The parameter $k$ behaves differently in this case as compared to the one in the case \ref{S2}. As for the higher values of $a$, $\gamma$ and $\lambda$, the shape of the shadows changes significantly with increase in $k$.

\section{Distortion and Energy Emission Rate}
The rotating BH shadow results carry some useful information apart from its shape and size. We apply these shadow results in determining the deviation in the shape of shadow from a circular shadow image of the static BH, termed as distortion. This is because the rotating BH shadow is flattened on one side. We also apply the shadow results in estimating the energy emission rate that describes evaporation rate of the BH.

The shadow radius \cite{80,81} is given as
\begin{equation}
R_{sh}=\frac{D_c^2+\beta_t^2}{2|D_c|}, \label{84}
\end{equation}
where, $D_c=\alpha_t-\alpha_r$. It is required in order to define the distortion. We consider that the points $(\alpha_t,\beta_t)$, $(\alpha_b,\beta_b)$ and $(\alpha_r,0)$ of the shadow coincides with the imaginary circle. The indices $b$, $r$ and $t$ correspond to the bottom, right and top most points on the shadow, respectively. One can see the Fig. $\textbf{9}$ in \cite{81} for a detailed view of the image describing these points. The relation (\ref{84}) does not apply in determining the radius of static BH shadows. The distortion is mathematically expressed as
\begin{equation}
\delta_s=\frac{|\tilde{D}_c|}{R_{sh}}, \label{85}
\end{equation}
where, $\tilde{D}_c=\bar{\alpha}_l-\alpha_l$. The point $(\alpha_l,0)$ lies on the shadow and $(\bar{\alpha}_l,0)$ lies on the hypothetical circle. The index $l$ represents the points on the shadow on the left side of the vertical axis, whereas the index $\bar{l}$ represents the points on the imaginary circle on the left side of the vertical axis.

Classically, an object or a particle cannot escape out, once it falls in a BH. However, quantum mechanically, a BH emits energy in the form of radiation. The quantum fluctuations in a BH creates or annihilates particles, generating an enormous pressure caused by the rise in density, resulting in a substantial quantity of energy. Hence, the particles that possess positive energy, escape out of the BH due to a process known as quantum tunneling. Hence, the BH evaporates as the energy is released carried by the particles. It is measured as an absorption process called absorption cross section which is a measure of the probability. The BH shadow can be useful in determining the absorption cross section at high energies. The geometric area of the shadow is approximately equal to $\sigma_{lim}\approx\pi R_{sh}^2$. The energy emission rate is written as \cite{82,83,84}
\begin{equation}
\mathcal{E}_{\omega t}:=\frac{d^2\mathcal{E}(\omega)}{dtd\omega} =\frac{2\pi^2\omega^3\sigma_{lim}}{e^{\frac{\omega}{T_H}}-1}\approx\frac{\pi^3\omega^3}{2\big(e^{\frac{\omega}{T_H}}-1\big)}\frac{\big((\alpha_t-\alpha_r)^2+\beta_t^2\big)^2}{|\alpha_t-\alpha_r|^2}. \label{86}
\end{equation}
In above Eq. (\ref{86}), the parameter $\omega$ denotes the photon's angular frequency and $T_H=\frac{\kappa}{2\pi}$ defines the Hawking temperature. The relation
\begin{equation}
\kappa(r_h)=\lim\limits_{\theta=0, r\rightarrow r_h}\frac{\partial_r\sqrt{g_{tt}}}{\sqrt{g_{rr}}} \label{87}
\end{equation}
determines the surface gravity at the event horizon of a rotating BH \cite{85}. For the metric (\ref{8}) the surface gravity reduces to
\begin{equation}
\kappa(r_h)=\frac{\Delta'(r_h)}{2(r_h^2+a^2)} \label{88}
\end{equation}
for the rotating metric. By using Eqs. (\ref{85}) and (\ref{86}), we plot the behavior of distortion and BH evaporation rate for the plasma cases as discussed in the previous sections. It is believed that the KR field is a quantum field that has some obvious connections with String Theory. In this regard, such a quantum field may have a great impact on a BH in KR gravity with various underlying quantum effects. We can possibly encounter such quantum effects in studying the energy emission rate that corresponds to the BH evaporation rate. It has already been mentioned that the quantum tunneling process enables the particles carrying energy to escape away from a BH and we are intended to measure this release of energy in this work. Inside a BH, due to sufficiently high pressure and density, many quantum processes take place during the creation and annihilation of particles. This is one reason to believe that a BH in KR gravity is highly preferable to study the quantum aspects of spacetime and BHs. We also know that a plasma medium is a super hot ionized gaseous matter made up of particles that are ions. Due to high temperature of the medium and therefore the pressure, there may arise various quantum processes because of electrons splitting away from the nuclei. In this work, we are studying the BH in KR gravity that is immersed in plasma medium. Therefore, it is most likely that the quantum processes take place in the vicinity of the BH and it might fascinate the people working in quantum gravity and quantum field theory in curved spacetimes.

\subsection{Case I}
This case corresponds to the plasma distribution with frequency given by Eq. (\ref{21}). For the functions $f_r(r)$ and $f_\theta(\theta)$, we further divide this case into two subsections.

\subsubsection{Case Ia}\label{d1a}
\begin{figure}[t!]
	\begin{center}
		\subfigure{\includegraphics[width=0.41\textwidth]{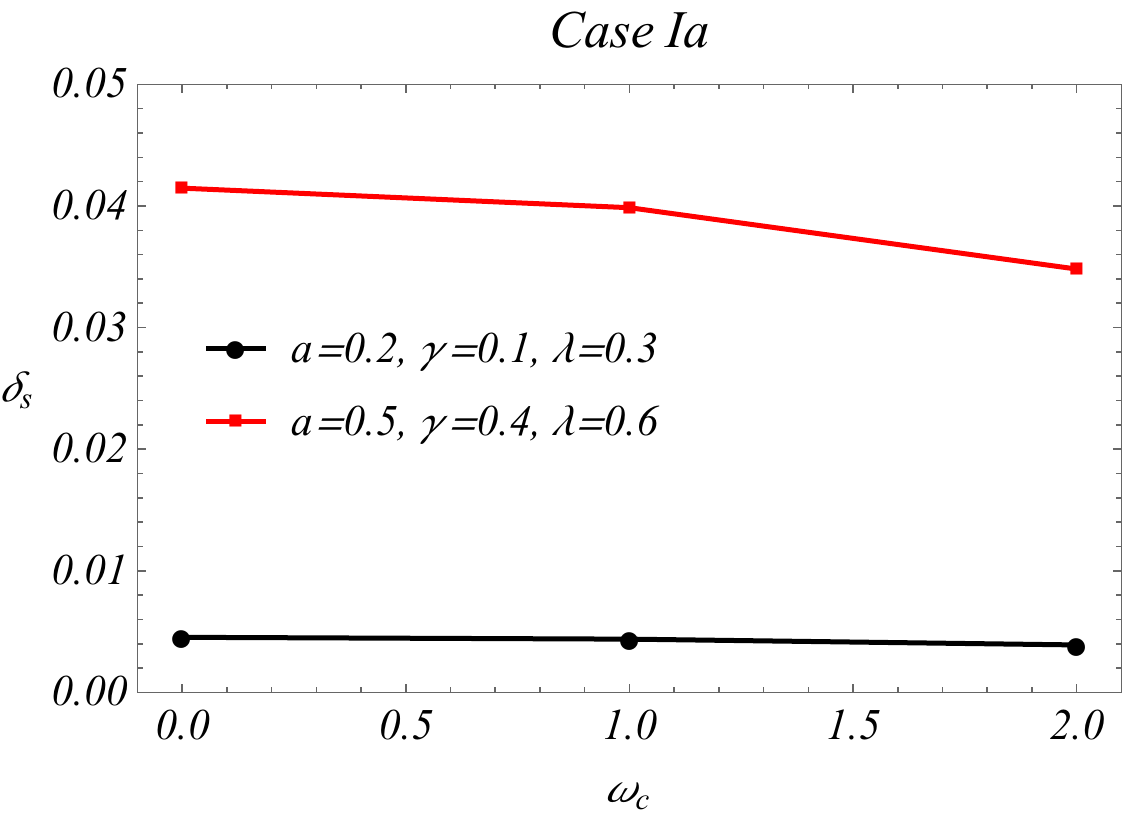}}~~~
		\subfigure{\includegraphics[width=0.41\textwidth]{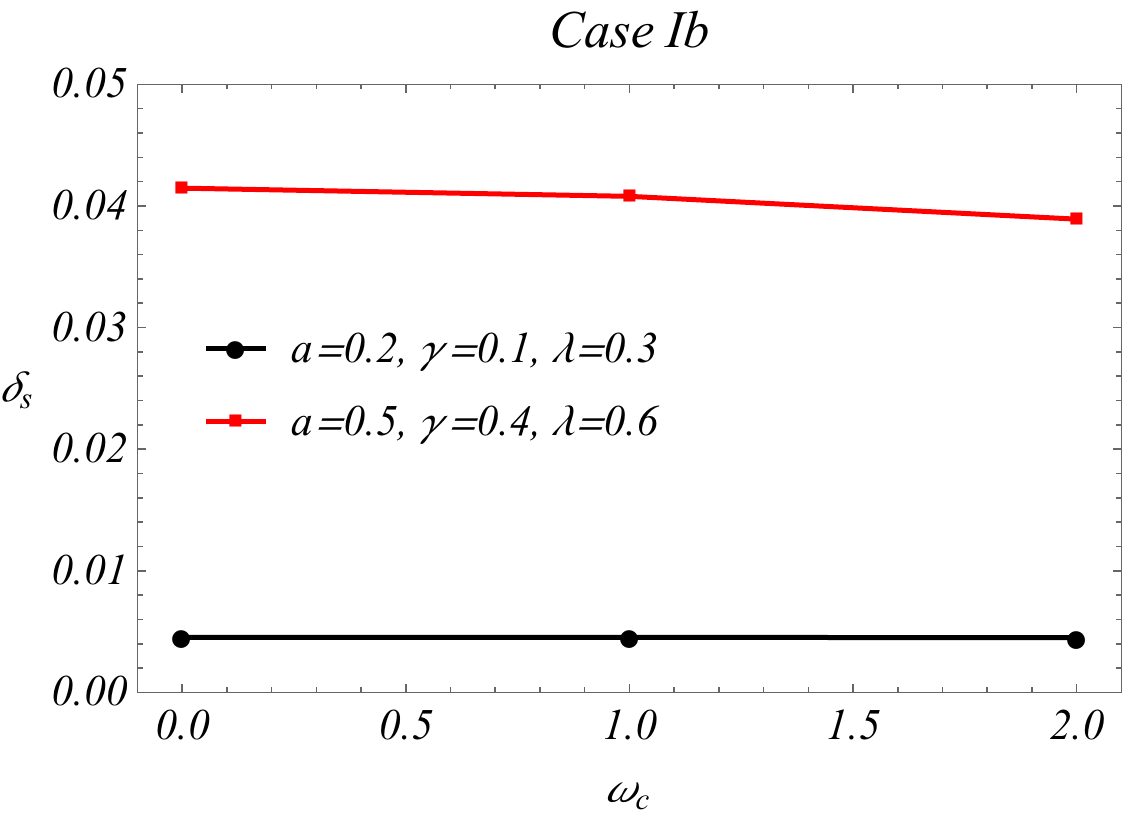}}
		\subfigure{\includegraphics[width=0.41\textwidth]{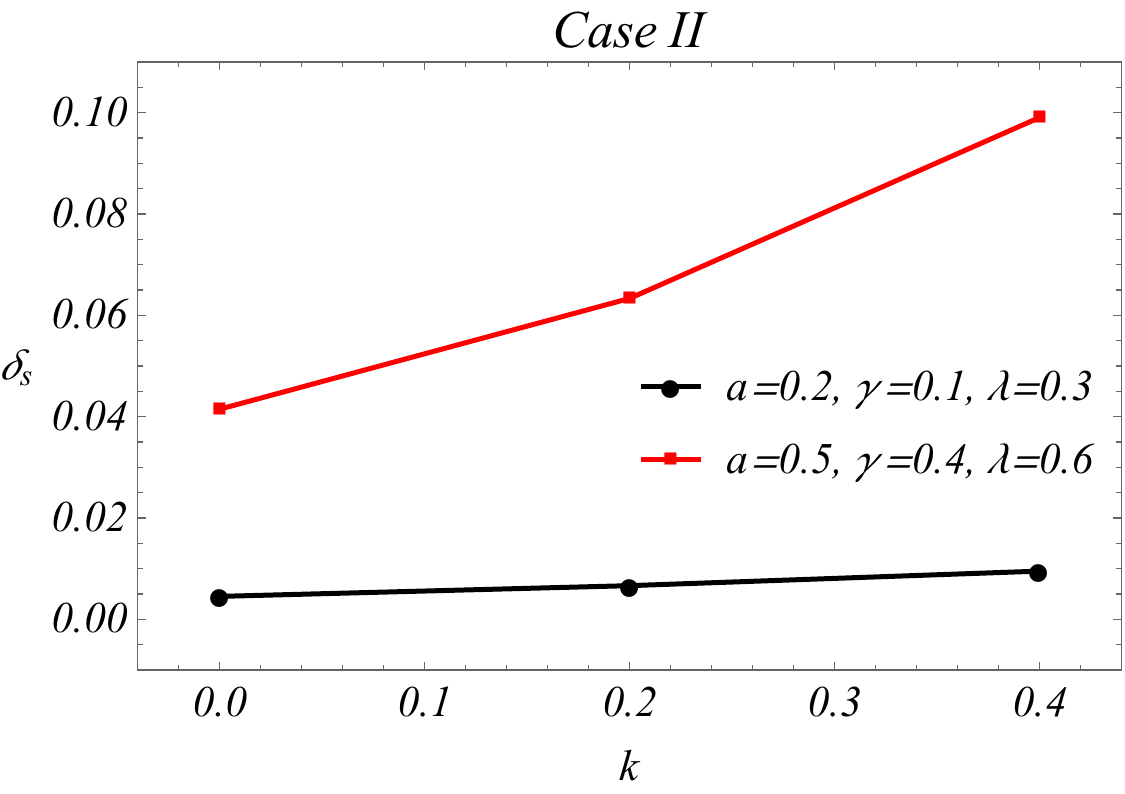}}~~~
		\subfigure{\includegraphics[width=0.41\textwidth]{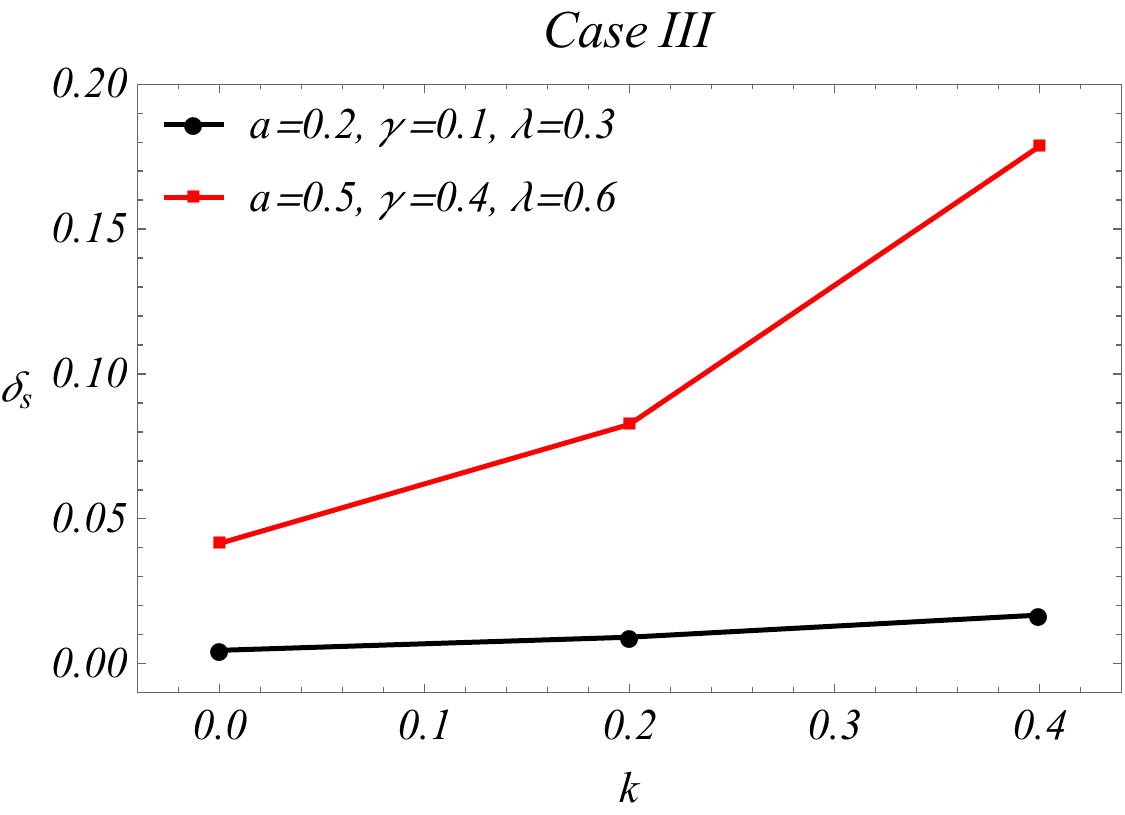}}
	\end{center}
	\caption{The distortion behavior for different values of plasma parameters and fixed $a$, $\lambda$ and $\gamma$. \label{dis}}
\end{figure}

\begin{figure}[t!]
	\begin{center}
		\subfigure{\includegraphics[width=0.41\textwidth]{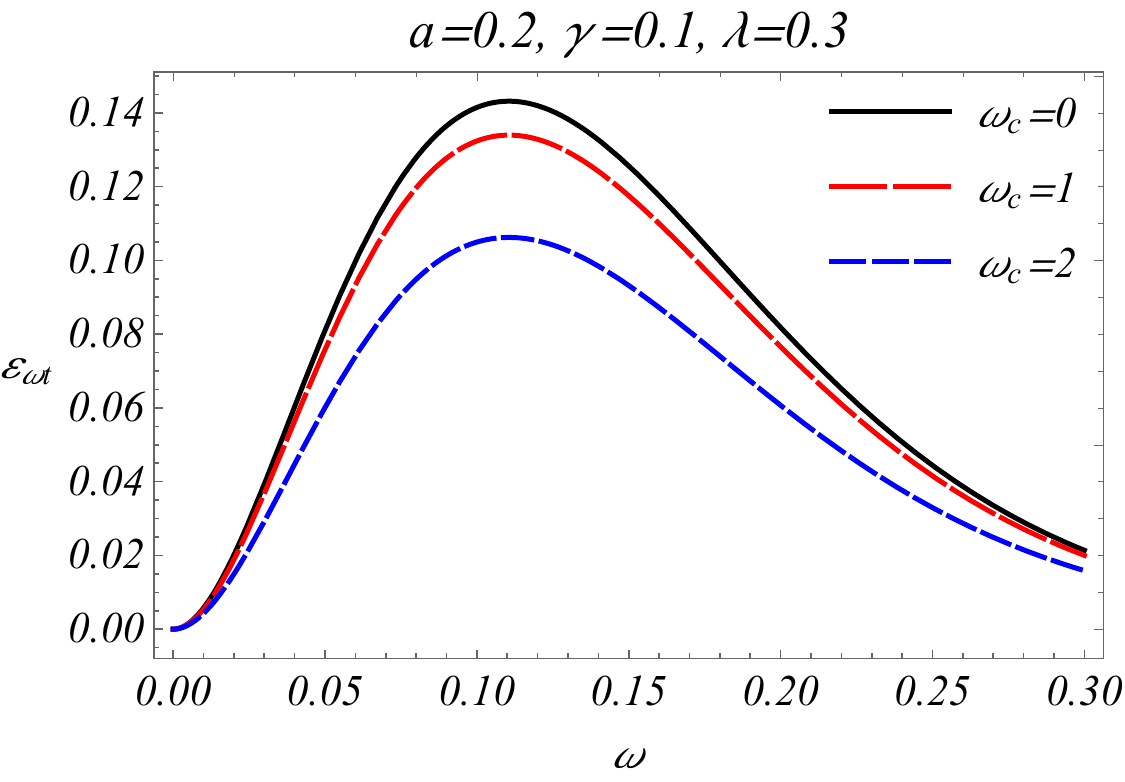}}~~~
		\subfigure{\includegraphics[width=0.41\textwidth]{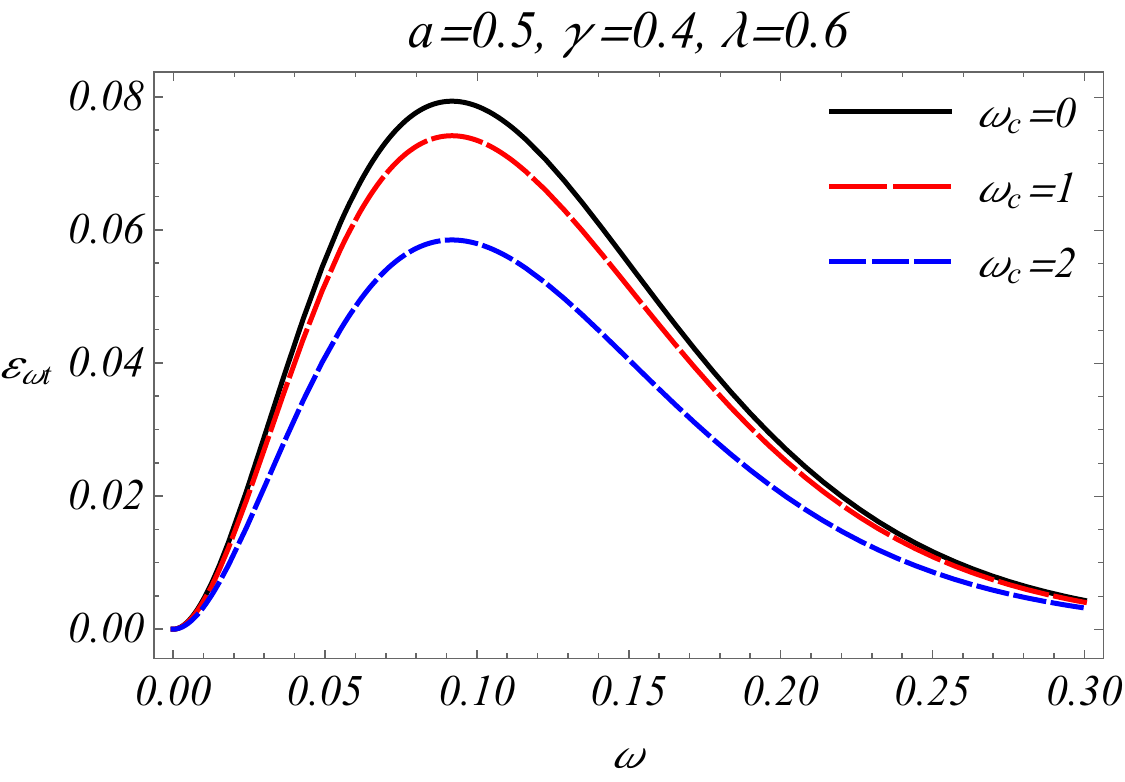}}
		\subfigure{\includegraphics[width=0.41\textwidth]{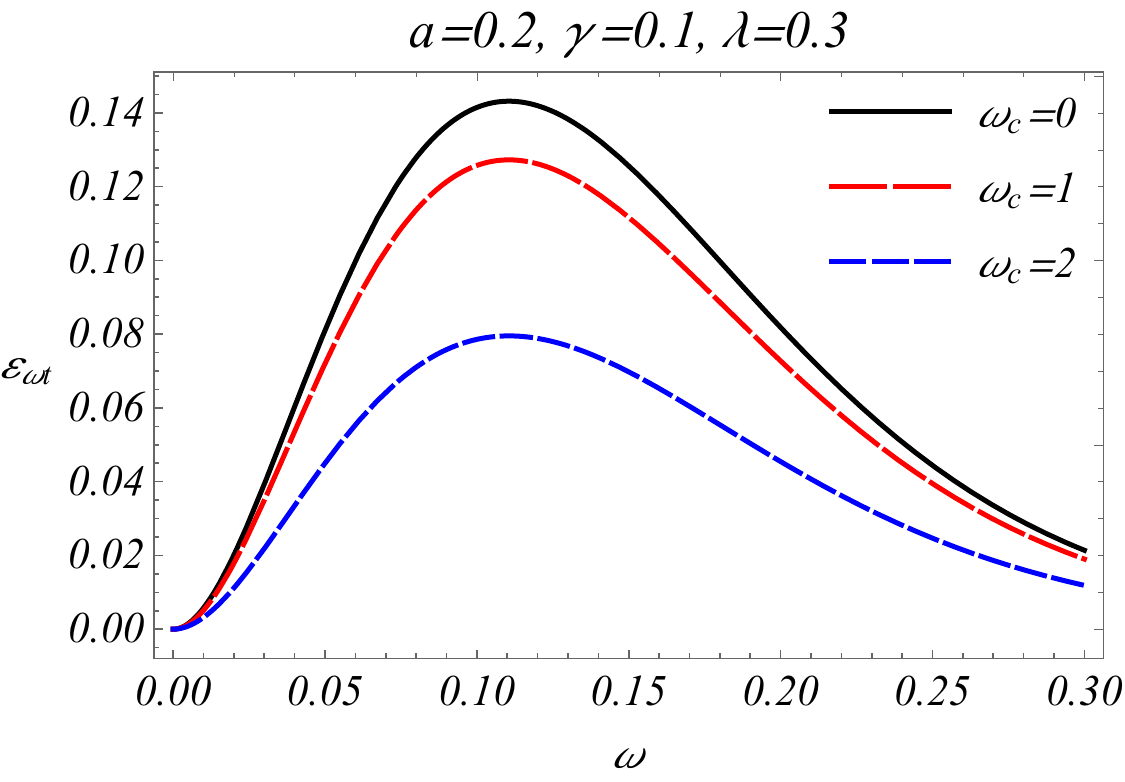}}~~~
		\subfigure{\includegraphics[width=0.41\textwidth]{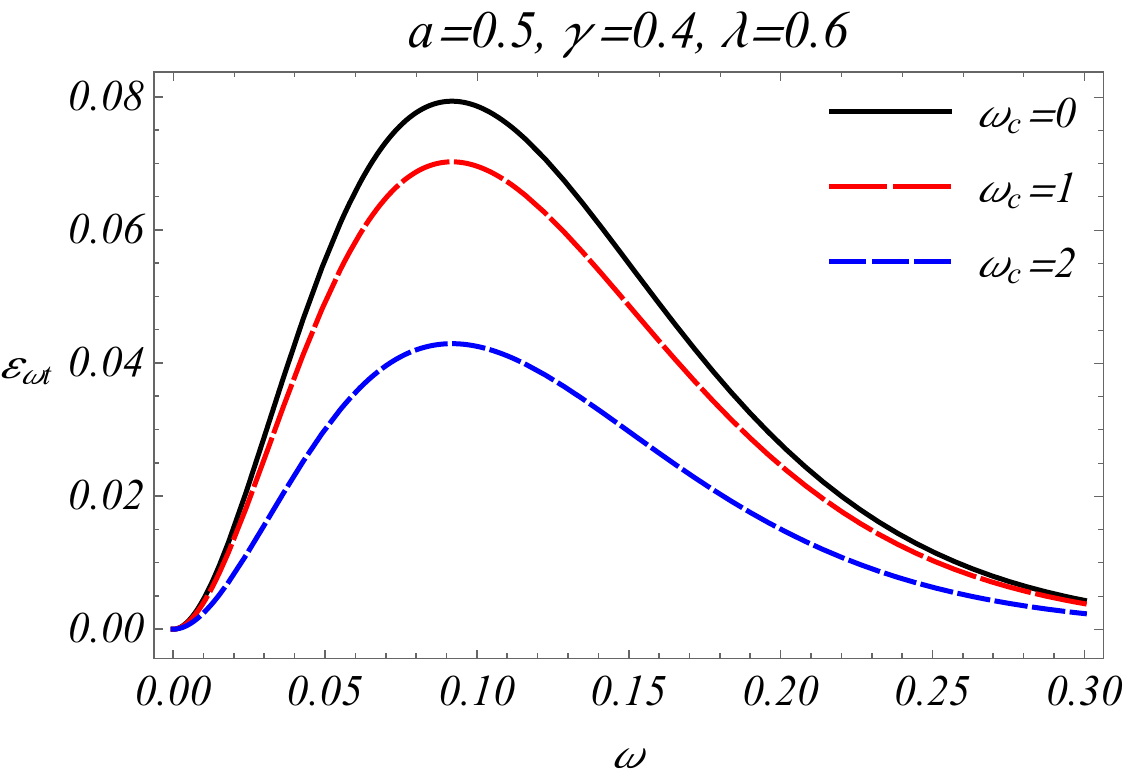}}
		\subfigure{\includegraphics[width=0.41\textwidth]{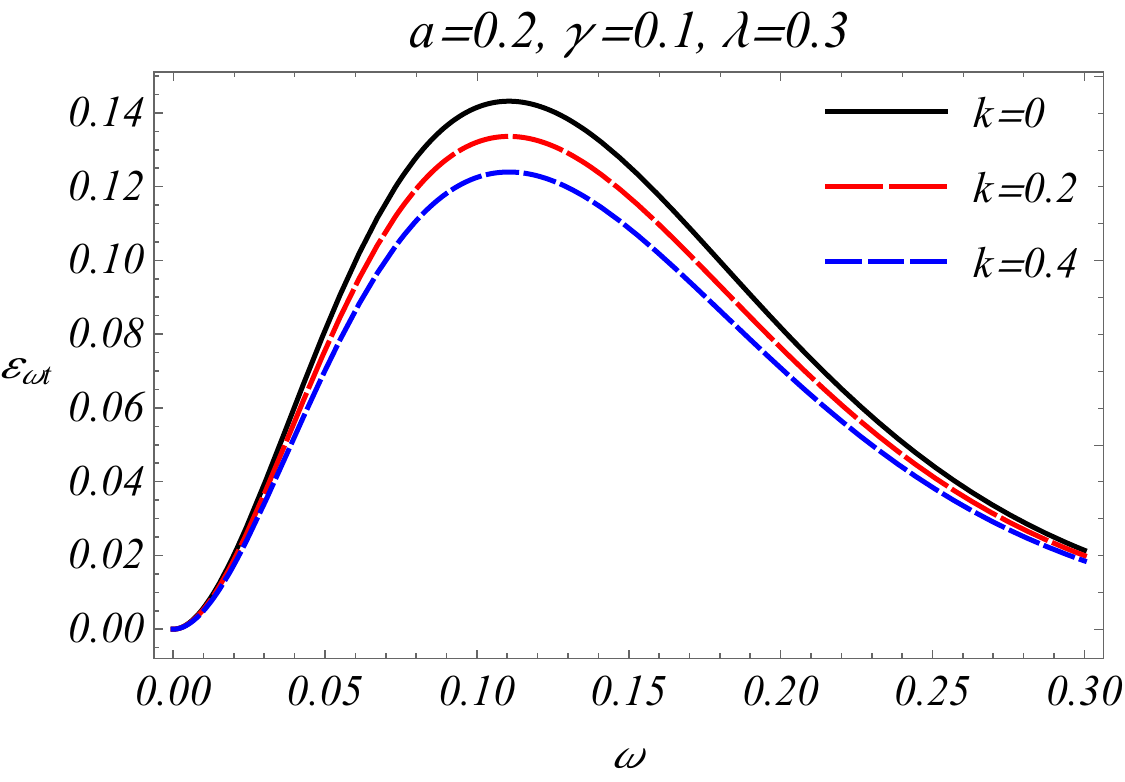}}~~~
		\subfigure{\includegraphics[width=0.41\textwidth]{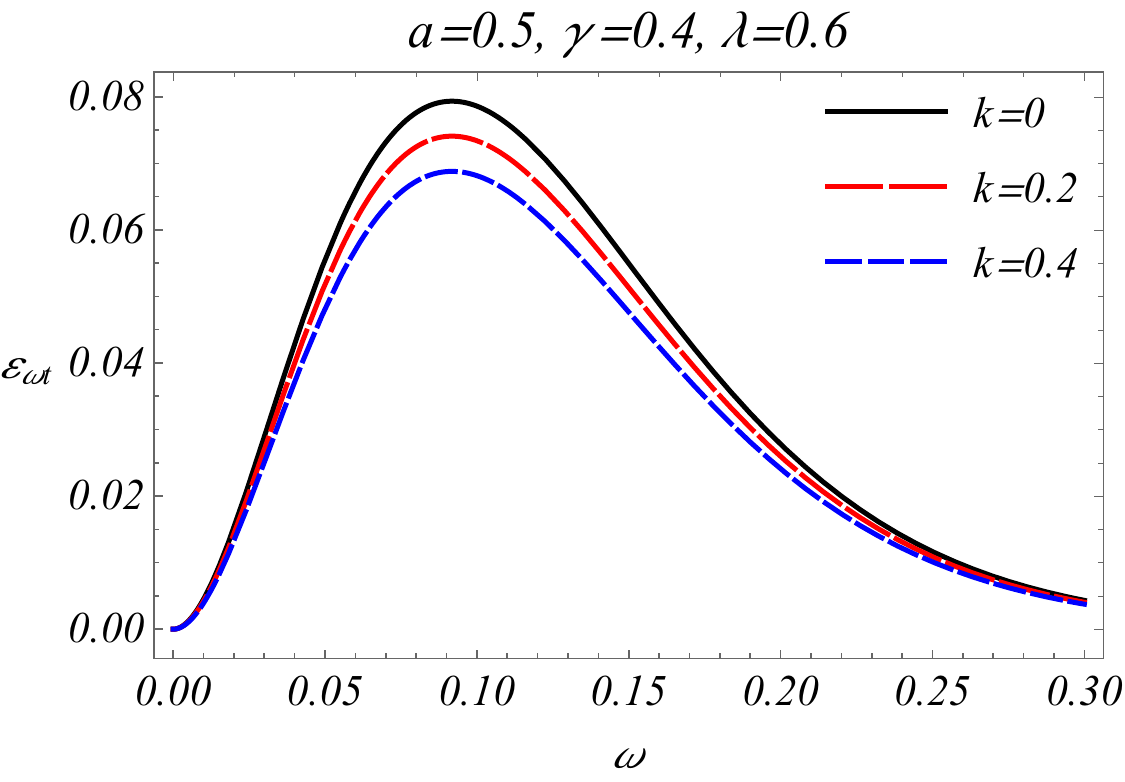}}
		\subfigure{\includegraphics[width=0.41\textwidth]{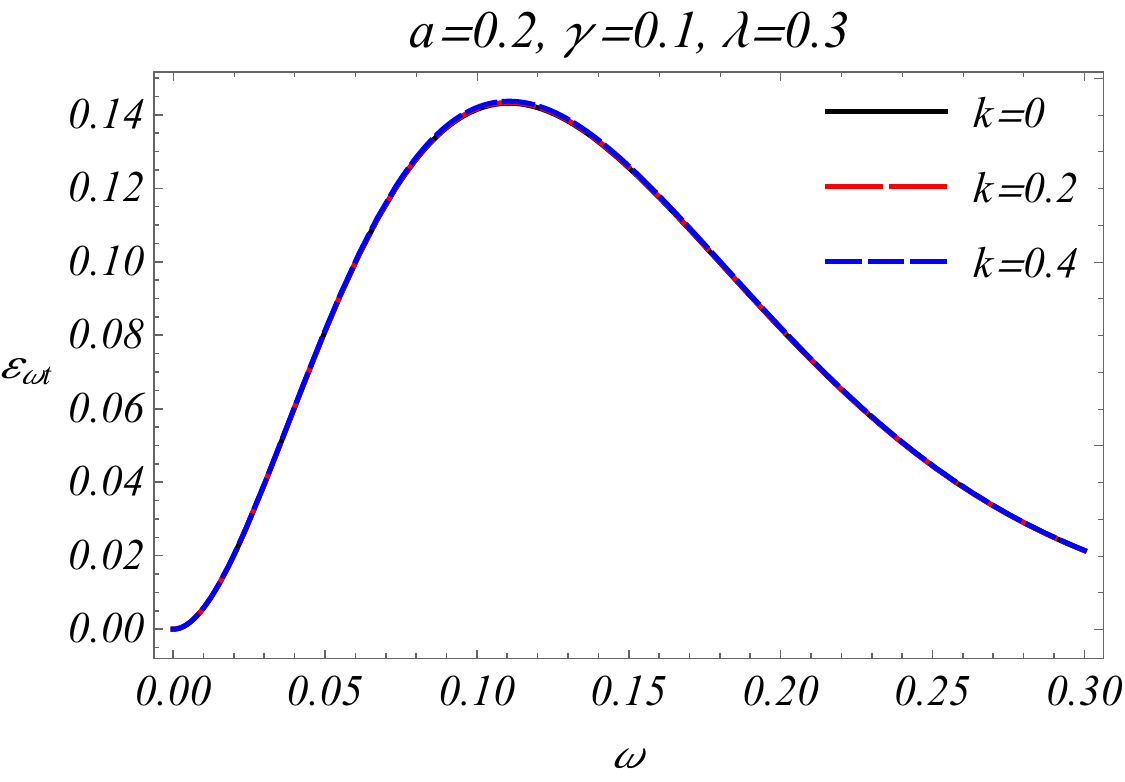}}~~~
		\subfigure{\includegraphics[width=0.41\textwidth]{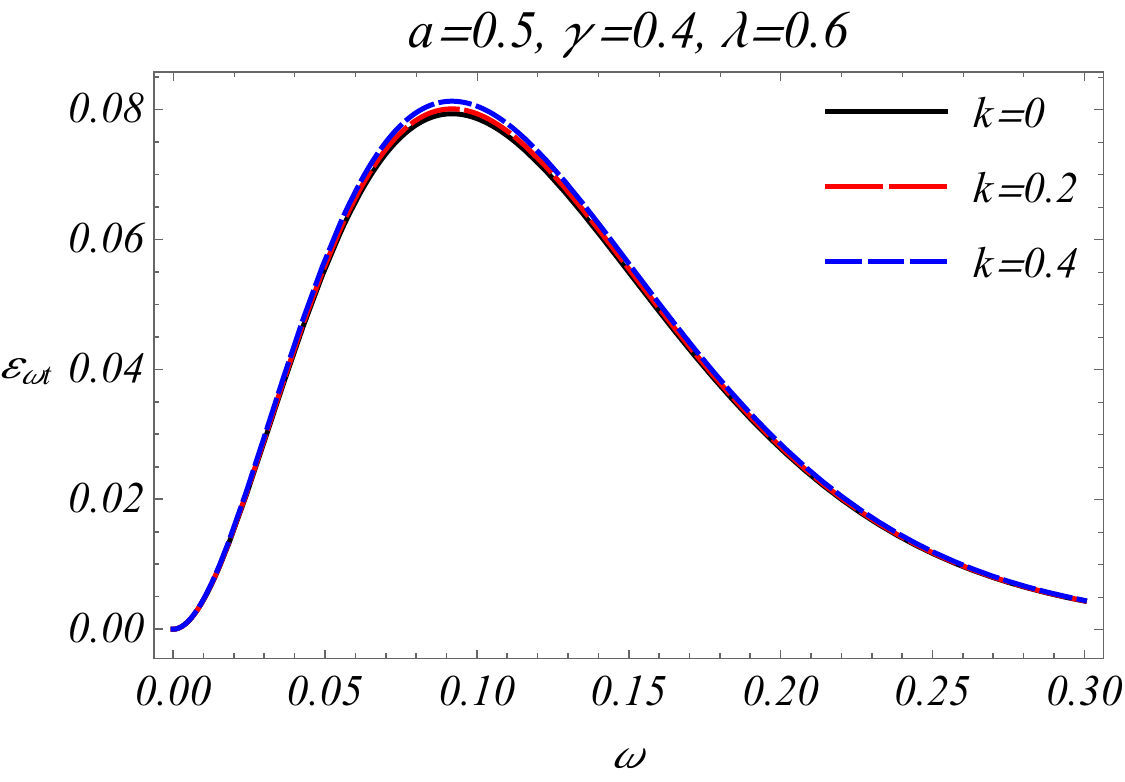}}
	\end{center}
	\caption{The graphical variation in the energy emission rate for different values of plasma parameters and fixed $a$, $\lambda$ and $\gamma$. \label{eer}}
\end{figure}

For the case when $f_r(r)$ and $f_\theta(\theta)$ are given by the Eq. (\ref{32}), the upper right plot in the Fig. \ref{dis} depicts the variation in the distortion. The smaller values of the BH parameters generate a constant behavior of the distortion as the plasma parameter $\omega_c$ grows larger. When we increase the values of the BH parameters, the distortion decreases as the value of $\omega_c$ increases. Moreover, for the higher values of the BH parameters, the distortion is significantly higher for a fixed value of $\omega_c$. The energy emission rate for this case is plotted in the top panel of Fig. \ref{eer}. It is found that for the smaller values of the BH parameters, the energy emission rate and hence the BH evaporation rate decreases with increase in $\omega_c$. Similar behavior of the BH evaporation rate has been observed when the values of the BH parameters are increased. However, for the fixed values of $\omega_c$, the BH evaporation slows down with increase in the BH parameters.

\subsubsection{Case Ib}\label{d1b}
Considering the values of $f_r(r)$ and $f_\theta(\theta)$ as given by the Eq. (\ref{33}), we plotted the distortion in the Fig. \ref{dis} as indicated by the corresponding label. We found that the smaller values of the BH parameters give the a constant behavior of the distortion as we increase $\omega_c$. When we increase the values of the BH parameters, the distortion decreases as the value of $\omega_c$ increases. Moreover, for the higher values of the BH parameters, the distortion is significantly higher for a fixed value of $\omega_c$. This shows that there is no significant change in the distortion behavior as compared to that in the case \ref{d1a}. The energy emission rate for this case is plotted in the second panel of Fig. \ref{eer}. We found that for the smaller values of the BH parameters, the BH evaporation rate decreases with increase in $\omega_c$. By increasing the values of the BH parameters, a similar kind of behavior of the BH evaporation rate has been observed. However, for the fixed values of $\omega_c$, the BH evaporation slows down with increase in the BH parameters. It also shows that in terms of BH evaporation, $\omega_c$ is more sensitive in this case as compared to the case \ref{d1a}.

\subsection{Case II}\label{d2}
This case corresponds to the plasma distribution defined in terms of refractive index as given in Eq. (\ref{49a}). We plotted the distortion in the lower left plot in Fig. \ref{dis} as indicated by the corresponding label. We can see that the distortion increases at a very slow rate with respect to increase in the value of $k$ and for the smaller values of the BH parameters. When we increased the values of the BH parameters, the distortion increases rapidly as the value of $k$ increases. Moreover, for a fixed value of $k$, the distortion is significantly higher for the higher values of the BH parameters. This shows a converse behavior of distortion as compared to the previous two subcases \ref{d1a} and \ref{d1b}. The energy emission rate for this case is plotted in the third panel of Fig. \ref{eer}. The plots show that for the smaller values of the BH parameters, the BH evaporation slows down with increase in $k$. By increasing the values of the BH parameters, the variation in the BH evaporation rate is same as for the smaller values of the BH parameters. However, for the fixed values of $k$, the BH evaporation is significantly delayed with increase in the BH parameters.

\subsection{Case III}\label{d3}
For this case, the plasma distribution is defined as the constant value of refractive index given as in Eq. (\ref{49b}). We plotted the distortion in the Fig. \ref{dis} as indicated by the corresponding label. We can see that for when BH parameters are kept smaller, there is a slow elevation in distortion with increase in $k$. When the values of the BH parameters are elevated, a rapid growth in the distortion is observed as the value of $k$ grows larger. Moreover, when the value of $k$ is kept fixed, the distortion is significantly higher for the higher values of the BH parameters. This shows an identical behavior of distortion as compared to the case \ref{d2}. The energy emission rate for this case is plotted in the bottom panel of Fig. \ref{eer}. As the BH parameters take on smaller values in the left plot, the rate of BH evaporation remains constant with the ascending values of $k$. Elevating the BH parameters leads to a minor fluctuation in the rate of BH evaporation. However, this variation shows that with increase in the value of $k$, the peak rises and the BH evaporation rate increases. This is an entirely different behavior of the BH evaporation rate among all of the discussed cases.

\section{Constraints on Black Hole Parameters}
Considering the BH in plasma medium, we will establish the bounds on the BH parameters $\gamma$, $a$ and $\lambda$ in this section by utilizing the data acquired through the observations at EHT for M87* and Sgr A*. Therefore, we establish a comparative analysis of the angular radii of the shadows of BH defined by the metric(\ref{8}) in plasma with M87* and Sgr A*. The BH parameters are said to be constrained when the BH shadow falls within 1-$\sigma$ uncertainty. Corresponding to these limits on the parameters, the BH (\ref{8}) immersed in plasma is considered as M87* or Sgr A*. This work exclusively focuses on rotating BH because supermassive BHs, by nature, exhibit rotational characteristics. It ensures a feasible and rigorous comparative study. A coordinate-independent formalism, vaguely known as Kumar-Ghosh method \cite{86,87}, is employed in which the shadow area $A_{sh}$ will be used that is defined as
\begin{eqnarray}
A_{Sh}&=&2\int_{r_-}^{r_+}dx^r\beta(r)\partial_r\alpha(r). \label{89}
\end{eqnarray}
The values $r_+$ and $r_-$ represent the size of the retrograde and prograde stable circular orbits as measured from the origin, respectively. Let us consider, the BH and the observer are separated by the linear distance $d$, then the diameter of the BH shadow is measured by using the relation for arc length as follows \cite{88,89}
\begin{eqnarray}
\theta_d=\frac{2R_{A_{sh}}}{d}, \label{90}
\end{eqnarray}
where $R_{A_{sh}}=\sqrt{\frac{A_{sh}}{\pi}}$ is the shadow radius. By employing the relations (\ref{89}) and (\ref{90}), the angular diameter of the BH shadow is expressed as a function involving the BH parameters and $\theta_0$. For the comparison of the shadows, we need to know the distance $d$ of Earth from M87* and Sgr A*, the mass $M$ and the shadow size $\theta_d$ of M87* and Sgr A*. For M87*, we get \cite{53,57,58}
\begin{eqnarray}
d=16.8Mpc, \qquad M=6.5\times10^9M_\odot, \qquad \theta_d=42\pm3\mu as, \label{m87}
\end{eqnarray}
where $M_\odot$ denotes the solar mass. For simplicity, we have disregarded uncertainties in the measurements of mass and distance. For Sgr A*, we get \cite{59,90}
\begin{eqnarray}
d=8kpc, \qquad M=4\times10^6M_\odot, \qquad \theta_d=48.7\pm7\mu as. \label{sgr}
\end{eqnarray}
Ultimately, the angular diameter of the shadow of the BH (\ref{8}) can be calculated. This calculated value can then be compared with the angular diameters of Sgr A* and M87* to assess constraints on the BH parameters. As we know that the largest amount of matter in the visible universe is plasma in the form of stars, nebulas and auroras etc. Therefore, due to strong gravity near a BH, it is quite certain that a plasma medium exists around a BH. Such a plasma medium is significantly crucial and influential on physics, especially the particle motion around the BH and various quantum aspects. Under the said comparative study, we mainly estimate the influence of plasma on the parametric bounds. From this analysis, we can deduce that what distribution of plasma and under what conditions it is likely to be present around the BH, if the BH (\ref{8}) immersed in plasma is one of M87* and Sgr A*. Therefore, the results obtained under this analysis might be useful for people working in observational astronomy and astroparticle physics, especially those who are interested in BH imaging. Note that the EHT observations were conducted at the inclination angles at $17^\circ$ for M87* and $<50^\circ$ for Sgr A*. Therefore, we perform the calculations for the shadows by considering the inclination angles of $17^\circ$ for M87* and $30^\circ$ for Sgr A*. Moreover, the black solid curves correspond to the values $\theta_d=42\mu as$ for M87* and $\theta_d=48.7\mu as$ for Sgr A*. Whereas, the dashed and dotted curves represent the boundaries of 1-$\sigma$ and 2-$\sigma$ uncertainty intervals.

\subsection{Case I}\label{e1}
Considering the plasma distribution with frequency given by Eq. (\ref{21}) and further assuming the two subcases for the functions $f_r(r)$ and $f_\theta(\theta)$, we determined the angular diameters of the BH shadows in terms of the parametric spaces.

\subsubsection{Case Ia}\label{e1a}
\begin{figure}[t!]
	\begin{center}
		\subfigure{\includegraphics[width=0.41\textwidth]{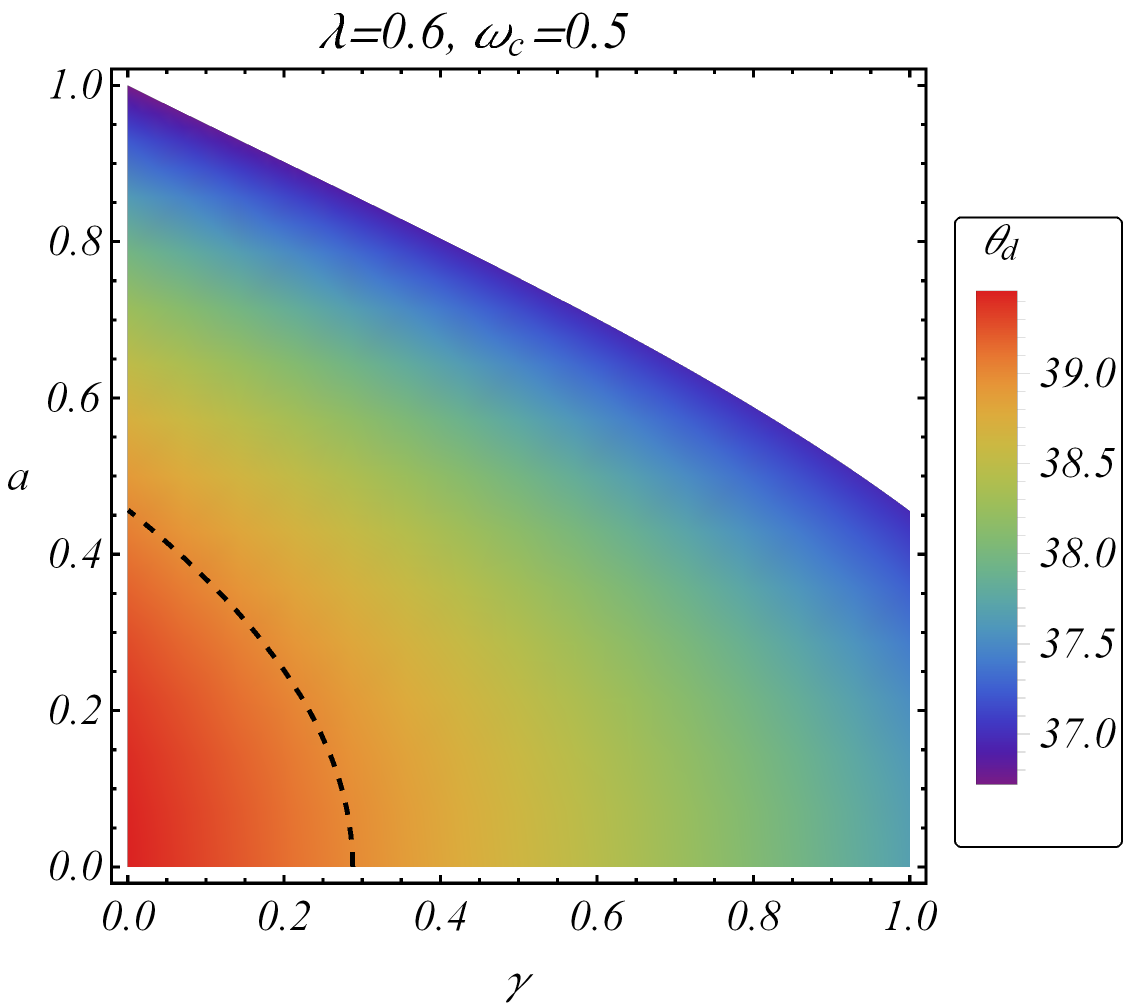}}~~~
		\subfigure{\includegraphics[width=0.41\textwidth]{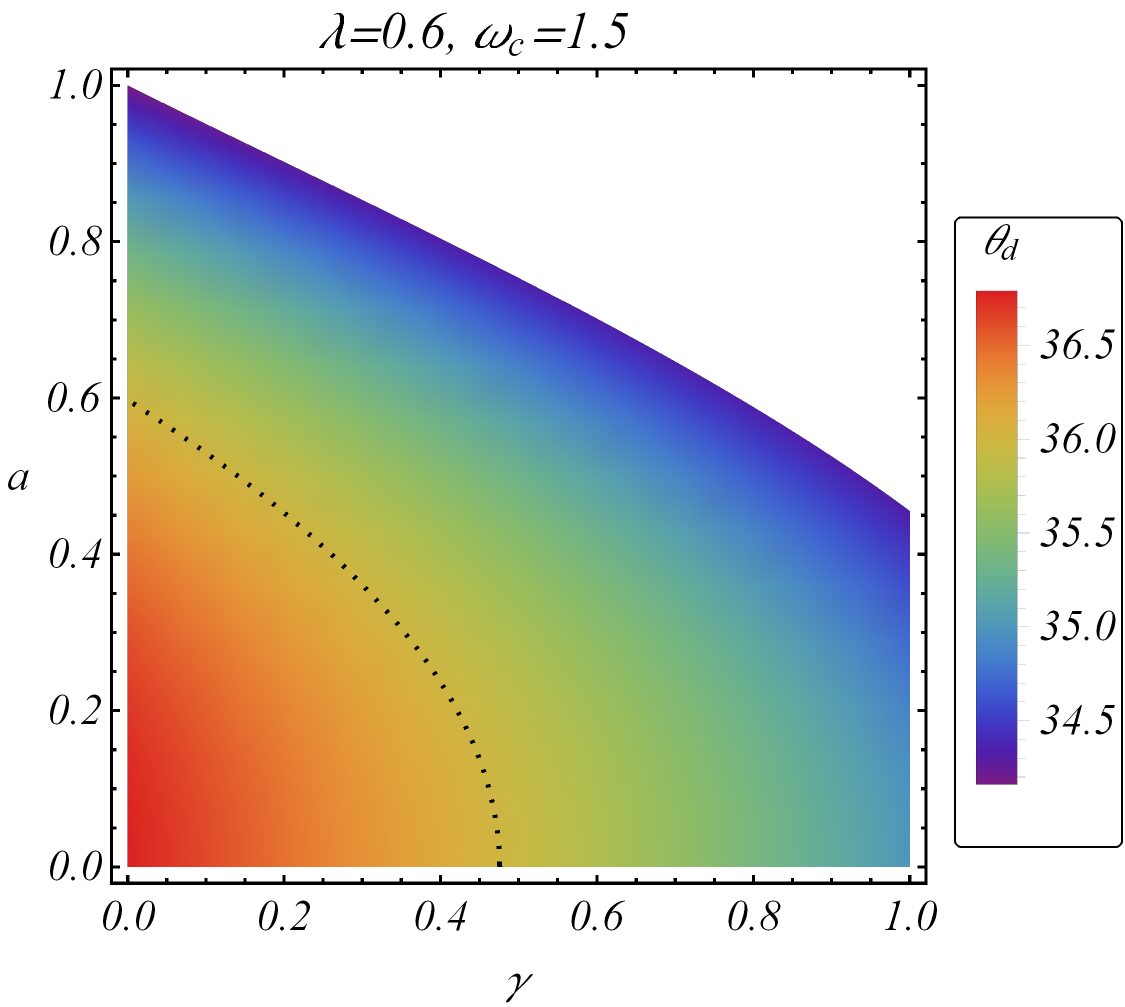}}
		\subfigure{\includegraphics[width=0.41\textwidth]{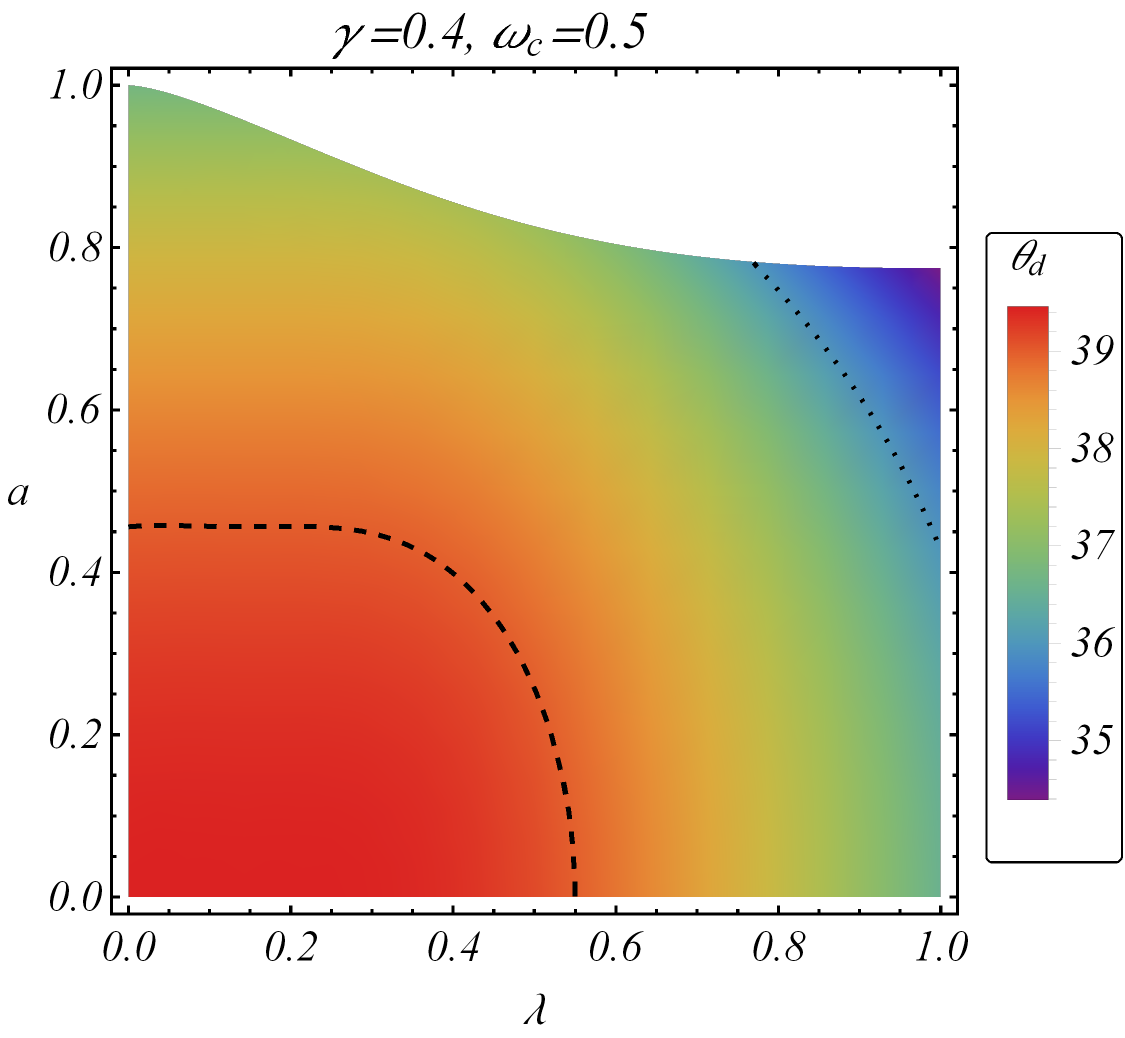}}~~~
		\subfigure{\includegraphics[width=0.41\textwidth]{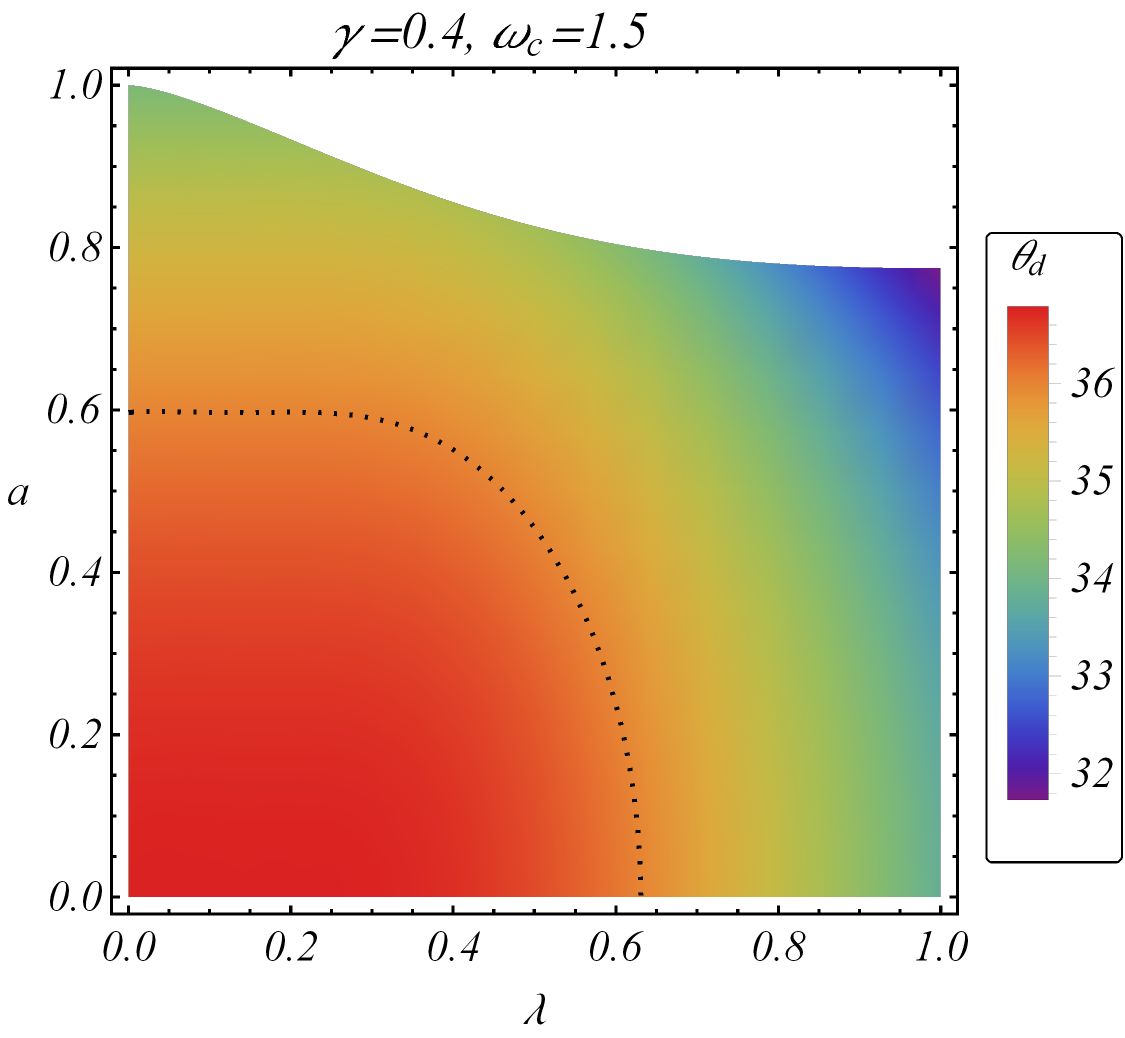}}
		\subfigure{\includegraphics[width=0.41\textwidth]{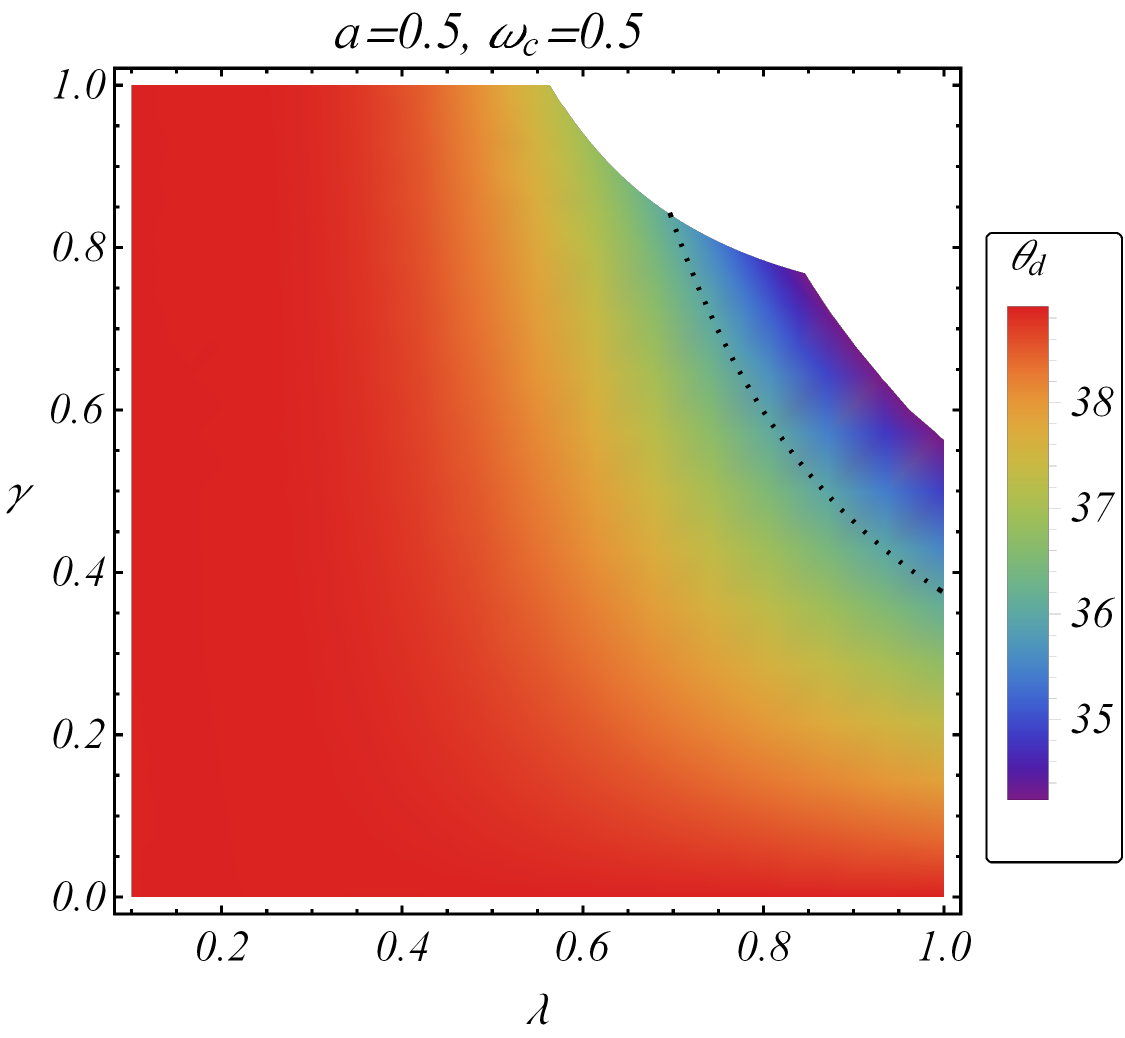}}~~~
		\subfigure{\includegraphics[width=0.41\textwidth]{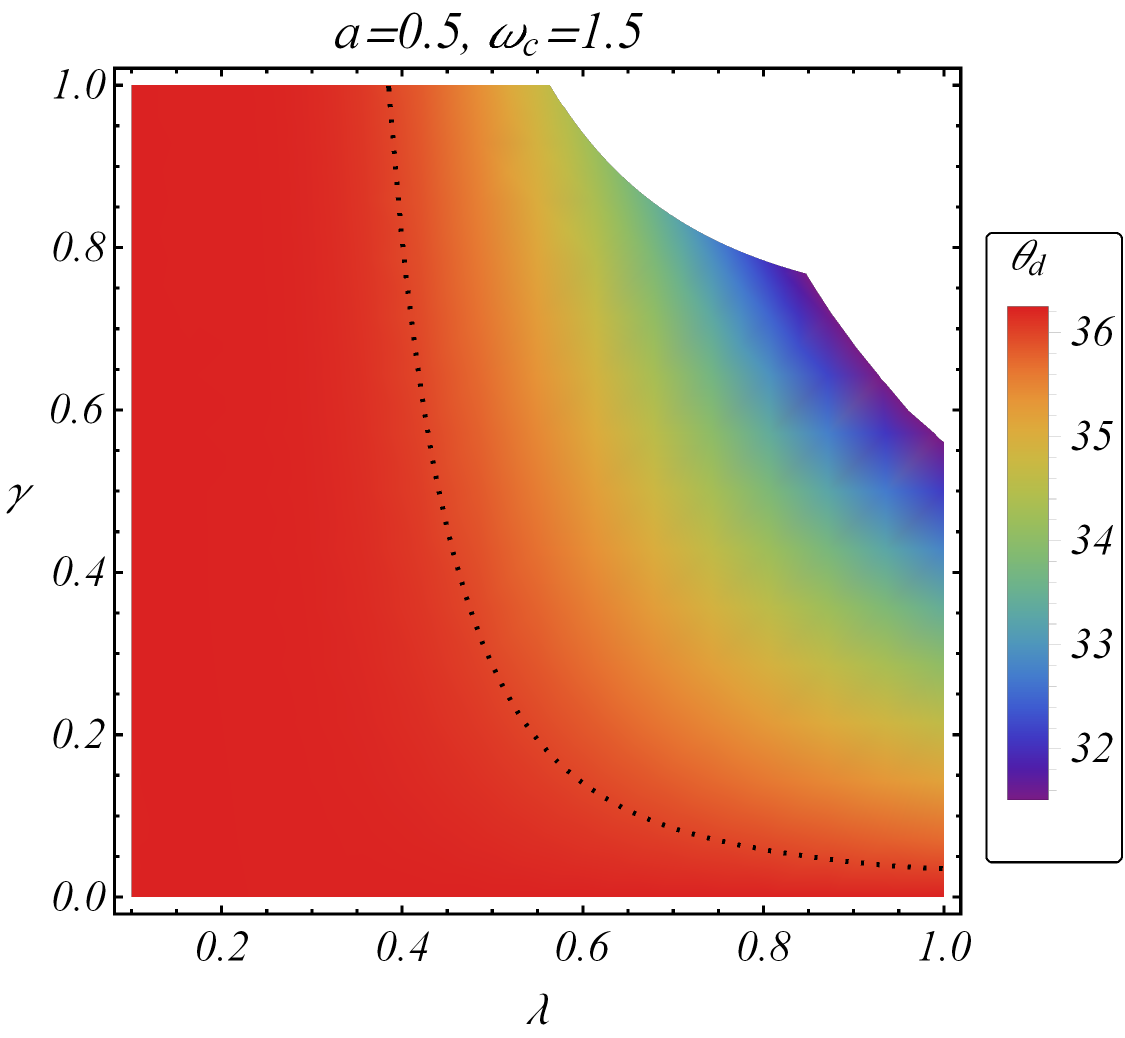}}
	\end{center}
	\caption{The density plot depicting the variation of angular size of the shadow of M87* in terms of parametric spaces for various values of plasma parameters for the case described by Eq. (\ref{32}). \label{m1}}
\end{figure}

The density plots depicted in the Fig. \ref{m1} illustrate the angular diameter across parametric spaces for M87*. These plots correspond to the specific case outlined by the Eq. (\ref{32}). The top panel corresponds to the parametric space $a$-$\gamma$ with $\lambda=0.6$ and two different values of $\omega_c$. In the left plot, we can find that for the given values of $\lambda$ and $\omega_c$, the limits of $a$ and $\gamma$ are shown by the dashed curve in the parametric space. For the specified parametric values below the dashed curve, the size of the shadow falls within the 1-$\sigma$ error level. However, in the right plot, it becomes evident that elevating the value of $\omega_c$ results in the dotted curve delineating the boundaries for both $a$ and $\gamma$. This indicates that the shadow size is confined inside the 2-$\sigma$ uncertainty level. Therefore, the BH (\ref{8}) immersed in plasma medium can be considered as M87* for the smaller value of $\omega_c$ and the corresponding parametric values under the dashed curve. The panel at the middle illustrates the parametric space $a$-$\lambda$ with $\gamma=0.4$ and two different values of $\omega_c$. In the left plot, the parametric space illustrates the constraints on $a$ and $\lambda$ for a given value of $\gamma$ and $\omega_c$, represented by the dashed and dotted curves. The size of the shadow falls within the 1-$\sigma$ error level for all values of the parameters situated below the dashed curve. Meanwhile, for parametric values below the dotted curve, the size of the shadow is within the 2-$\sigma$ uncertainty interval. The right plot shows that by increasing the value of $\omega_c$, the dotted curve determines the limits on $a$ and $\lambda$ which means that the shadow size is bounded within 2-$\sigma$ error level. Therefore, the BH (\ref{8}) in plasma background can be considered as M87* for the smaller value of $\omega_c$ and the corresponding parametric values bounded by the dashed curve. The lower panel corresponds to the parametric space $\gamma$-$\lambda$ with $a=0.5$ and two different values of $\omega_c$. In both plots, we find that for the given value of $a$ and $\omega_c$, the limits of $\gamma$ and $\lambda$ are shown by the dotted curves in the parametric space. The size of the shadow falls within the 2-$\sigma$ confidence level for all specified values of the parameters located below the dotted curves. Therefore, we exclude the consideration of the BH (\ref{8}) in plasma background as M87* for both values of $\omega_c$ and the corresponding parametric values.

\begin{figure}[t!]
	\begin{center}
		\subfigure{\includegraphics[width=0.41\textwidth]{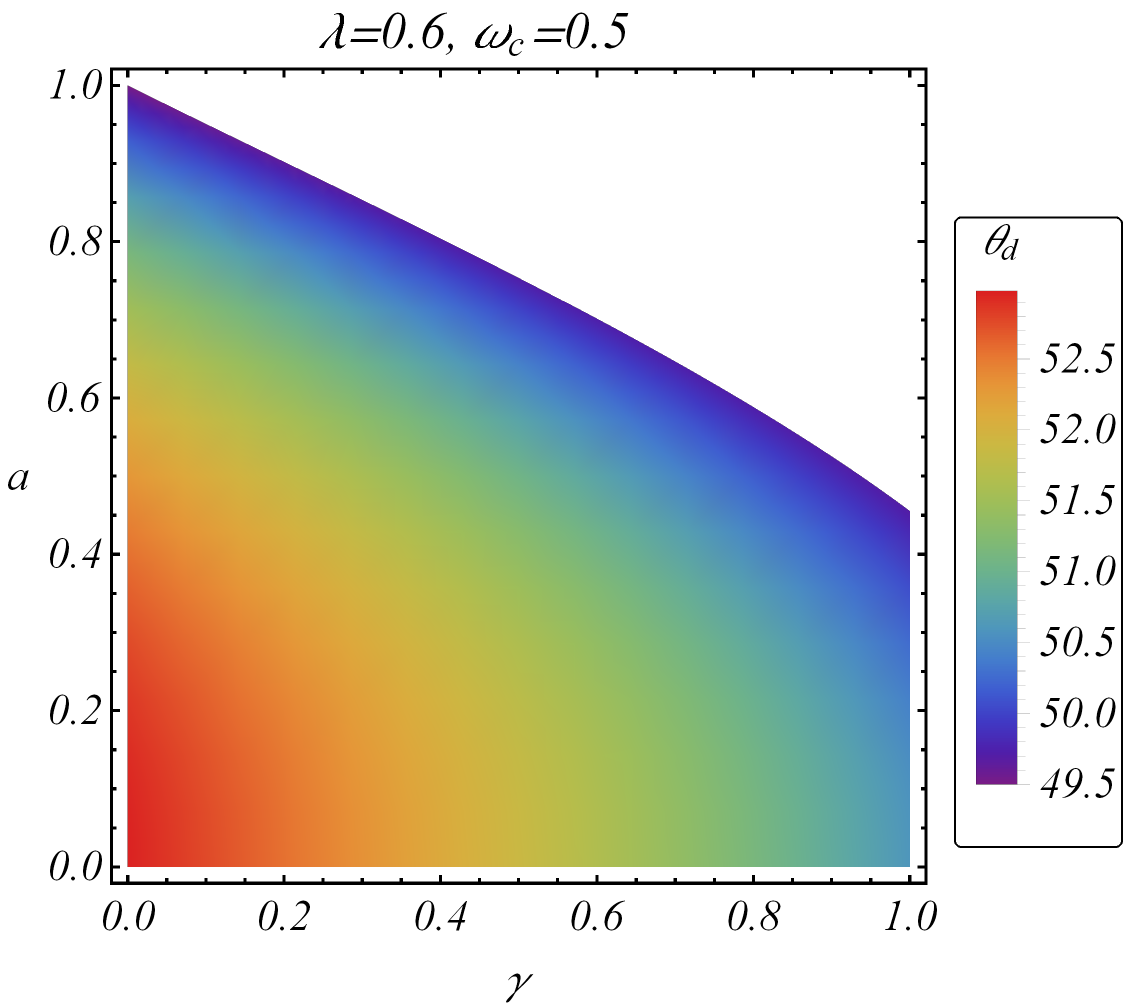}}~~~
		\subfigure{\includegraphics[width=0.41\textwidth]{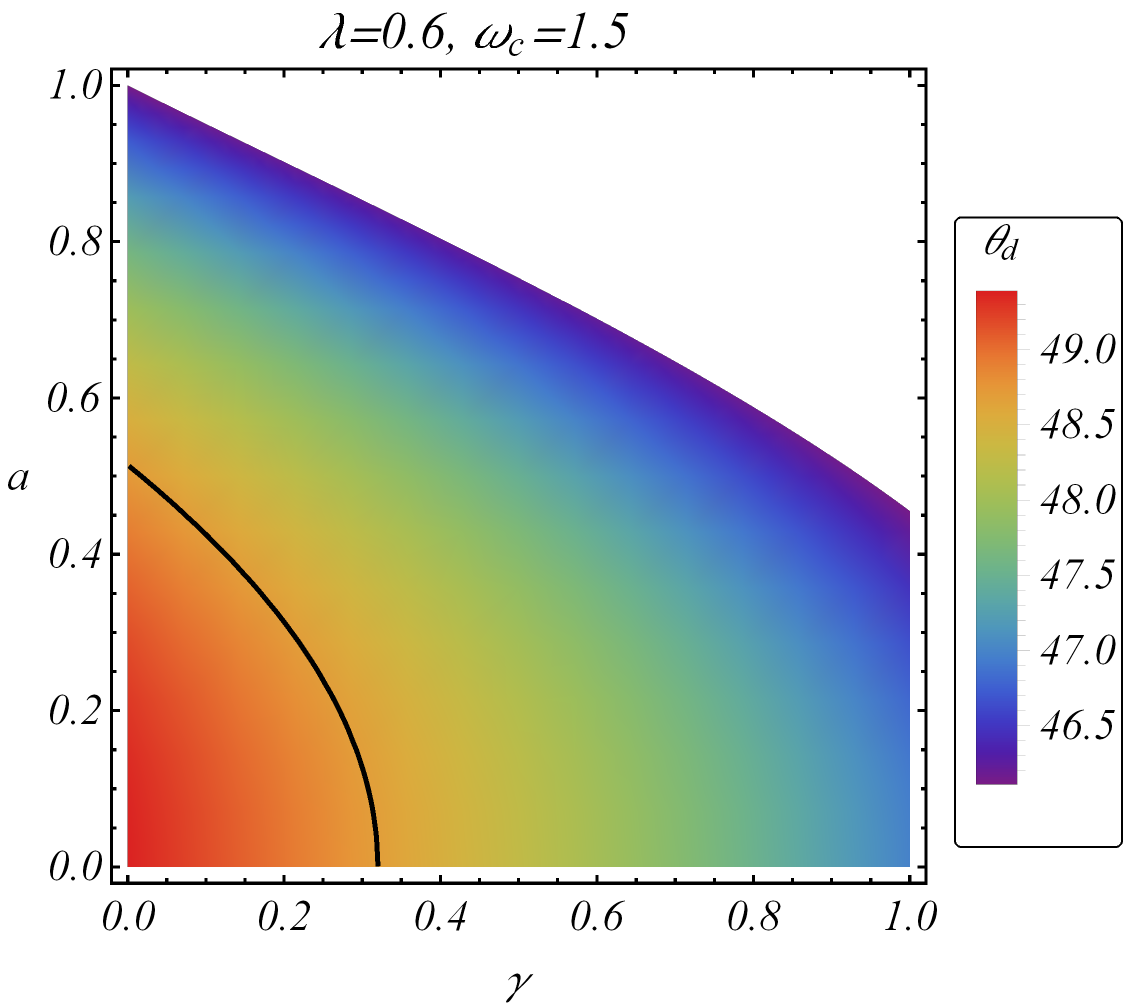}}
		\subfigure{\includegraphics[width=0.41\textwidth]{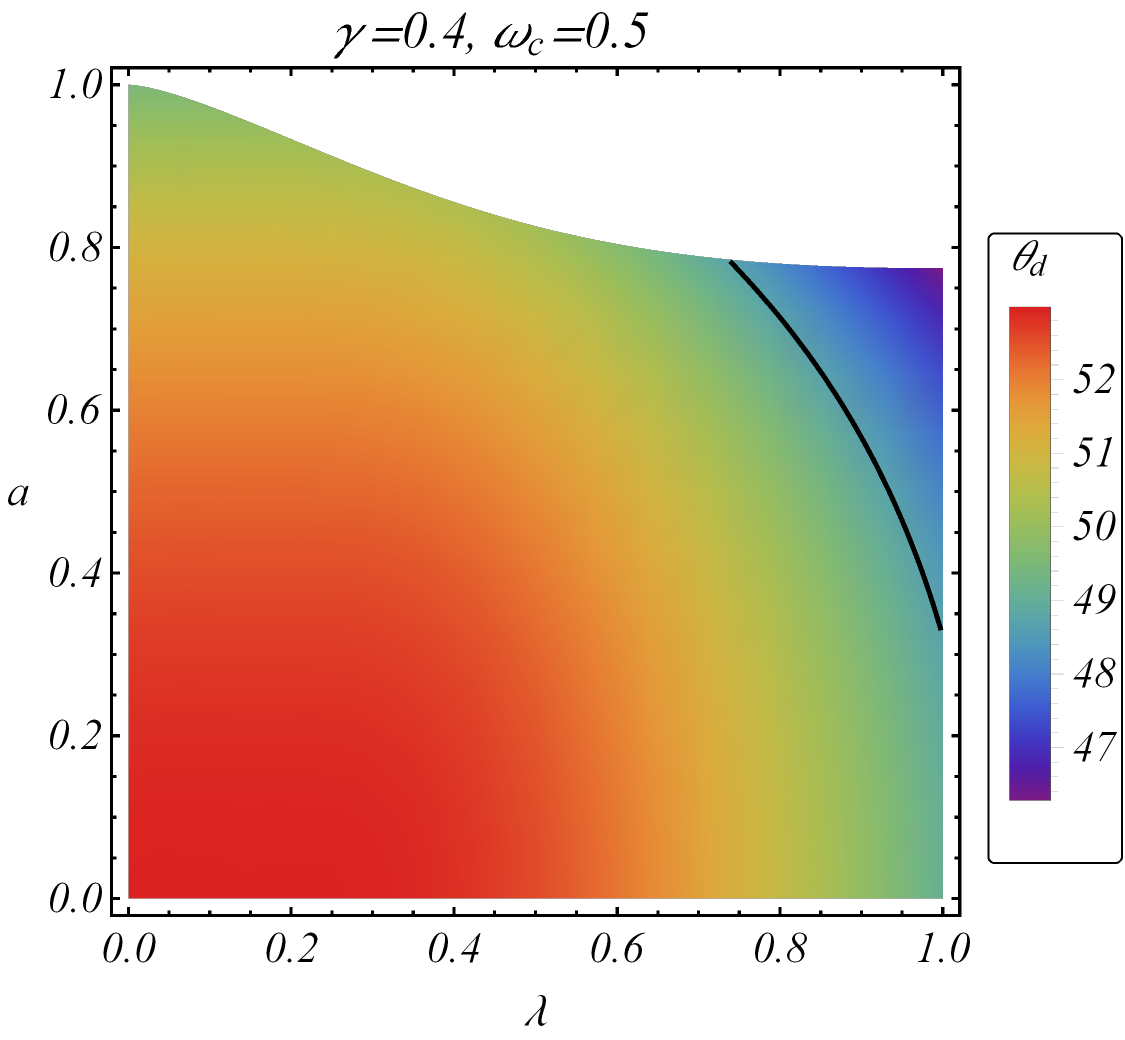}}~~~
		\subfigure{\includegraphics[width=0.41\textwidth]{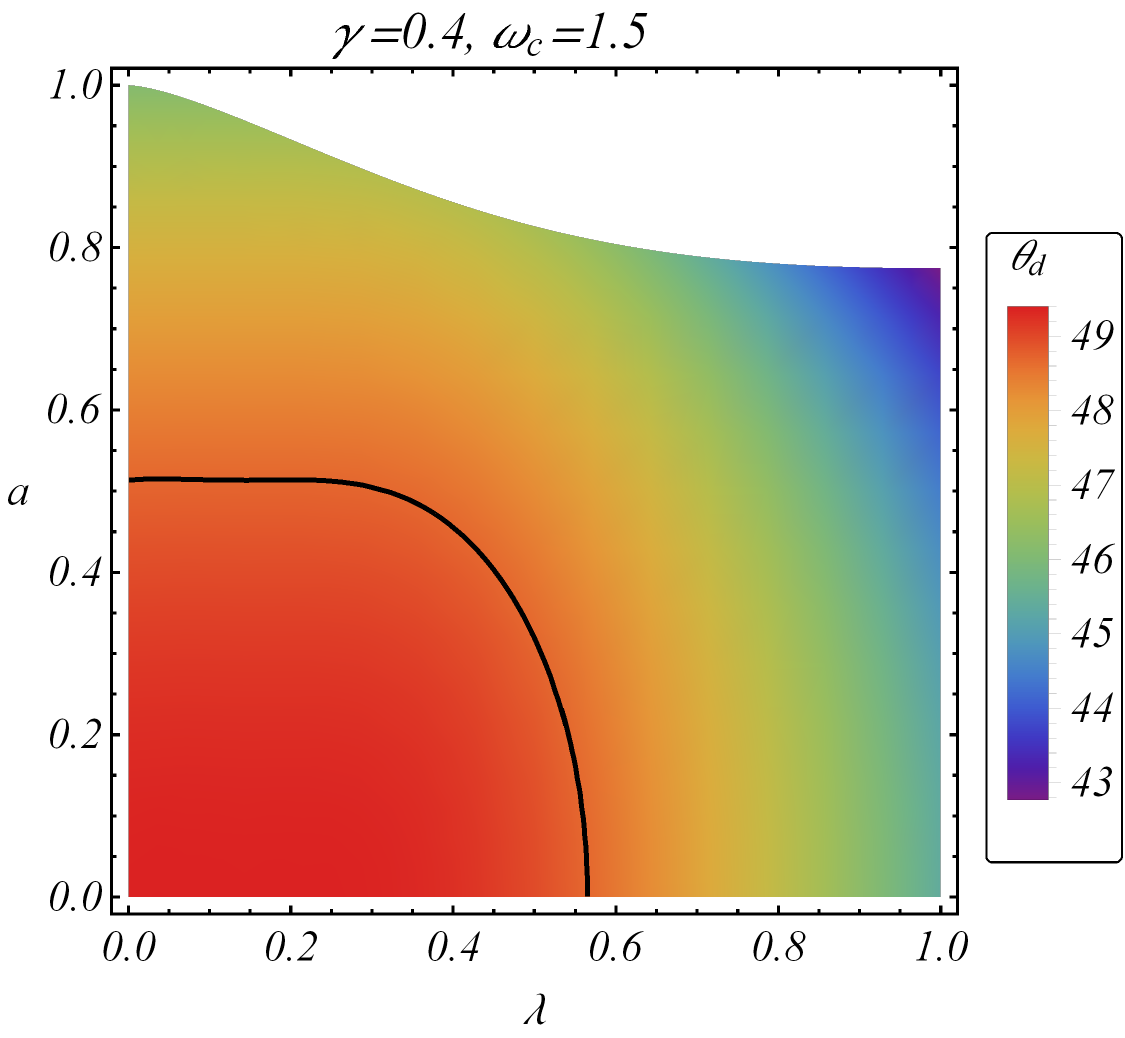}}
		\subfigure{\includegraphics[width=0.41\textwidth]{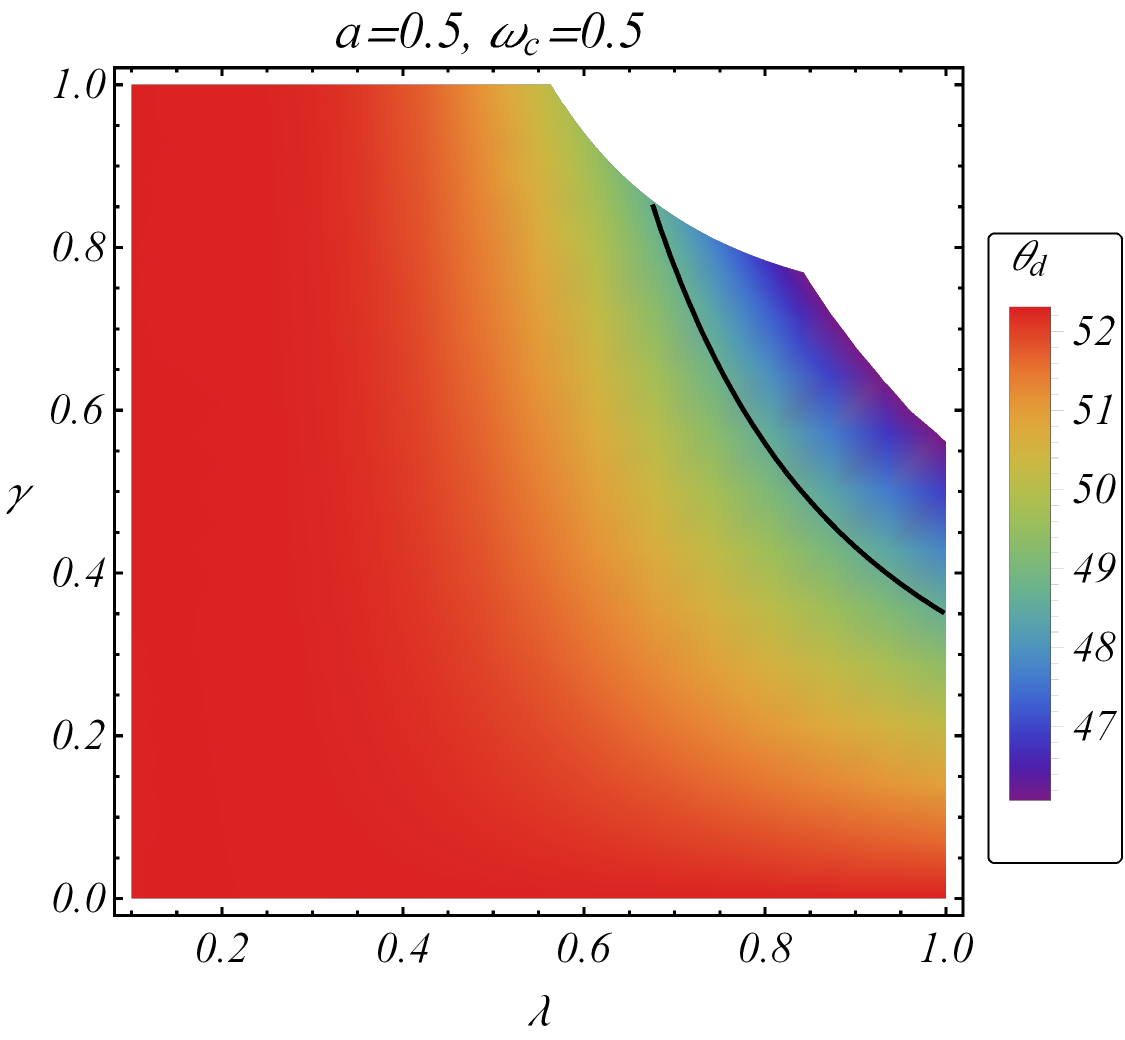}}~~~
		\subfigure{\includegraphics[width=0.41\textwidth]{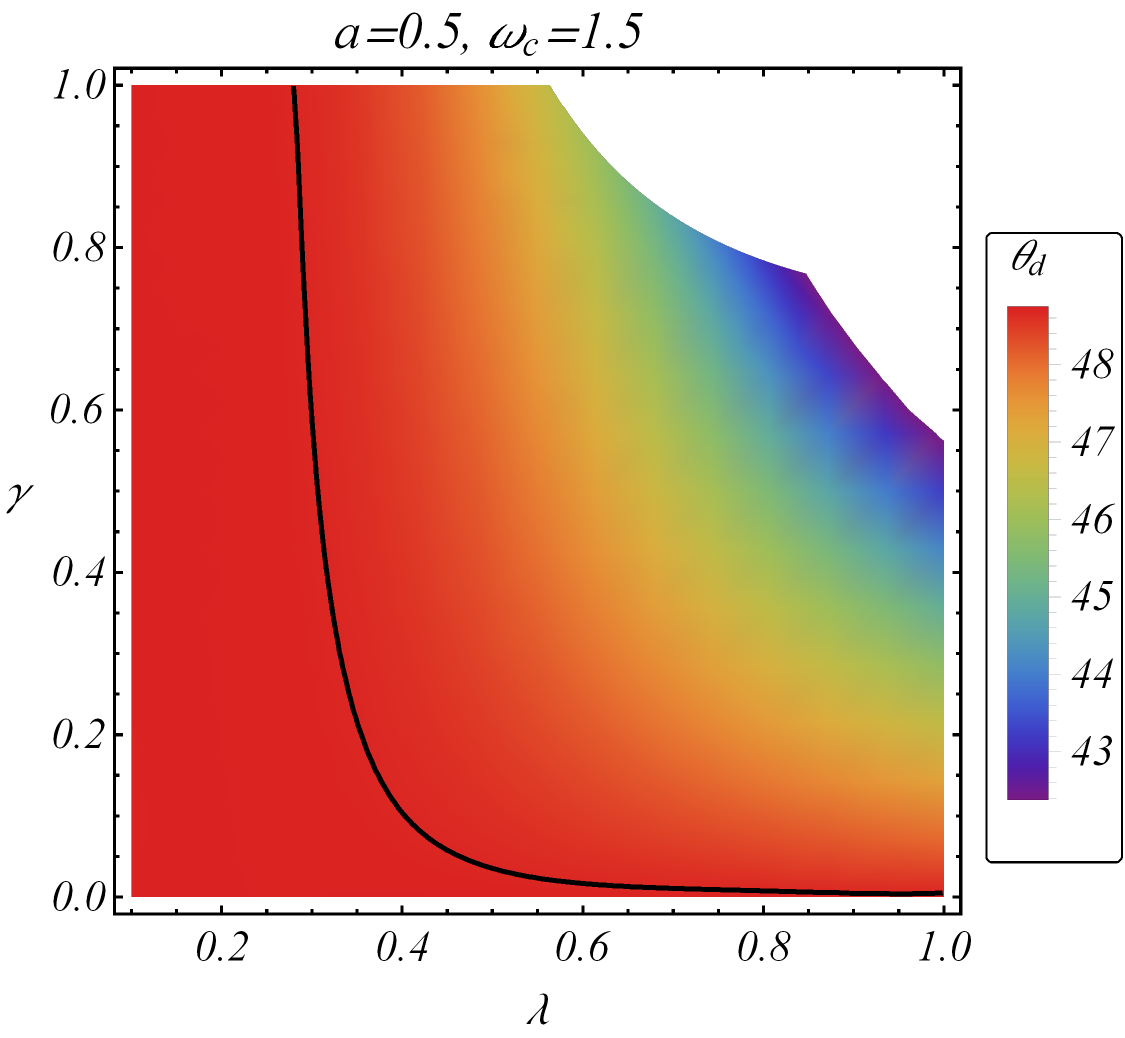}}
	\end{center}
	\caption{Plots showing the variation of angular size of the shadow of Sgr A* in terms of parametric spaces for various values of plasma parameters for the case described by Eq. (\ref{32}). \label{m2}}
\end{figure}

In the graphical representation denoted by Fig. \ref{m2}, the plots illustrate the angular diameter across various parametric spaces, specifically pertaining to Sgr A*, subjected to the case specified by Eq. (\ref{32}). The top panel corresponds to parametric space $a$-$\gamma$ with $\lambda=0.6$ and two different values of $\omega_c$. In both plots, we can find that for the given values of $\lambda$ and $\omega_c$, the angular diameter falls inside 1-$\sigma$ uncertainty level for all values of $a$ and $\gamma$ for which the density plot exists. Therefore, the BH (\ref{8}) immersed in plasma medium can be considered as Sgr A* for both values of $\omega_c$ and all of the corresponding parametric values for which the density plot exists. The middle panel corresponds to parametric space $a$-$\lambda$ with $\gamma=0.4$ and two different values of $\omega_c$. We found that for both values of $\omega_c$, the angular diameter falls inside 1-$\sigma$ level of error for each value of $a$ and $\lambda$ for which the density plot exists. Therefore, the BH (\ref{8}) in plasma background can be considered as Sgr A* for both values of $\omega_c$ and all of the corresponding parametric values for which the density plot exists as in the upper panel. The lower panel corresponds to the parametric space $\gamma$-$\lambda$ with $a=0.5$ and two different values of $\omega_c$. For the given value of $a$ and $\omega_c$, both plots show that the angular diameter lies in 1-$\sigma$ level of uncertainty for every value of $\gamma$ and $\lambda$ for which the density plot exists. Therefore, the BH (\ref{8}) in plasma background can be considered as Sgr A* for both values of $\omega_c$ and all of the corresponding parametric values for which the density plot exists.
\begin{figure}[t!]
	\begin{center}
		\subfigure{\includegraphics[width=0.41\textwidth]{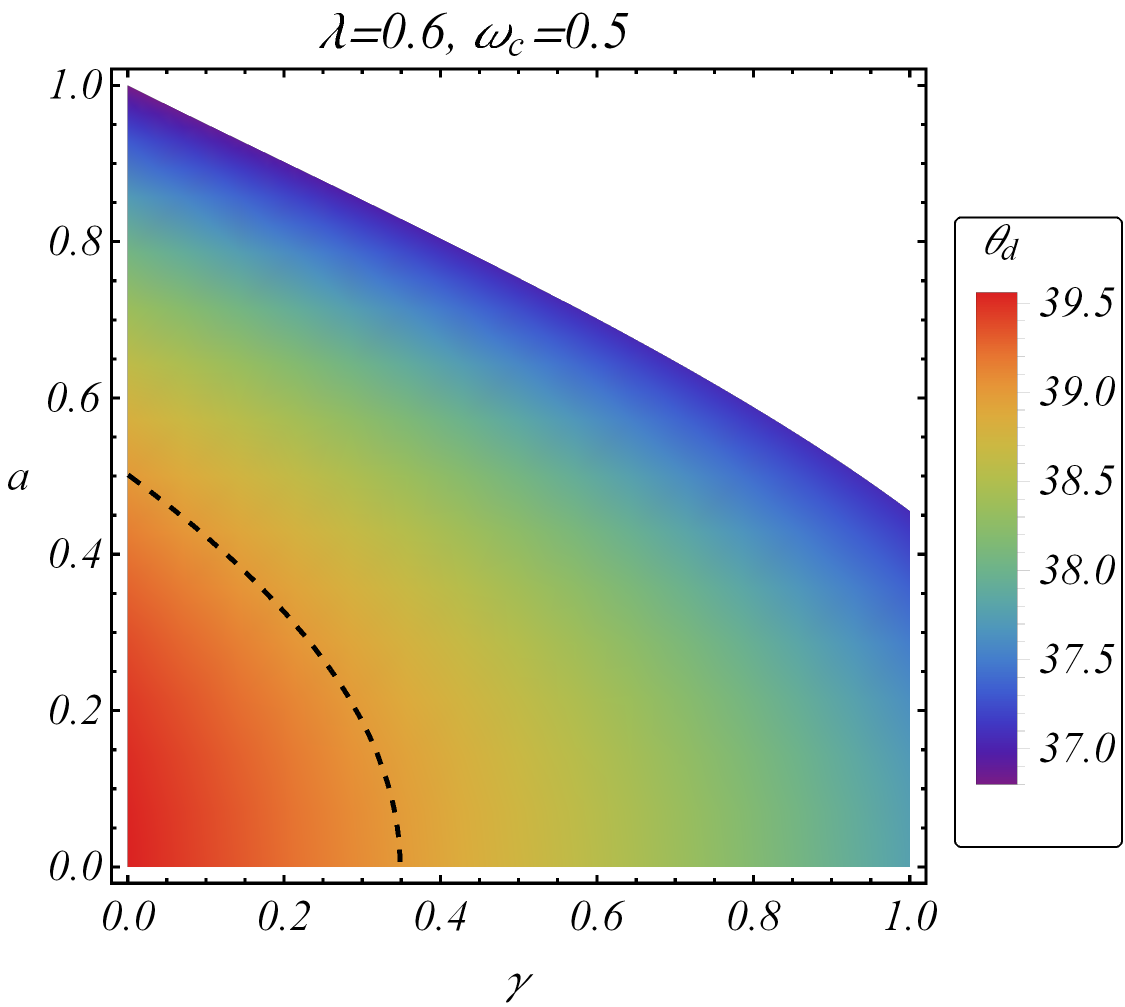}}~~~
		\subfigure{\includegraphics[width=0.41\textwidth]{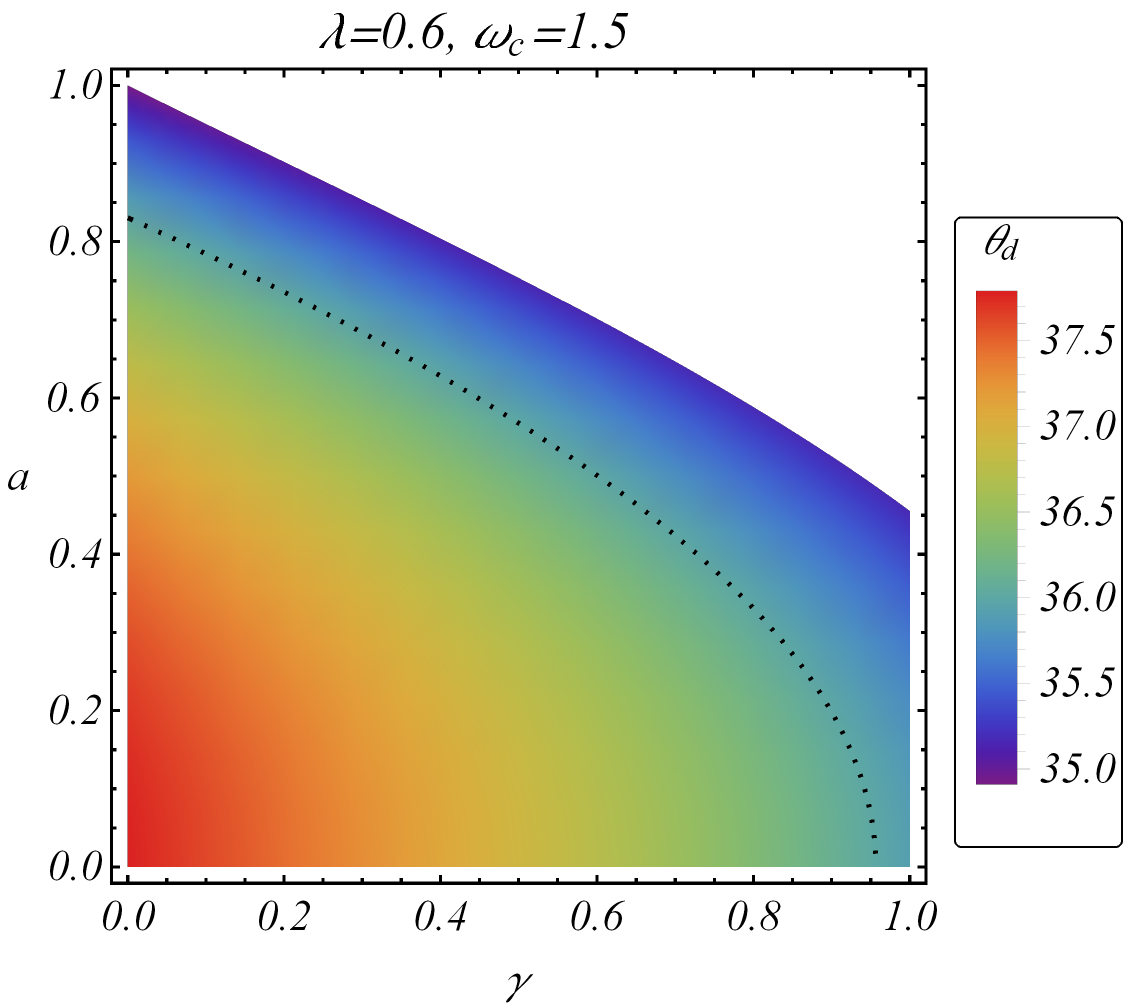}}
		\subfigure{\includegraphics[width=0.41\textwidth]{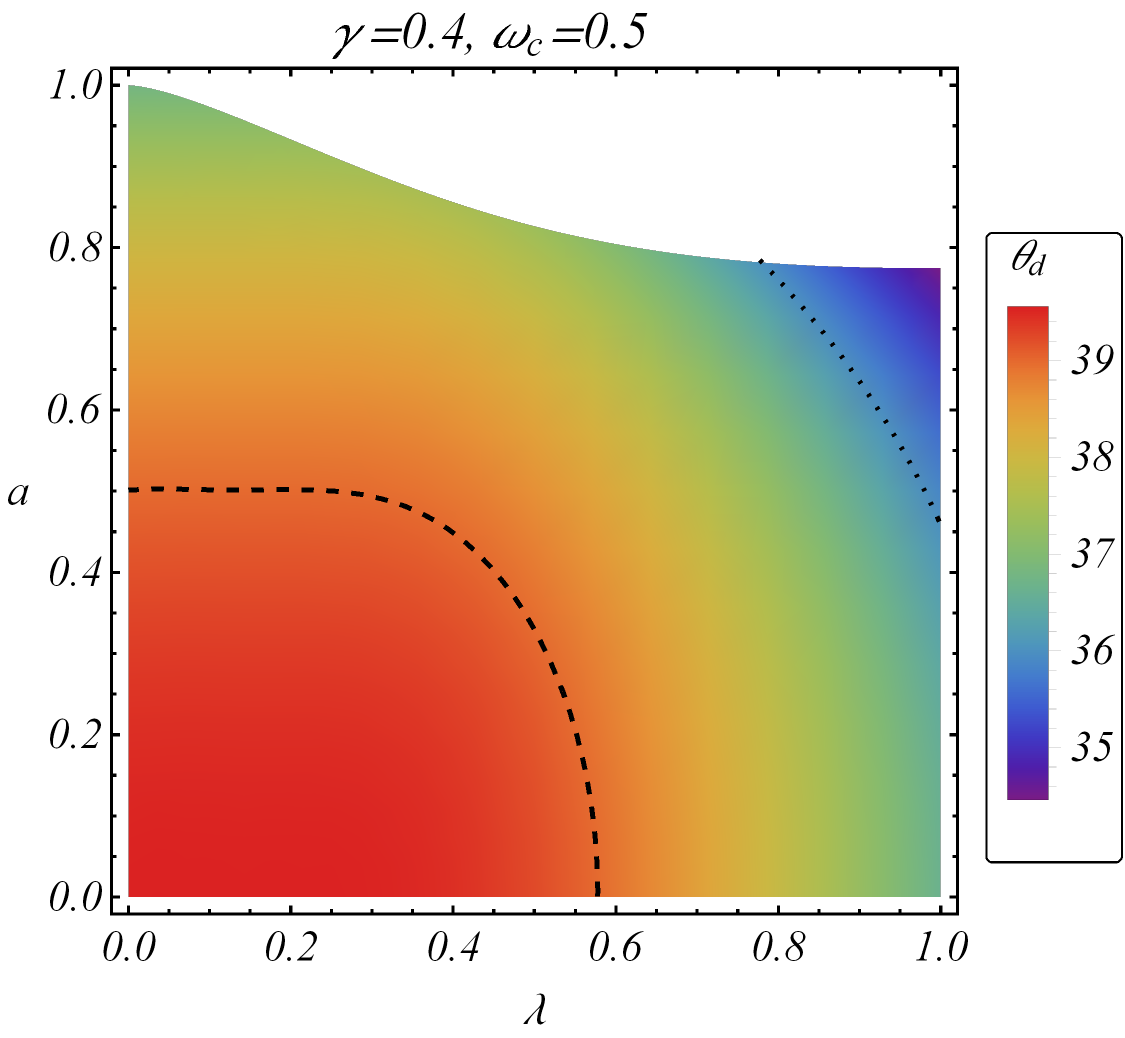}}~~~
		\subfigure{\includegraphics[width=0.41\textwidth]{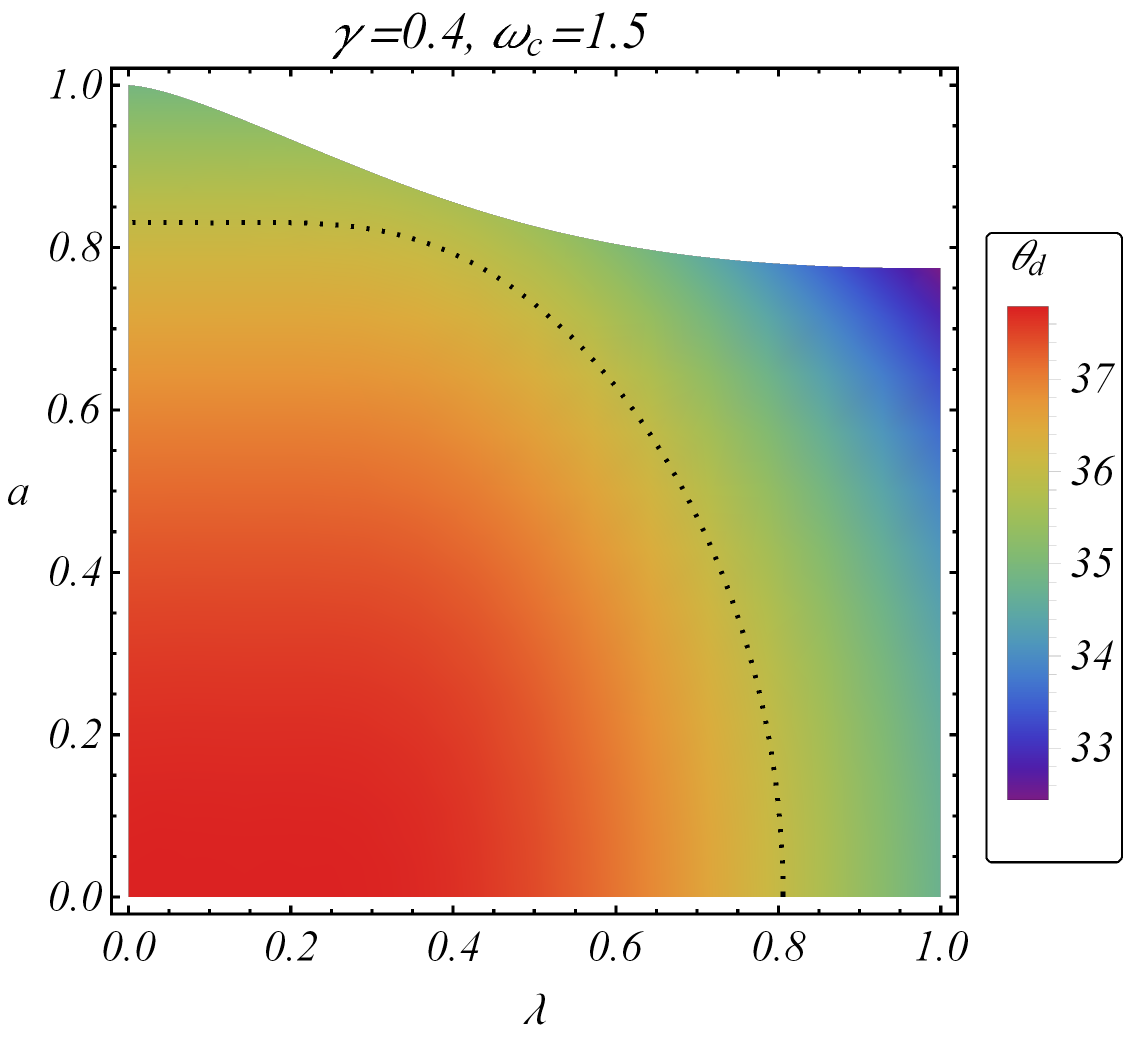}}
		\subfigure{\includegraphics[width=0.41\textwidth]{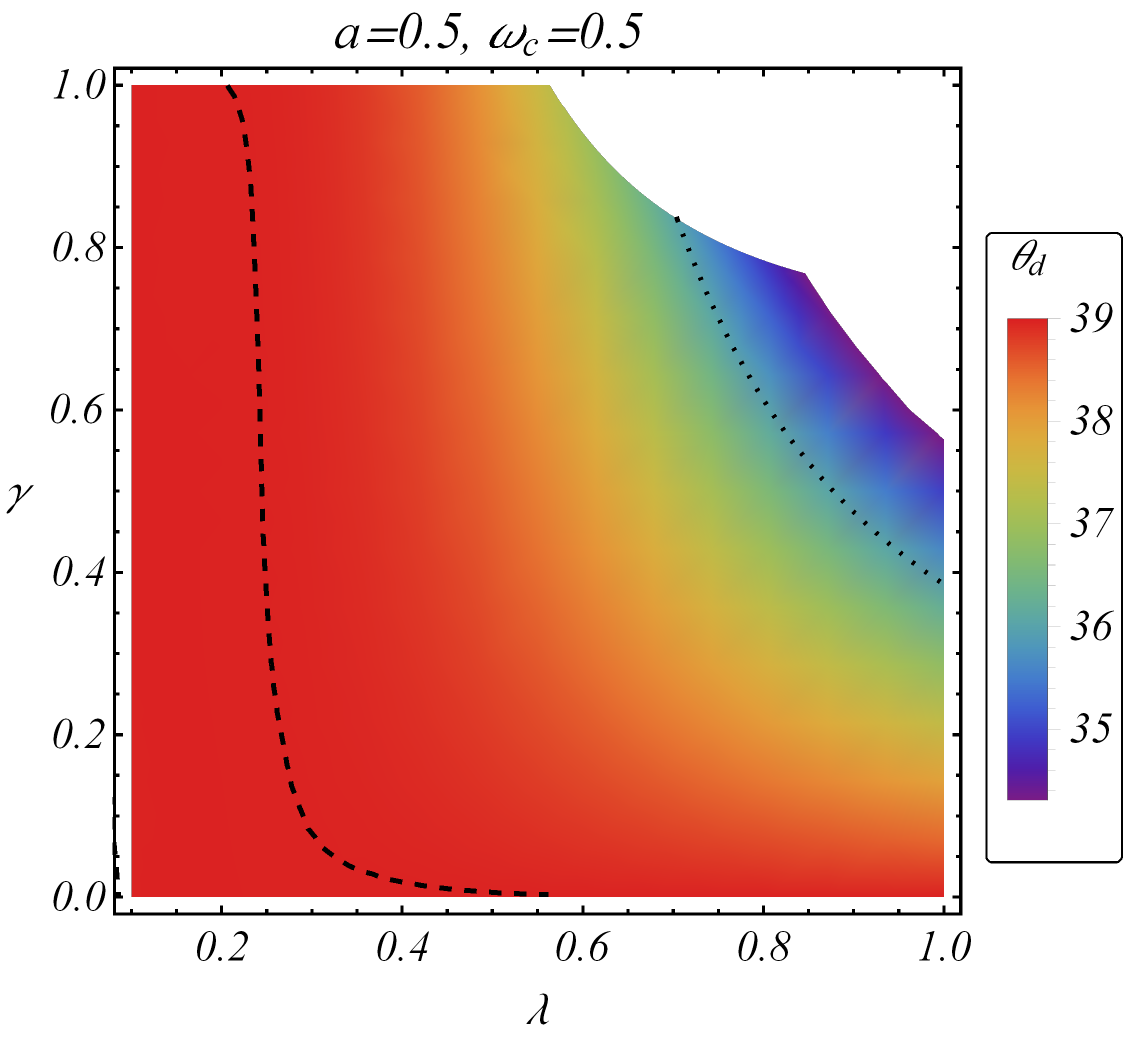}}~~~
		\subfigure{\includegraphics[width=0.41\textwidth]{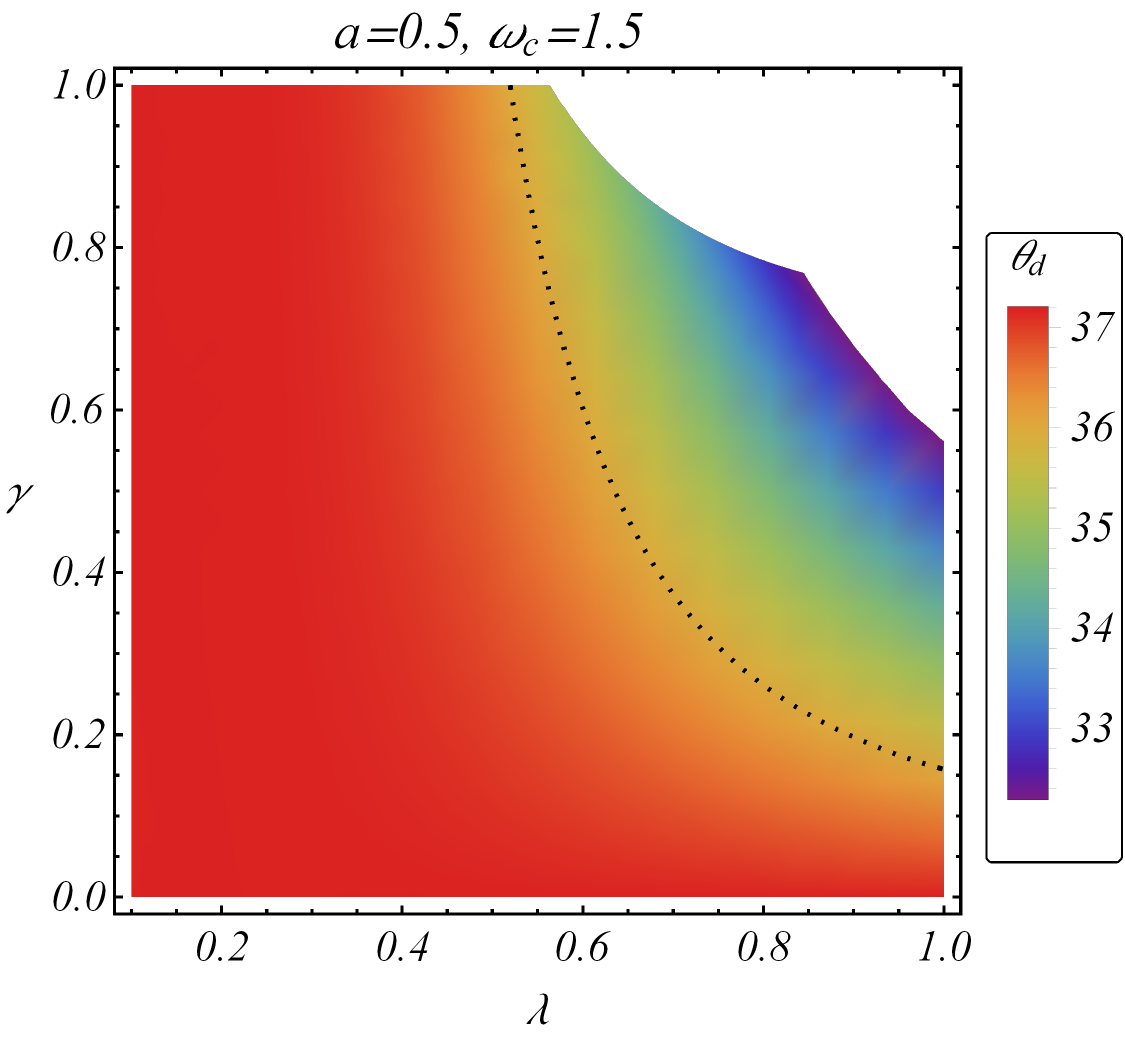}}
	\end{center}
	\caption{The variation of angular size of the shadow of M87* in terms of parametric spaces for various values of plasma parameters for the case described by Eq. (\ref{33}). \label{m3}}
\end{figure}
\begin{figure}[t!]
	\begin{center}
		\subfigure{\includegraphics[width=0.41\textwidth]{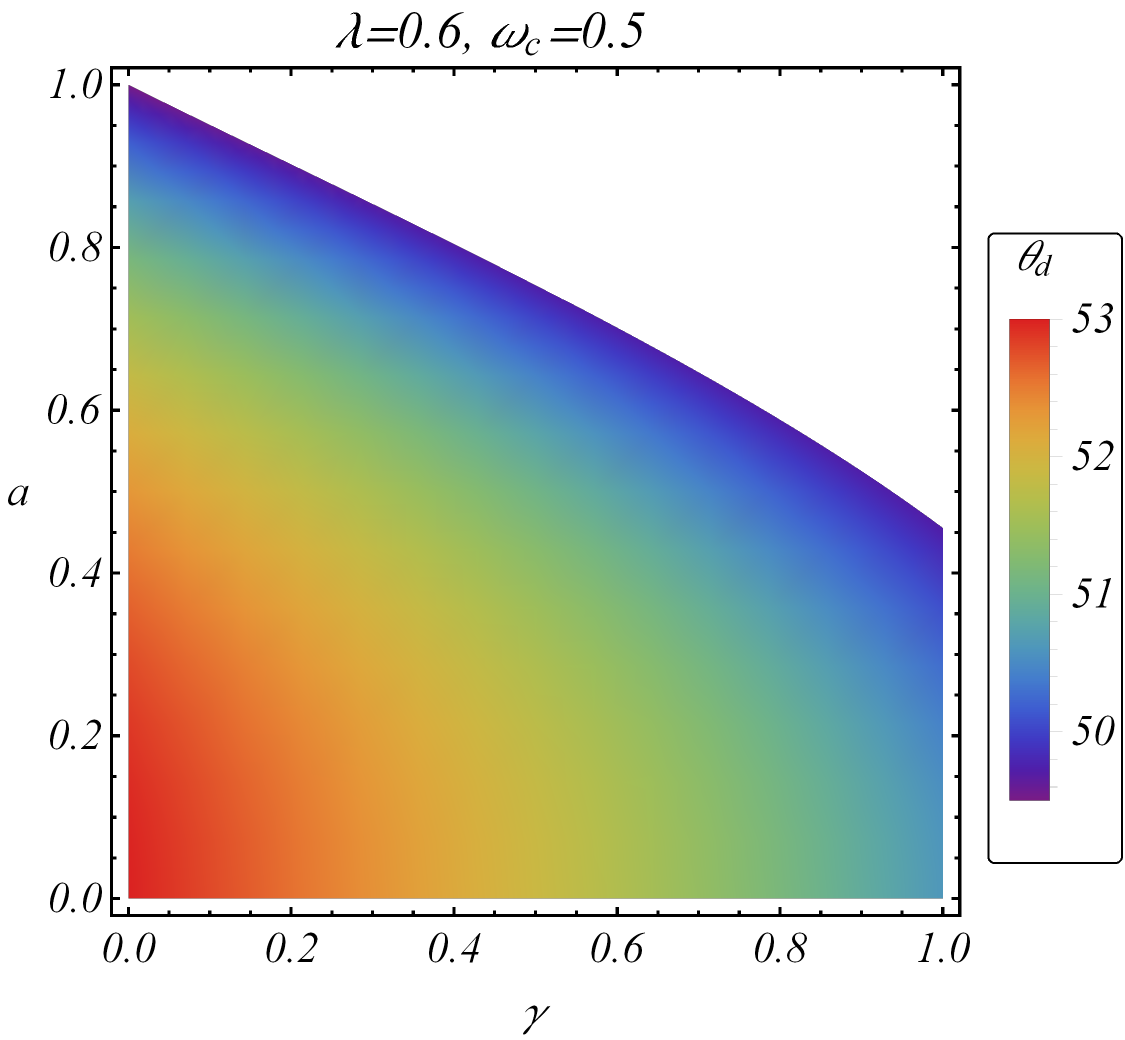}}~~~
		\subfigure{\includegraphics[width=0.41\textwidth]{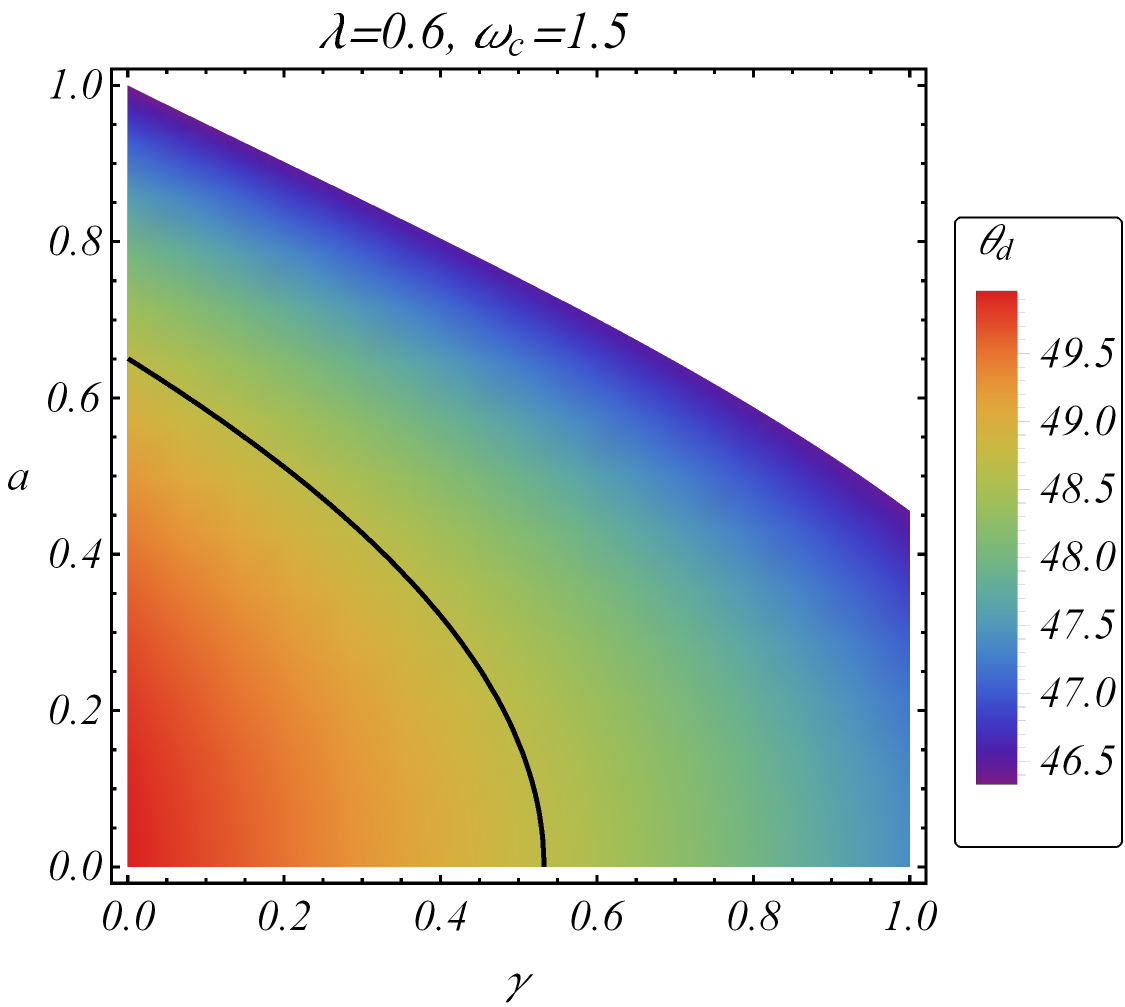}}
		\subfigure{\includegraphics[width=0.41\textwidth]{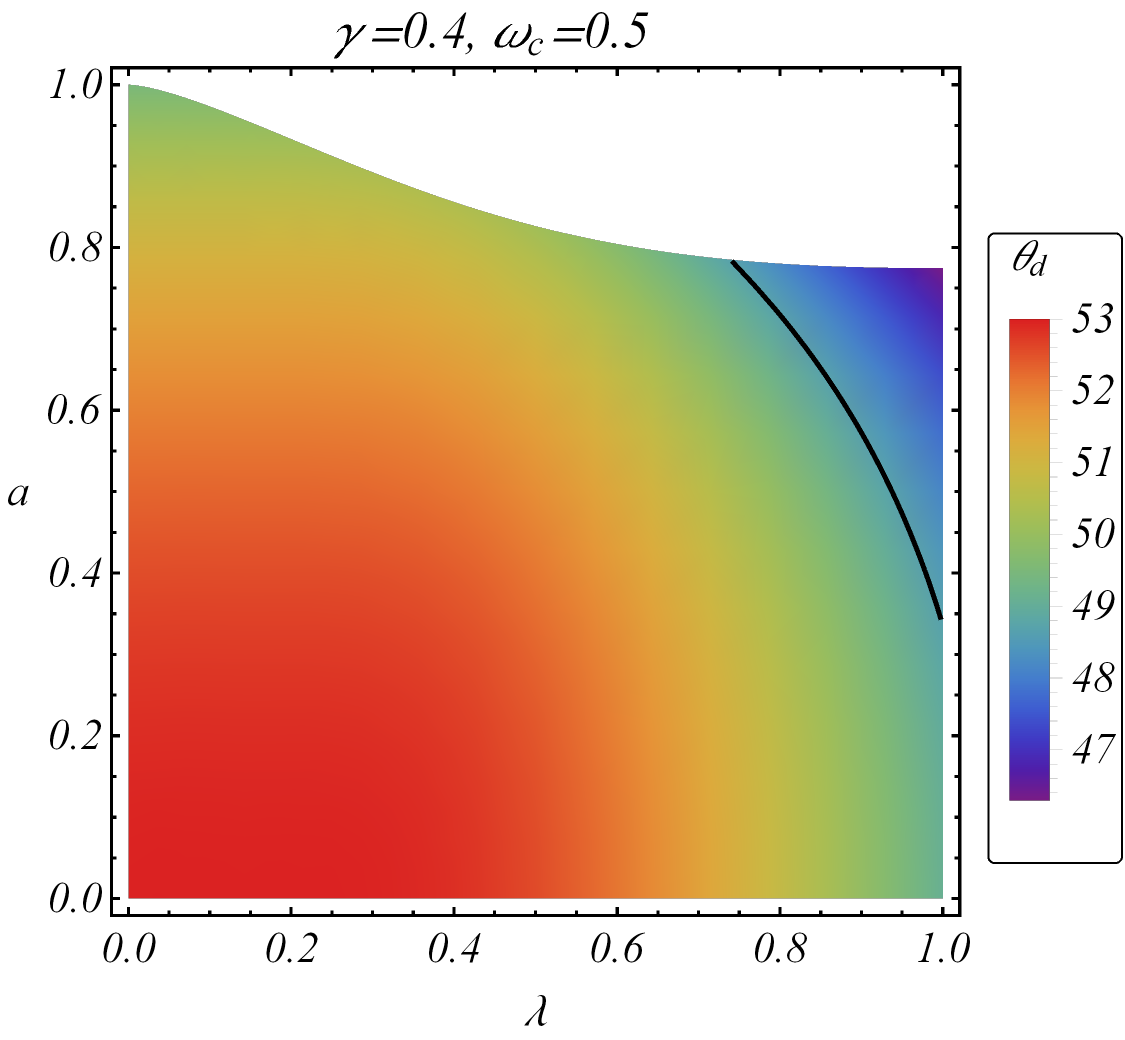}}~~~
		\subfigure{\includegraphics[width=0.41\textwidth]{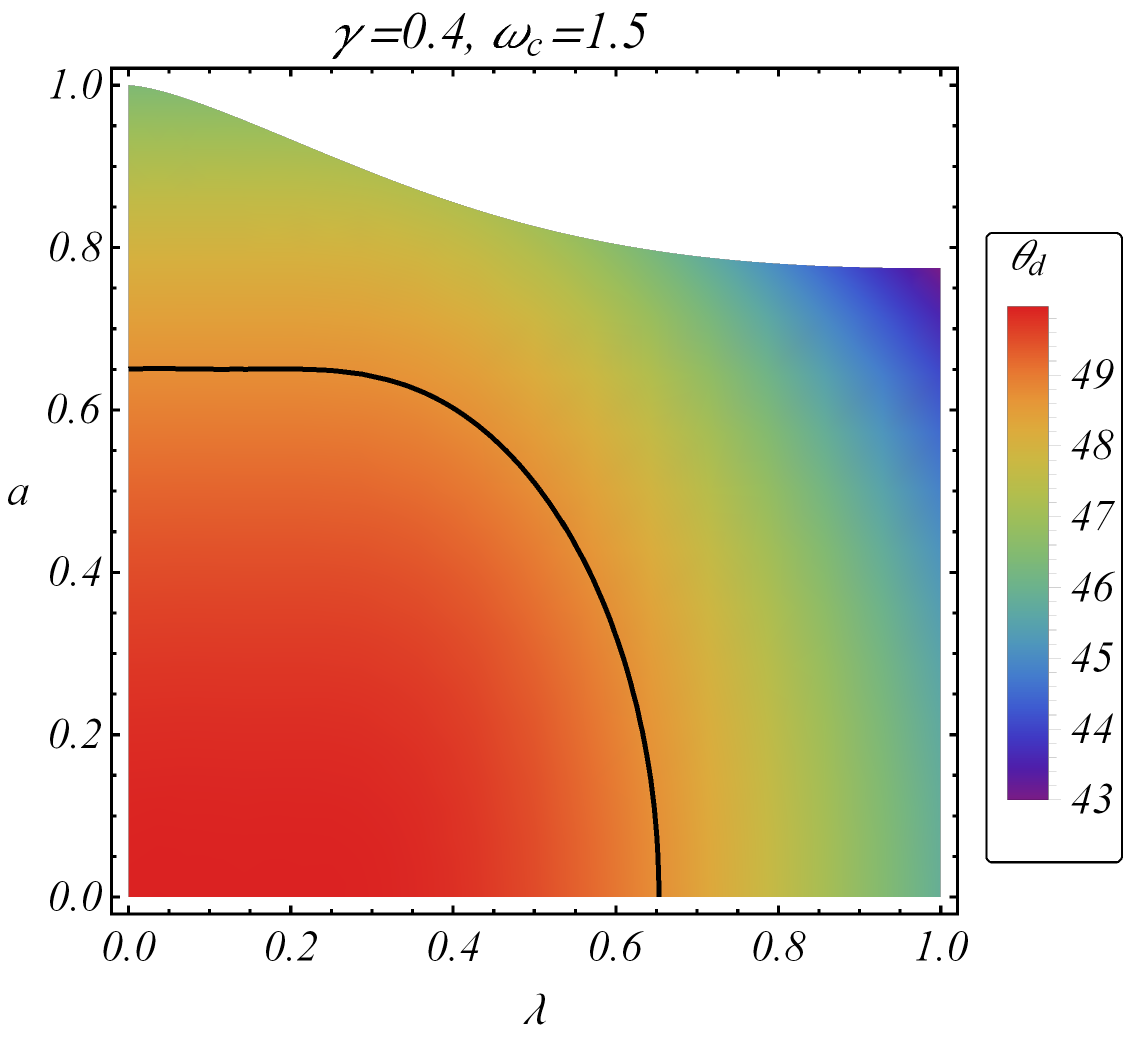}}
		\subfigure{\includegraphics[width=0.41\textwidth]{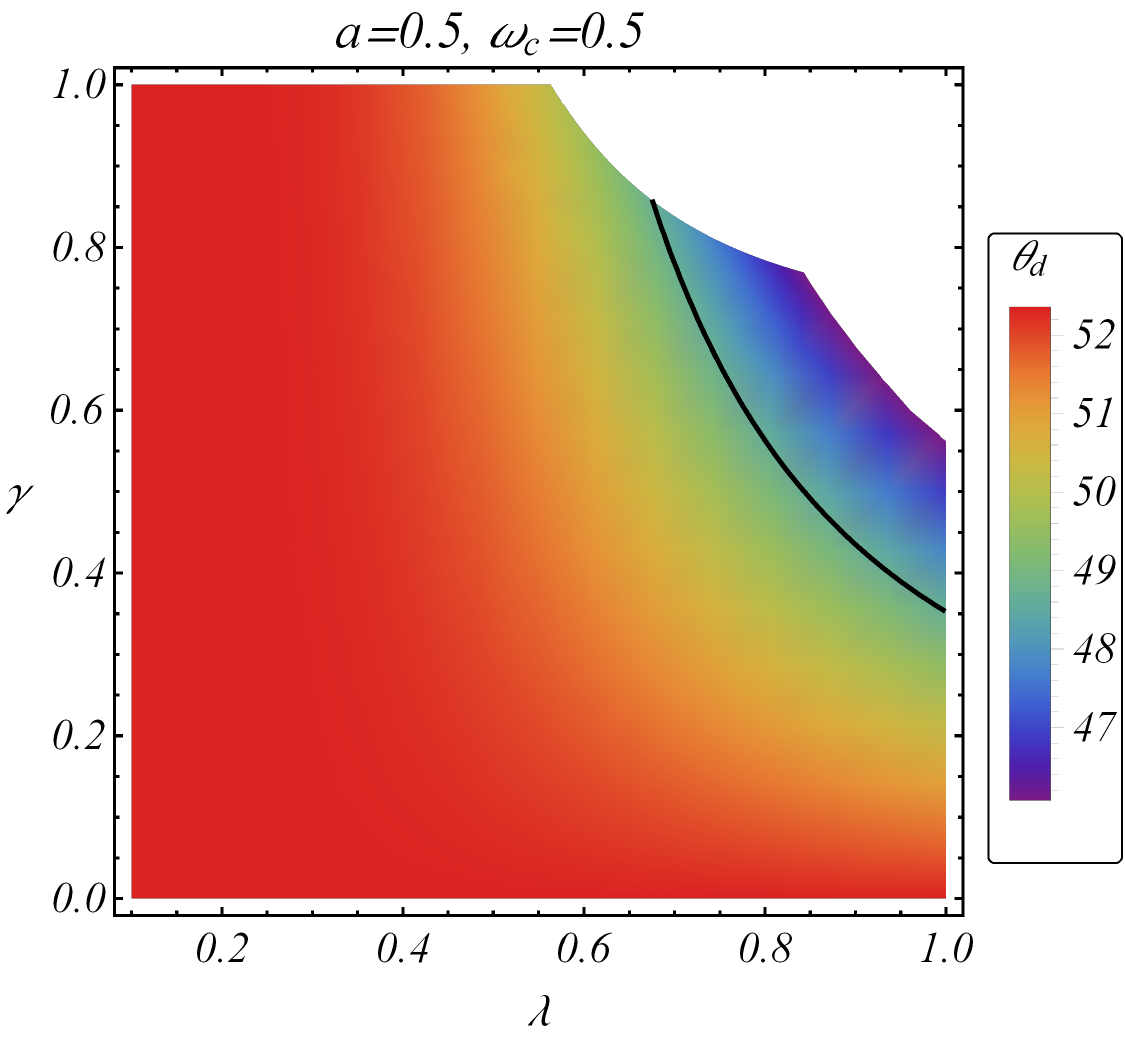}}~~~
		\subfigure{\includegraphics[width=0.41\textwidth]{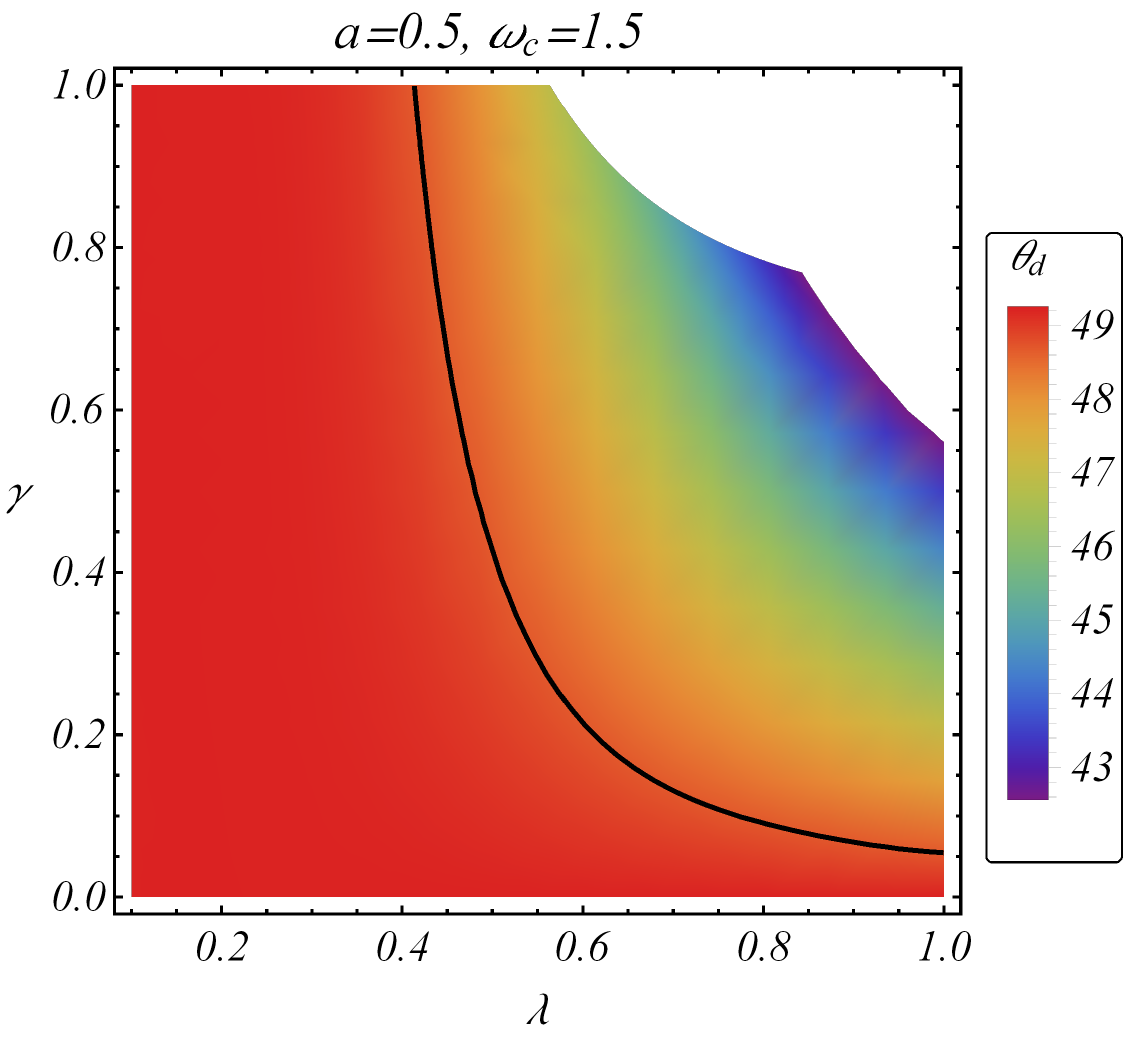}}
	\end{center}
	\caption{The behavior of angular size of the shadow of Sgr A* in terms of parametric spaces for various values of plasma parameters for the case described by Eq. (\ref{33}). \label{m4}}
\end{figure}
\begin{figure}[t!]
	\begin{center}
		\subfigure{\includegraphics[width=0.41\textwidth]{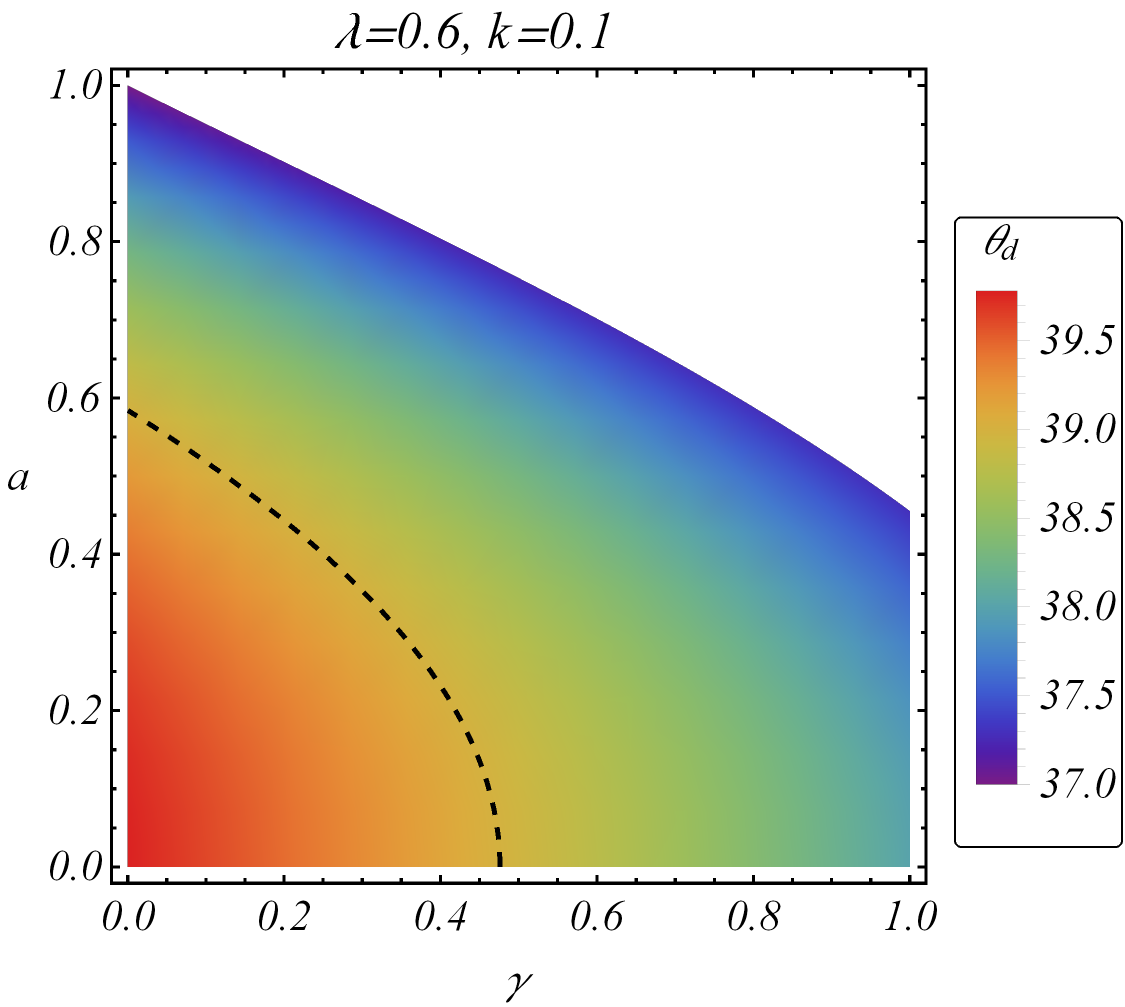}}~~~
		\subfigure{\includegraphics[width=0.41\textwidth]{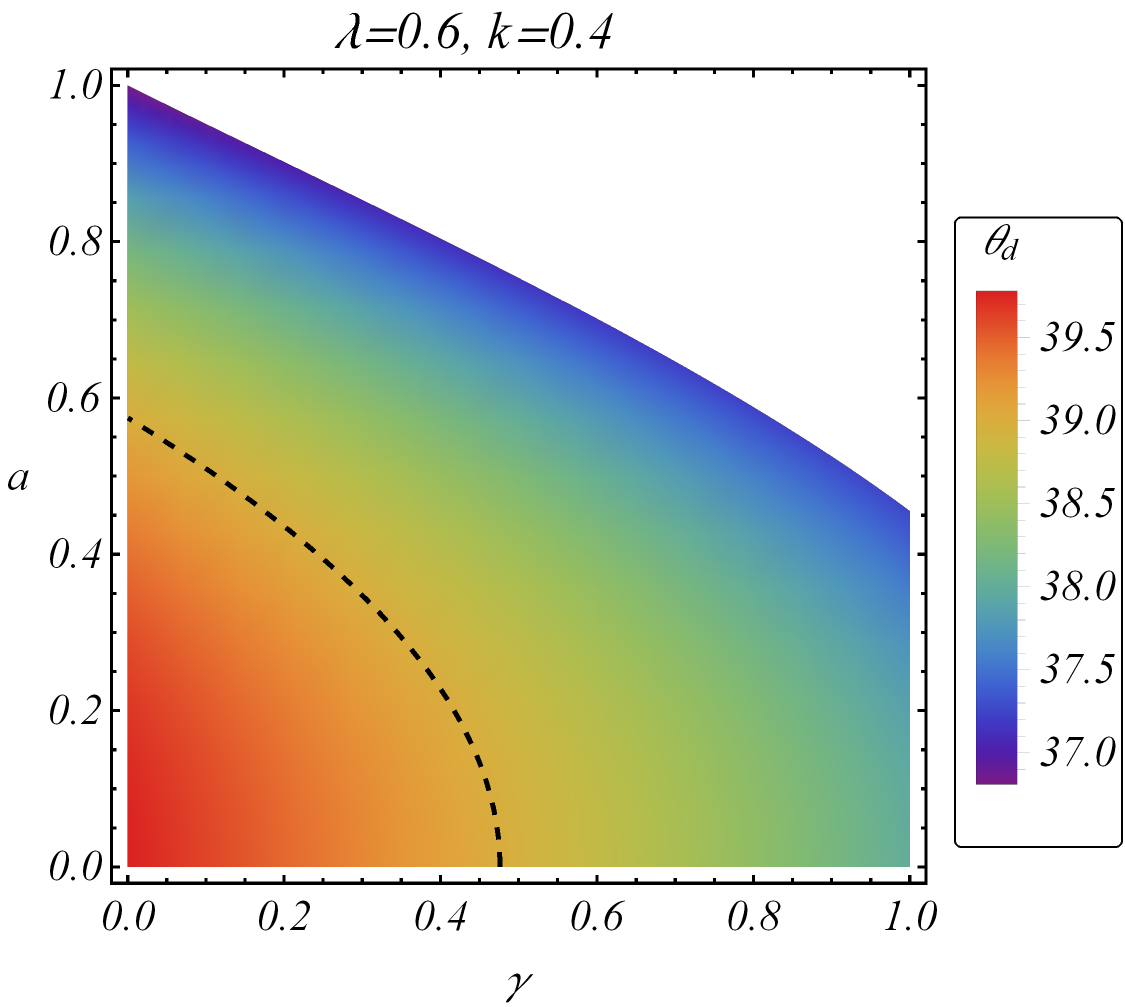}}
		\subfigure{\includegraphics[width=0.41\textwidth]{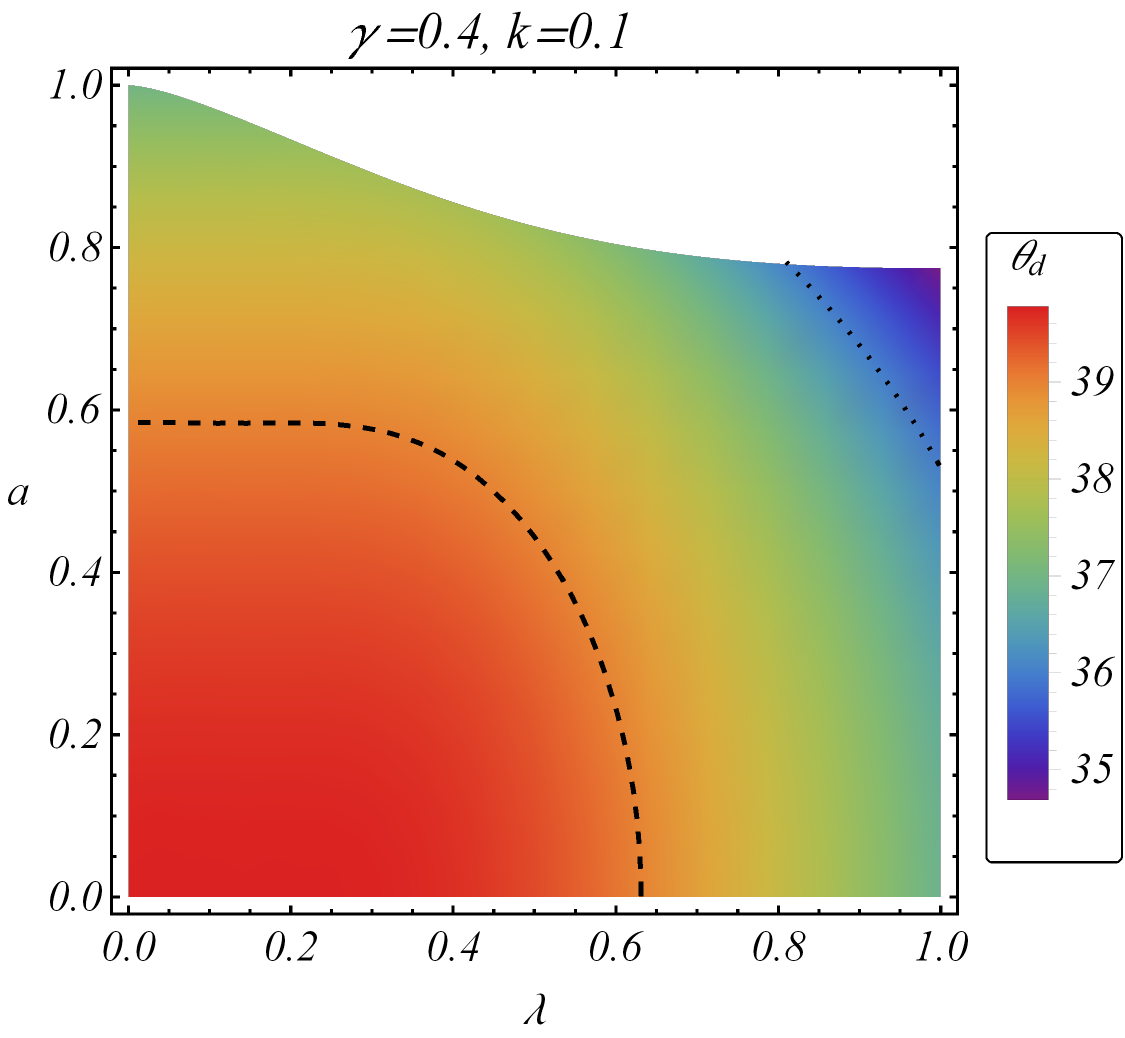}}~~~
		\subfigure{\includegraphics[width=0.41\textwidth]{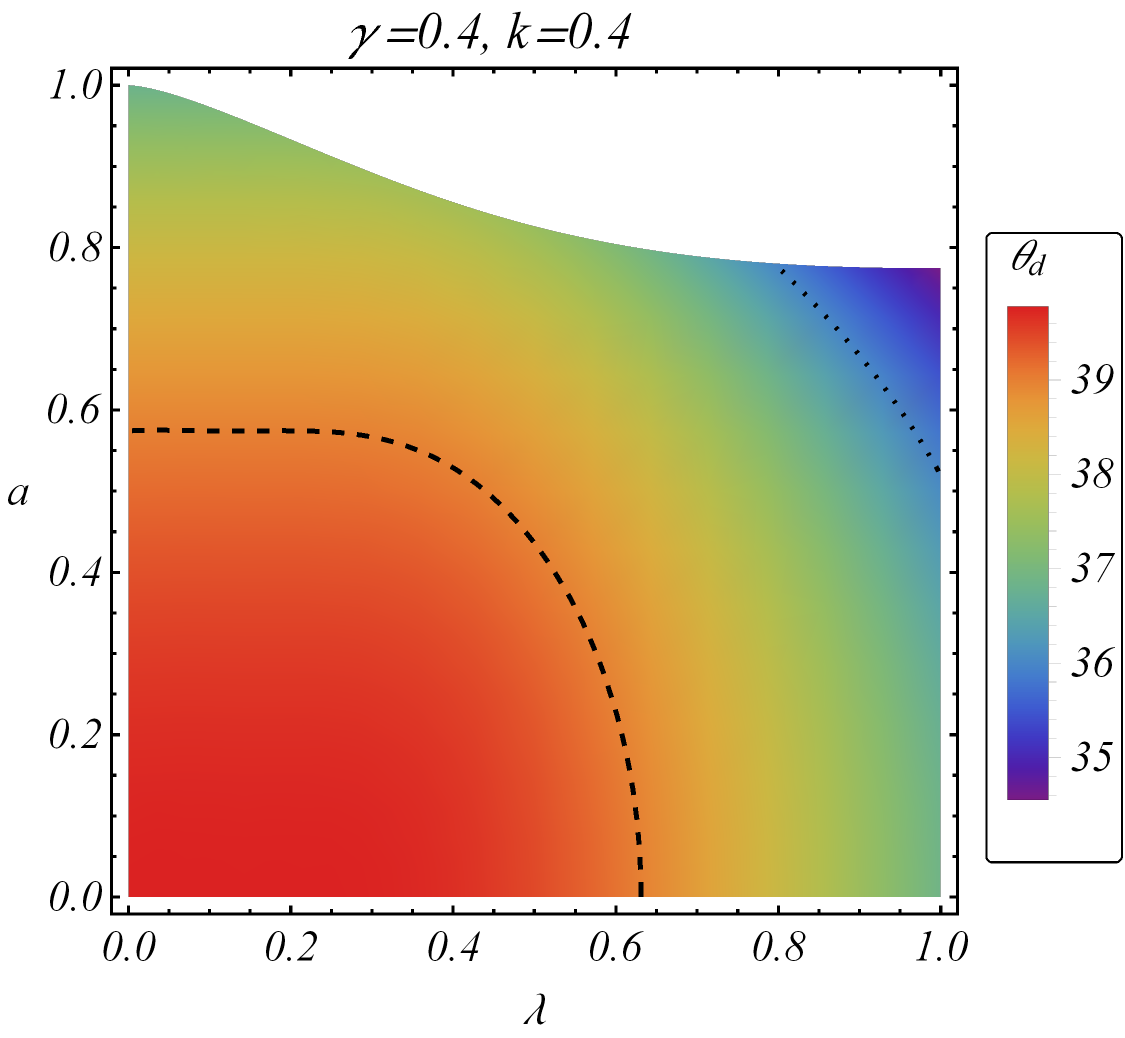}}
		\subfigure{\includegraphics[width=0.41\textwidth]{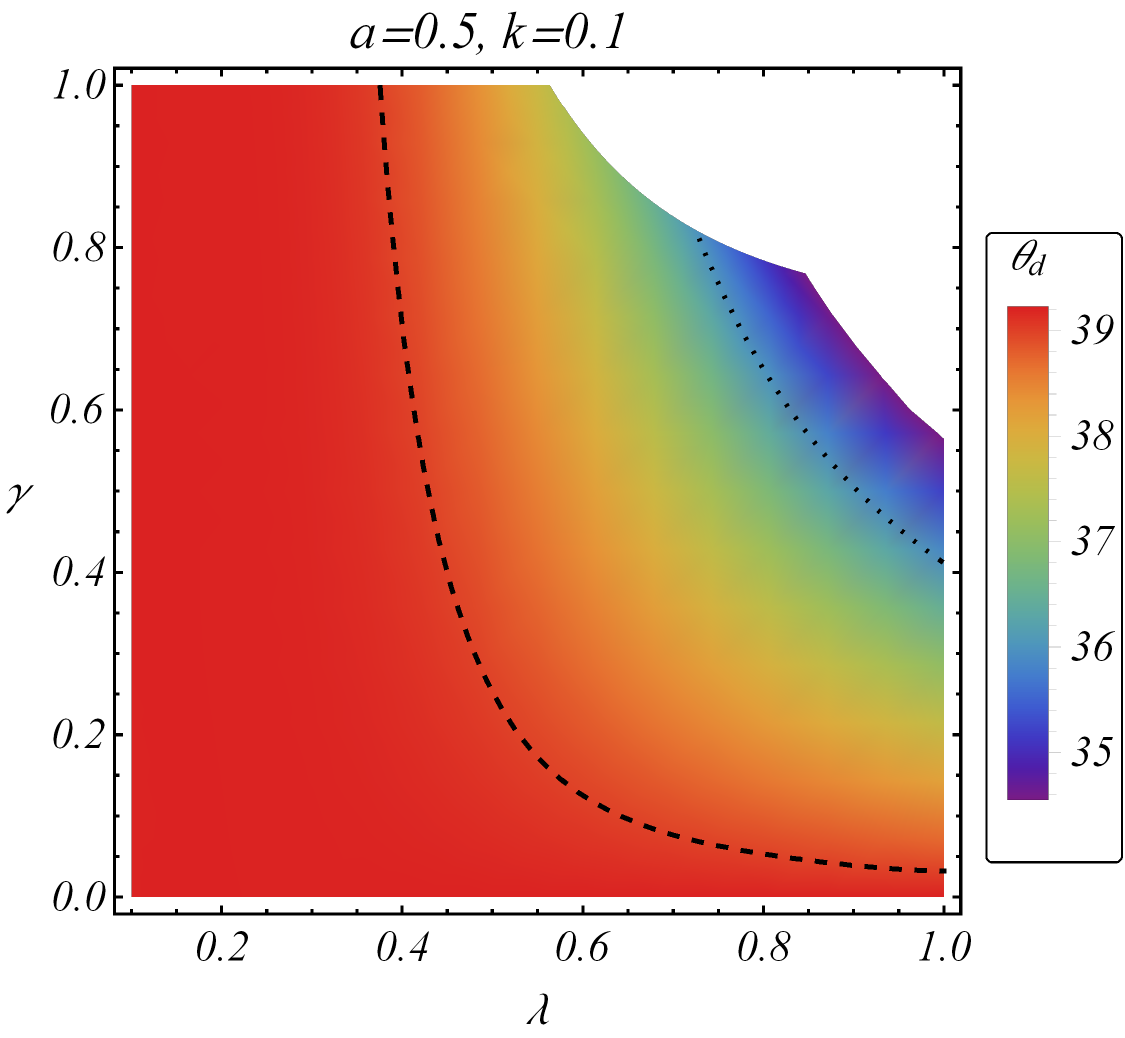}}~~~
		\subfigure{\includegraphics[width=0.41\textwidth]{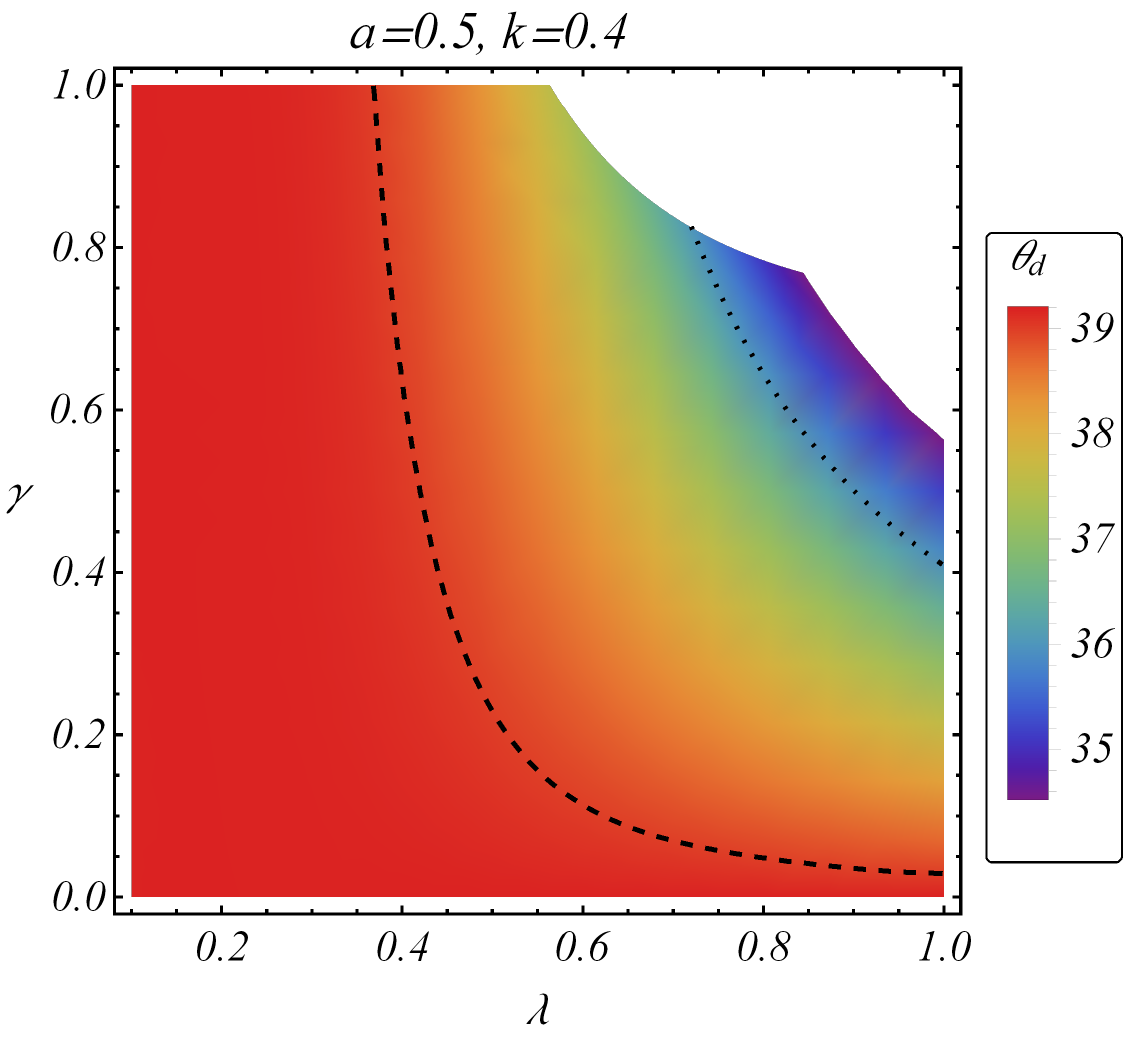}}
	\end{center}
	\caption{Plots showing the behavior of angular size of the shadow of M87* in terms of parametric spaces for various values of plasma parameters for the case described by Eq. (\ref{49b}). \label{m5}}
\end{figure}
\begin{figure}[t!]
	\begin{center}
		\subfigure{\includegraphics[width=0.41\textwidth]{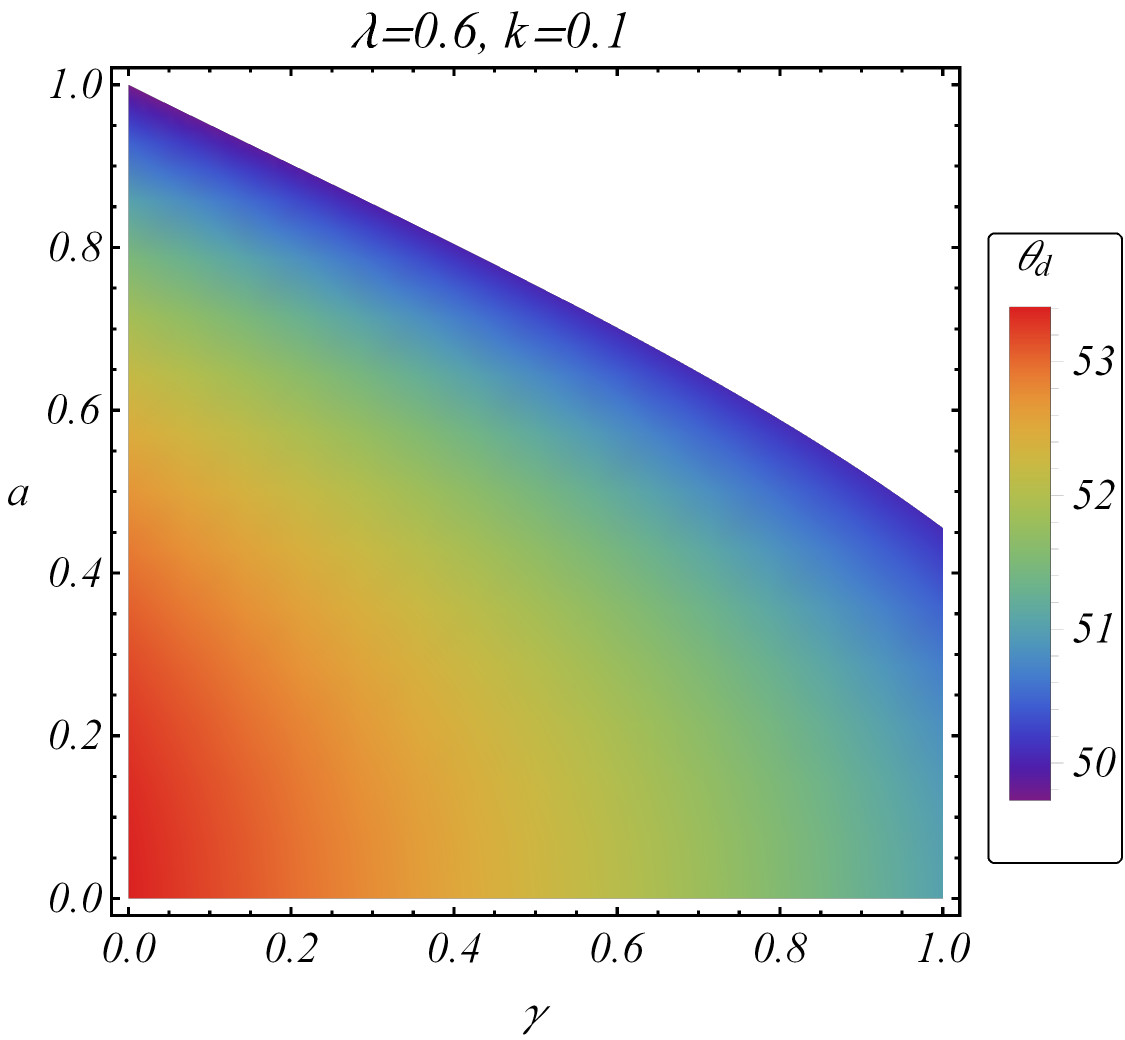}}~~~
		\subfigure{\includegraphics[width=0.41\textwidth]{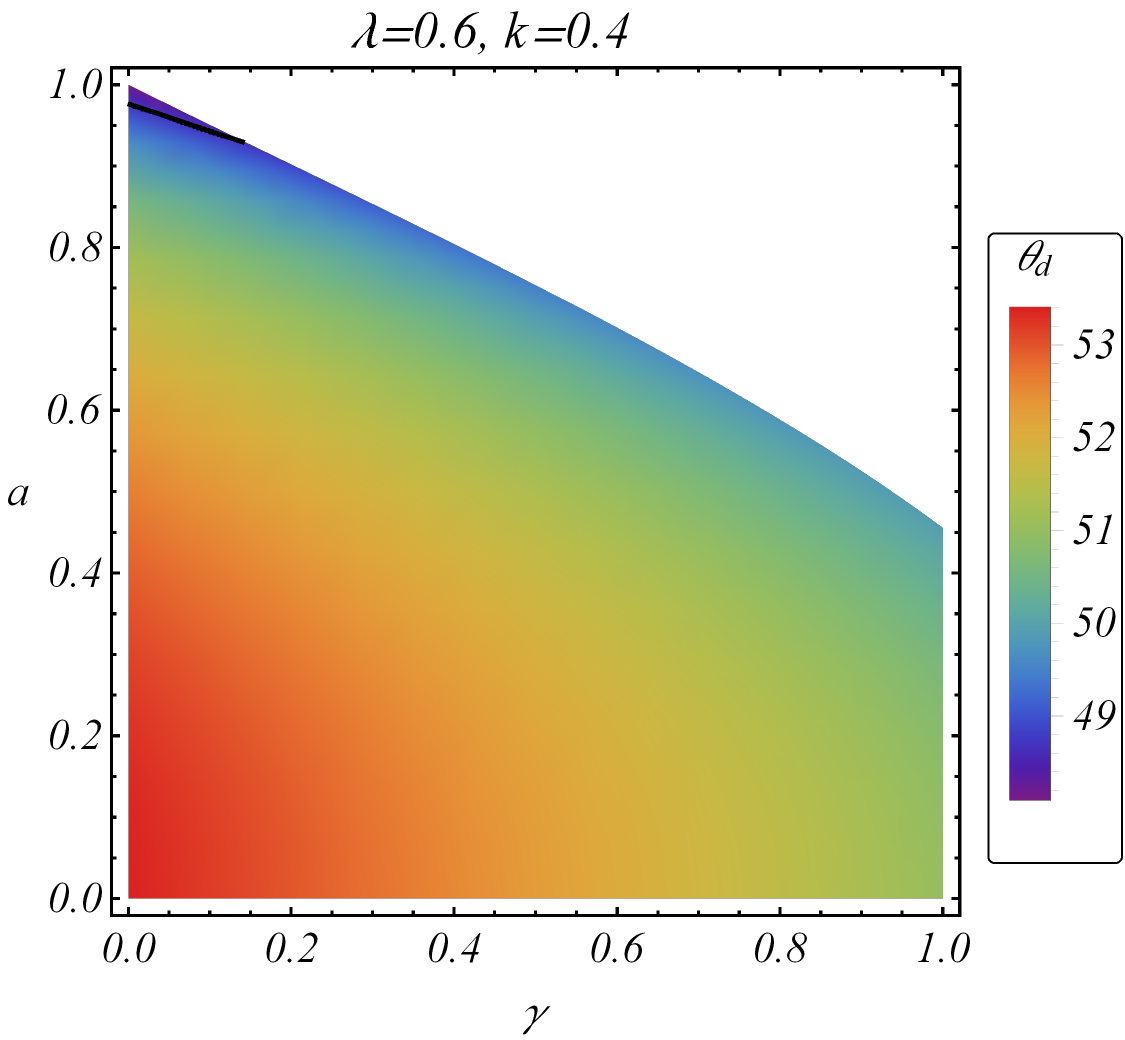}}
		\subfigure{\includegraphics[width=0.41\textwidth]{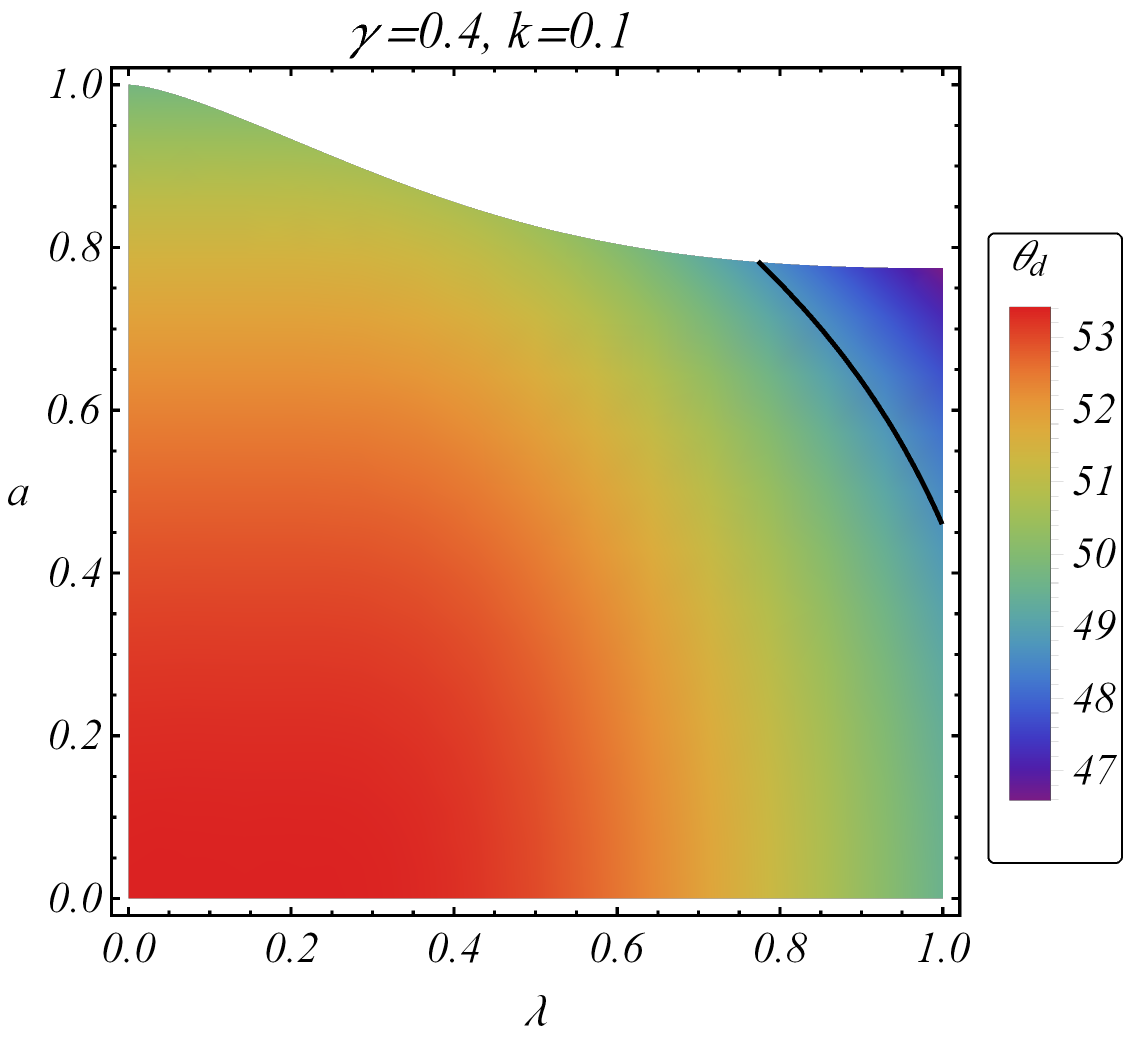}}~~~
		\subfigure{\includegraphics[width=0.41\textwidth]{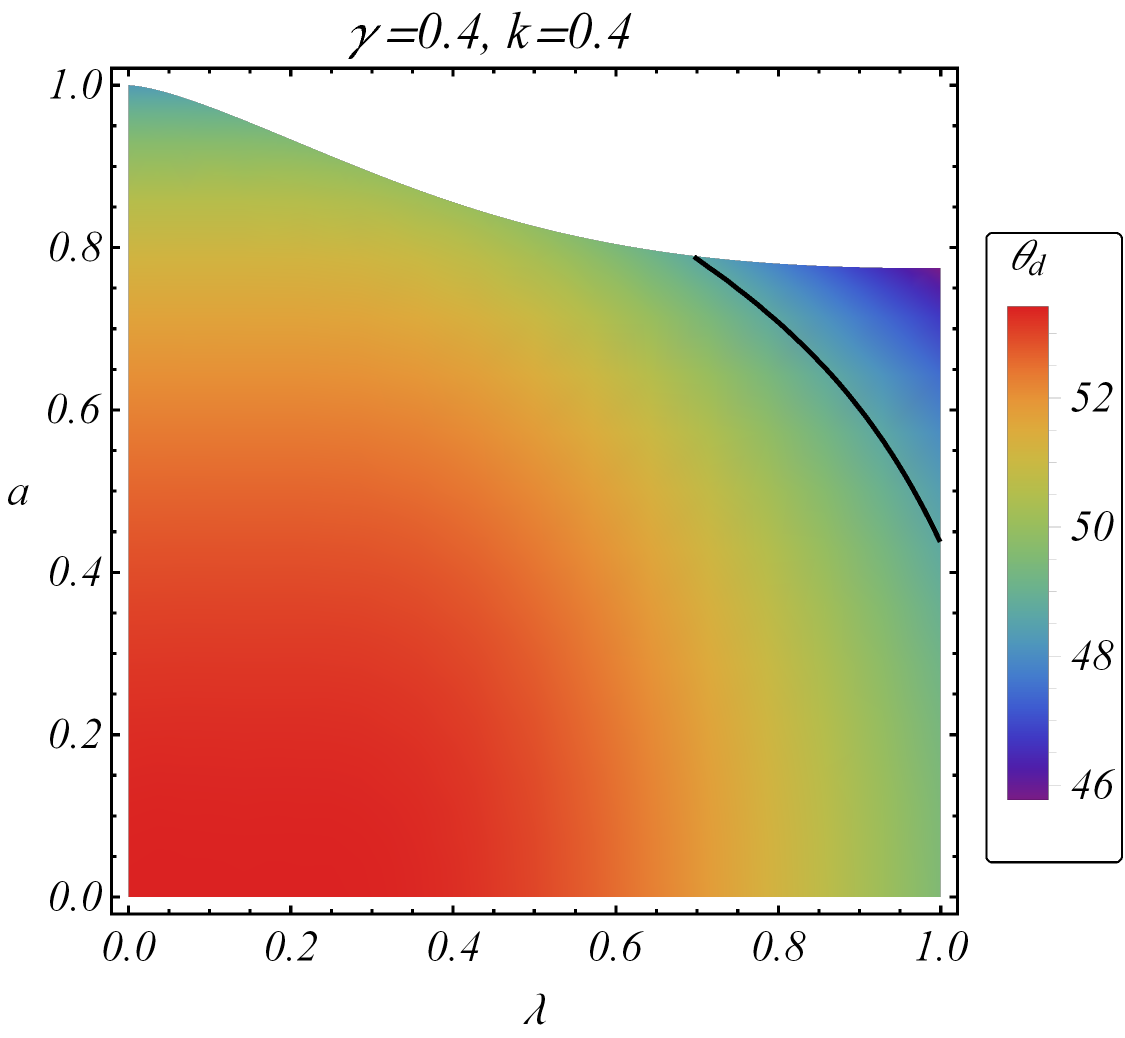}}
		\subfigure{\includegraphics[width=0.41\textwidth]{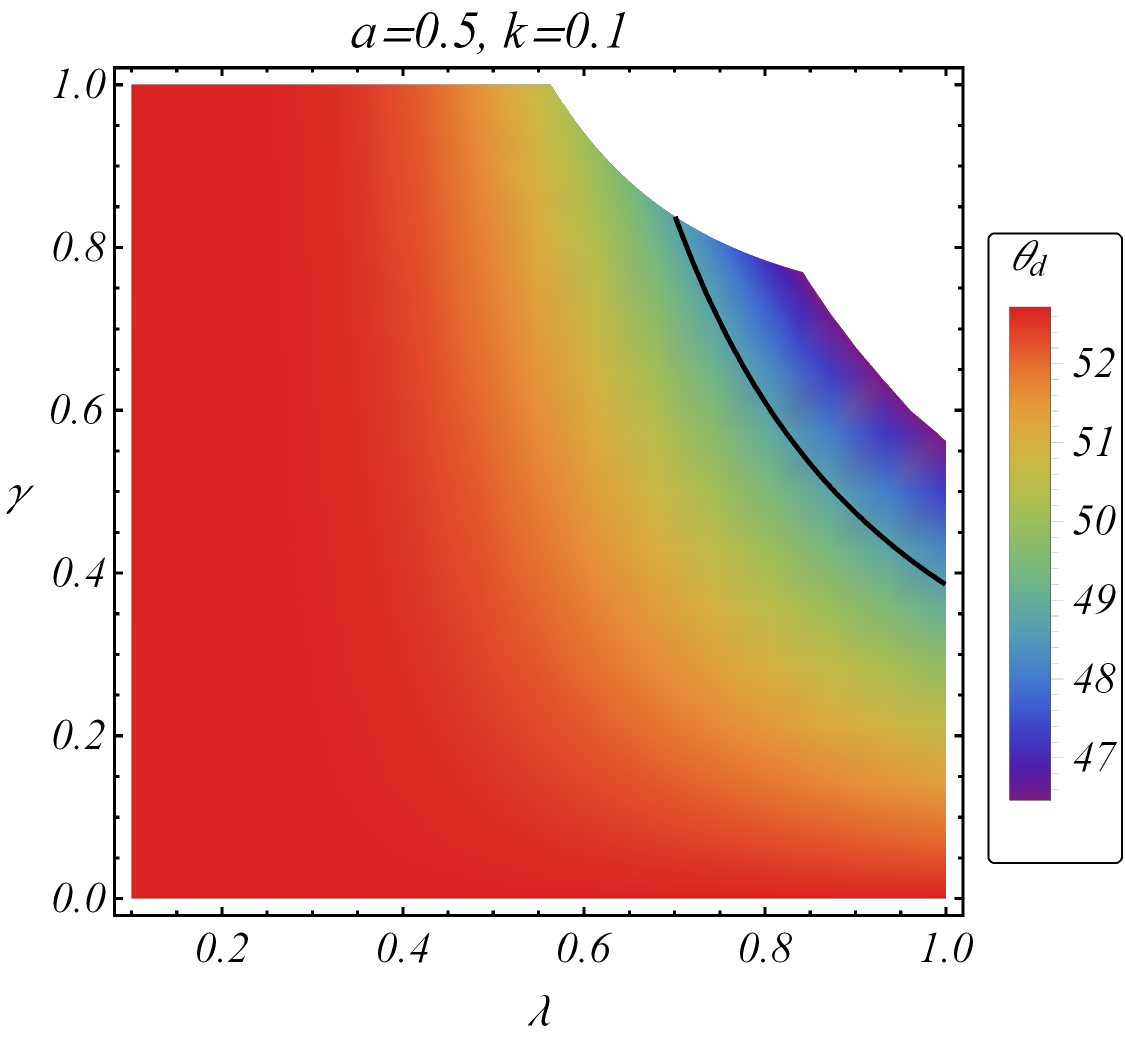}}~~~
		\subfigure{\includegraphics[width=0.41\textwidth]{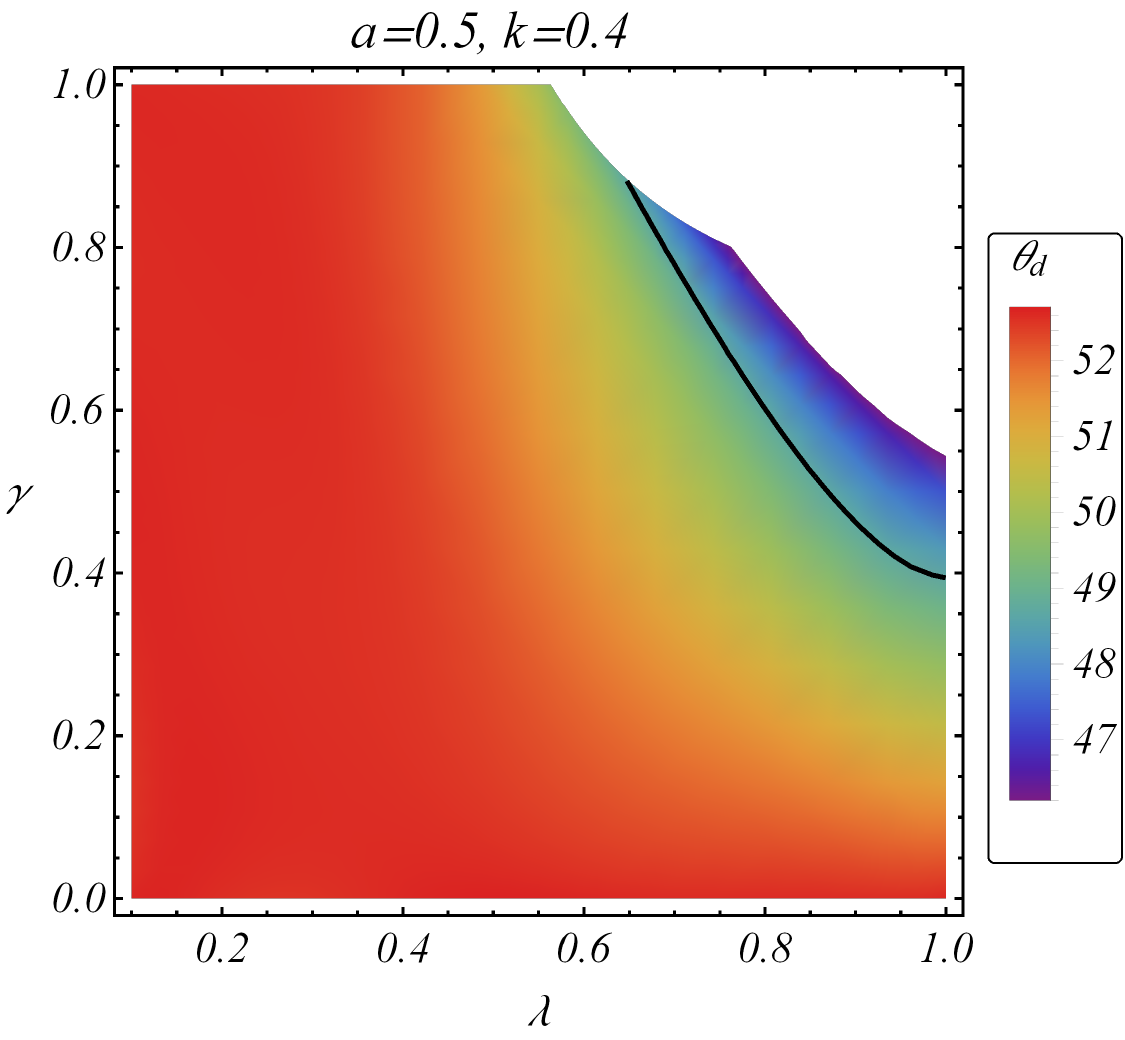}}
	\end{center}
	\caption{The density plots showing the angular size of the shadow of Sgr A* in terms of parametric spaces for various values of plasma parameters for the case described by Eq. (\ref{49b}). \label{m6}}
\end{figure}

\subsubsection{Case Ib}\label{e1b}
The parametric spaces in Fig. \ref{m3} correspond to the plots for angular diameter of M87* for the case described by the Eq. (\ref{33}). The top panel corresponds to parametric space $a$-$\gamma$ with $\lambda=0.6$ and two different values of $\omega_c$. The left plot is associated with a lower $\omega_c$ value and given the $\lambda$ value, it demonstrates that for all parametric values situated below the dashed curve, the shadow size falls within the 1-$\sigma$ error interval. Nevertheless, in the right plot, it becomes apparent that with an increase in the $\omega_c$ value, the parametric values confined by the dotted curve indicate that the angular diameter of the shadow falls within the 2-$\sigma$ level. Consequently, we consider the BH (\ref{8}) immersed in plasma medium as M87* for only the smaller value of $\omega_c$ together with the corresponding parametric values under the dashed curve. It is also obvious that the interval of $a$ and $\gamma$ bounded by the dashed curve is larger for this case as compared to that in the Fig. \ref{m1}. The panel at the middle of the figure is associated to the parametric space $a$-$\lambda$ with $\gamma=0.4$ and two different values of $\omega_c$. The left plot shows that for the given values of $\gamma$ and $\omega_c$, the intervals of $a$ and $\lambda$ are bounded by the dashed and dotted curves in the parametric space. On the other hand, in the right plot, elevating the value of $\omega$ reveals that solely the dotted curve delineates the limits on both $a$ and $\lambda$, ensuring that the shadow size resides within the 2-$\sigma$ interval. In this way, the BH (\ref{8}) in plasma background is considered as M87* only for the smaller value of $\omega_c$ corresponding to parametric values under the dashed curve. It is also observed that for this case, the interval of $a$ and $\lambda$ bounded by the dashed curve is larger as compared to that in the Fig. \ref{m1}. The lower panel corresponds to parametric space $\gamma$-$\lambda$ with $a=0.5$ and two different values of $\omega_c$. Corresponding to the given values of $a$ and $\omega_c$ in the left plot, the limits of $\gamma$ and $\lambda$ are shown by the dashed and dotted curves in the parametric space. However, in the right plot, it can be seen that by increasing the value of $\omega_c$, the bounds on $\gamma$ and $\lambda$ lies only within 2-$\sigma$ interval. Therefore, we consider the BH (\ref{8}) in plasma background as M87* for the smaller value of $\omega_c$ corresponding to the parametric values under the dashed curve. For the corresponding case in Fig. \ref{m1}, we did not find any bound on the BH parameters.

Corresponding to the Eq. (\ref{33}), the parametric spaces in Fig. \ref{m4} show the angular diameter for Sgr A*. The parametric space $a$-$\gamma$ with $\lambda=0.6$ and two different values of $\omega_c$ describe the top panel. In both plots, for the given values of $\lambda$ and $\omega_c$, the angular diameter falls inside 1-$\sigma$ level of error for every value of $a$ and $\gamma$ for which the density plot exists. Thus, the BH (\ref{8}) immersed in plasma medium can be considered as Sgr A* for both values of $\omega_c$ and all of the corresponding parametric values for which the density plots exist. The middle panel corresponds to parametric space $a$-$\lambda$ with $\gamma=0.4$ and two different values of $\omega_c$ for which the angular diameter resides inside the error interval 1-$\sigma$ for all values of $a$ and $\lambda$ for which the density plots exist. Likewise, in the lower panel in which the parametric space $\gamma$-$\lambda$ with $a=0.5$ and two different values of $\omega_c$, the angular diameter is confined inside error interval 1-$\sigma$ for every value of $\gamma$ and $\lambda$ for which the density plots exist. Therefore, the BH (\ref{8}) immersed in plasma can be considered as Sgr A* for both values of $\omega_c$ and all of the corresponding parametric values for which the density plots exist in middle and lower panels.

\subsection{Case II}\label{e2}
To construct a comparison considering the plasma distribution with refractive index given by Eq. (\ref{49b}), we determined the angular diameters of the BH shadows in terms of the parametric spaces as in the previous case.

The parametric spaces in Fig. \ref{m5} show the angular diameter for M87* and the bounds on the parameters. The top panel corresponds to the parametric space $a$-$\gamma$ with $\lambda=0.6$ and two different values of $k$. It is found that for the given values of $\lambda$ and $k$ in both plots, the limits of $a$ and $\gamma$ are shown by the dashed curve in the parametric space corresponding to which the size of shadow lies within the 1-$\sigma$ level of error. Therefore, for all these parametric values, the BH (\ref{8}) immersed in plasma is considered to be M87*. The panel lying at the middle refers to the parametric space $a$-$\lambda$ with $\gamma=0.4$ and two different values of $k$. In both plots, we can find that for the given value of $\gamma$ and $k$, the limits of $a$ and $\lambda$ are shown by the dashed and dotted curves in the parametric spaces. The 1-$\sigma$ error level encompasses the shadow size for all parametric values positioned beneath the dashed curves. Therefore, the BH (\ref{8}) in plasma background can be considered as M87* for both values of $k$ and the corresponding parametric values under the dashed curve. The lower panel corresponds to parametric space $\gamma$-$\lambda$ with $a=0.5$ and two different values of $k$. The plots in this panel exhibit an exactly identical behavior as compared to the middle panel. Therefore, the spinning BH (\ref{8}) in plasma background can be considered as M87* for both values of $k$ and the corresponding parametric values under the dashed curve. There is no significant variation in the plots with the change in the value of $k$ in all panels. Moreover, the intervals of the BH parameters bounded by the dashed curves in this case are larger than those in the all previous cases for M87*.

As a final case, we discuss the density plots in Fig. \ref{m6} that delineates the angular diameter through the parametric spaces for Sgr A*. The top, middle and lower panels corresponds to parametric spaces $a$-$\gamma$, $a$-$\lambda$ and $\lambda$-$\gamma$ with fixed values of the third BH parameter, respectively and two different values of $k$ in each column. In each plot, we can find that the angular diameter falls inside 1-$\sigma$ error level for all respective parametric values. Therefore, we determine the bounds on the parameters of the BH and for both values of $k$ from which we can deduce that the BH (\ref{8}) immersed in plasma medium can be regarded as the analog of Sgr A*.

\section{Conclusion}
In this work, we mainly emphasized on the impact of different descriptions of plasma fields on the light trajectories and optical images of the BH (\ref{8}). A meticulous discussion is presented on the effect of various plasma parameters affect describing the corresponding plasma medium. We summarize as:

\begin{itemize}
\item The BH (\ref{8}) reduces to Kerr and Kerr-Newman BHs for $\lambda=0$ and $\lambda=1$, respectively. The event horizon goes down by elevating both $\lambda$ and $\gamma$ with consistent value of the other parameter. With this increase in parameters, the extremal value of $a$ also decreases.
\item The unstable null orbits are not affected by $\theta$ coordinate because the relation (\ref{31}) depends only on $r$. For the case \ref{Ia}, the size of unstable null orbits is elevated with raise in $\omega_c$. Whereas in the following two cases, the size of unstable orbits of light decrease with elevation in $k$. This shows an entirely different behavior of the plasma media on the unstable orbits.
\item The two subcases in \ref{aa1} reveal that the shadow shrinks for both subcases with a rise in $\omega_c$. However, for the subcase \ref{S1b}, the shrinking rate of the shadow is much more than the other subcase. This shows that $\omega_c$ is more sensitive for the case \ref{S1b} than for the case \ref{s1a}. In the cases \ref{S2} and \ref{S3}, the parameter $k$ has quite unusual behavior. For the case \ref{S2}, the shadow shrinks significantly with increase in $k$. Whereas, in the case \ref{S3}, the shadow size remains consistent, instead the distortion is highly increased for the larger values of the BH parameters.
\item The distortion is diminished with increase in $\omega_c$ for the cases \ref{d1a} and \ref{d1b}. Whereas, it increases rapidly with increase in $k$ for the cases \ref{d2} and \ref{d3}. Therefore we may infer that the parameter $k$ for both respective cases supports the deviation of shadows from circularity than $\omega_c$.
\item In analogy with the shadows, the BH evaporation slows down with increase in $\omega_c$ for both respective cases. However, for the second case, the parameter $\omega_c$ is more sensitive. Moreover, for the other cases described by the parameter $k$, the first of the two cases shows that the BH evaporation decelerates with elevating value of $k$. However, for the other case, there is almost no BH evaporation.
\item A rigorous analysis shows that the BH (\ref{8}) is more probable to be considered as Sgr A* rather than M87* because the parametric bounds for Sgr A* are obtained with larger intervals than for M87*. Moreover, both smaller and larger values of plasma parameters conform to the findings by the EHT collaboration for Sgr A*. However, for many cases, the larger value of plasma parameter did not support the EHT data for M87*. We also propose that if there exists a plasma medium around the BH described by the metric (\ref{8}) that is considered as Sgr A*, then this plasma medium is described the Eq. (\ref{32}). It is because for this case, the angular diameters of the shadow are more convergent in 1-$\sigma$ interval.
\end{itemize}

We conclude that all of the plasma fields have a significant impact on the unstable orbits, shadows and BH evaporation rate. The spinning BH in KR gravity can be considered as Sgr A* instead of M87*. Among all plasma fields considered in this work, it is more likely that the plasma field described by the Eq. (\ref{32}) is surrounding Sgr A*. However, by looking at the variation in shadows, distortion and BH evaporate rate, we may infer that the parameter $\omega_c$ described by the Eq. (\ref{33}) is more sensitive among all cases. Following this attempt, numerous potential research directions may be pursued in the future such as the observation and experimental verification of the type of the plasma medium present around Sgr A*. It can also be investigated whether our prediction on the BH being not considered as M87* is true or false. For the people working in areas of quantum aspects of gravity and high energy astroparticle physics, it can be useful to consider the KR field together with a plasma medium to discover and comprehend the subject.

\end{document}